\documentclass[12pt]{amsart}
\usepackage{latexsym,amsmath,amssymb,amscd,eucal}
\usepackage{xypic}
\usepackage{mathrsfs}
\usepackage[all]{xy}
\usepackage{color}

\title[Algebraic deformations of toric varieties II]
{Algebraic deformations of toric varieties II. \\[5pt] Noncommutative instantons}

\date{June 2011 \hfill{HWM--11--19 , \quad EMPG--11--18 }}

\author{Lucio Cirio}

\address{Grupo de Fisica Matem\'{a}tica da Universidade de Lisboa (GFM-UL),
Instituto para a Investiga\c{c}\~{a}o Interdisciplinar,
Av. Prof. Gama Pinto 2, 1649-003 Lisboa, Portugal}

\address{Max Planck Institute for Mathematics, Vivatsgasse 7, 53111 Bonn, Germany}

\email{lucio.cirio@gmail.com}

\author{Giovanni Landi}

\address{Dipartimento di Matematica e Informatica, Universit\`a di
 Trieste, Via A. Valerio 12/1, I-34127 Trieste, Italy and INFN,
 Sezione di Trieste, Trieste, Italy}

\email{landi@univ.trieste.it}

\author{Richard J. Szabo}

\address{Department of Mathematics, Heriot-Watt University, Colin
  Maclaurin Building, Riccarton, Edinburgh EH14 4AS, U.K. and Maxwell
  Institute for Mathematical Sciences, Edinburgh, U.K.}

\email{R.J.Szabo@ma.hw.ac.uk}

\newcommand{\gl}[1]{\mathfrak{gl}(#1)}

\newcommand{\Proof}[1]{\noindent\underline{\textsf{Proof}}: #1 \hfill
  $\blacksquare$\\}

\newcommand{\newsection}{\setcounter{equation}{0}\section}

\def\appendix#1{\addtocounter{section}{1}\setcounter{equation}{0}
\renewcommand{\thesection}{\Alph{section}}
\section*{Appendix \thesection\protect\indent \parbox[t]{11.715cm} {#1}}
\addcontentsline{toc}{section}{Appendix \thesection\ \ \ #1} }
\newcommand{\eq}{\begin{equation}}
\newcommand{\eqend}{\end{equation}}

\newbox\ncintdbox \newbox\ncinttbox
\setbox0=\hbox{$-$} \setbox2=\hbox{$\displaystyle\int$}
\setbox\ncintdbox=\hbox{\rlap{\hbox to
\wd2{\hskip-.125em\box2\relax \hfil}}\box0\kern.1em}
\setbox0=\hbox{$\vcenter{\hrule width 4pt}$}
\setbox2=\hbox{$\textstyle\int$}
\setbox\ncinttbox=\hbox{\rlap{\hbox to
\wd2{\hskip-.175em\box2\relax \hfil}}\box0\kern.1em}

\def\={\ =\ }

\newcommand{\Open}{{\sf Open}}

\newcommand{\Module}{{\mathscr{M}}}
\newcommand{\Homfun}{{\mathscr{H}}}
\newcommand{\coh}{{\sf coh}}
\newcommand{\Alg}{{\sf Alg}}
\newcommand{\Set}{{\sf Set}}
\newcommand{\tor}{{\sf tor}}
\newcommand{\gr}{{\sf gr}}
\newcommand{\qgr}{{\sf coh}}

\newcommand{\complex}{{\mathbb C}} 
\newcommand{\zed}{{\mathbb Z}} 
\newcommand{\nat}{{\mathbb N}} 
\newcommand{\real}{{\mathbb R}} 
\newcommand{\rat}{{\mathbb Q}} 
\newcommand{\hyper}{{\mathbb H}} 
\def\alg{{\mathcal A}}

\def\calg{{\mathcal C}}
\def\galg{{\mathcal G}}
\def\hil{{\mathcal H}}

\def\bun{{\mathcal E}}

\def\comp{{\mathcal K}}
\def\sheaf{{\mathcal O}}
\def\Qcal{{\mathcal Q}}
\def\Dcal{{\mathcal D}}
\def\Mcal{{\mathcal M}}
\def\Ncal{{\mathcal N}}
\def\Rcal{{\mathcal R}}
\def\Tcal{{\mathcal T}}
\def\Scal{{\mathcal S}}
\def\Qcal{{\mathcal Q}}

\def\Fcal{{\mathcal F}}
\def\Xcal{{\mathcal X}}

\def\Ical{{\mathcal I}}
\def\Iscr{{\mathscr I}}

\def\Jscr{{\mathscr J}}

\def\gl{{\mathfrak{gl}}}

\def\rank{{\rm rank}}

\def\P{{\mathbb{P}}}

\def\CP{{\mathbb{CP}}}

\def\U{{\rm U}}
\def\im{{\rm im}}
\def\coker{{\rm coker}}

\def\Ext{{\rm Ext}}
\def\Hom{{\rm Hom}}
\def\End{{\rm End}}
\def\Aut{{\rm Aut}}
\def\Homc{\underline{\mathcal{H}{\rm om}}}
\def\Extc{\underline{\mathcal{E}{\rm xt}}}

\def\ch{{\rm ch}}

\def\Id{{\rm id}}

\def\Hilb{{\sf Hilb}}
\def\Sym{{\sf Sym}}

\newcommand{\Tr}[1]{\:{\rm Tr}\,#1}

\def\be{\begin{equation}}
\def\ee{\end{equation}}
\def\bea{\begin{eqnarray}}
\def\eea{\end{eqnarray}}
\def\bd{\begin{displaymath}}
\def\ed{\end{displaymath}}

\def\dd{{\rm d}}

\def\ii{{\,{\rm i}\,}}

\makeatletter
\newdimen\normalarrayskip              
\newdimen\minarrayskip                 
\normalarrayskip\baselineskip
\minarrayskip\jot
\newif\ifold             \oldtrue            
\def\arraymode{\ifold\relax\else\displaystyle\fi} 
\def\@arrayskip{\ifold\baselineskip\z@\lineskip\z@
     \else
     \baselineskip\minarrayskip\lineskip2\minarrayskip\fi}
\def\@arrayclassz{\ifcase \@lastchclass \@acolampacol \or
\@ampacol \or \or \or \@addamp \or
   \@acolampacol \or \@firstampfalse \@acol \fi
\edef\@preamble{\@preamble
  \ifcase \@chnum
     \hfil$\relax\arraymode\@sharp$\hfil
     \or $\relax\arraymode\@sharp$\hfil
     \or \hfil$\relax\arraymode\@sharp$\fi}}
\def\@array[#1]#2{\setbox\@arstrutbox=\hbox{\vrule
     height\arraystretch \ht\strutbox
     depth\arraystretch \dp\strutbox
     width\z@}\@mkpream{#2}\edef\@preamble{\halign \noexpand\@halignto
\bgroup \tabskip\z@ \@arstrut \@preamble \tabskip\z@ \cr}%
\let\@startpbox\@@startpbox \let\@endpbox\@@endpbox
  \if #1t\vtop \else \if#1b\vbox \else \vcenter \fi\fi
  \bgroup \let\par\relax
  \let\@sharp##\let\protect\relax
  \@arrayskip\@preamble}
\makeatother

\newcommand{\beq}{\begin{eqnarray}}
\newcommand{\eeq}{\end{eqnarray}}

\newcommand{\GL}{{\rm GL}}
\newcommand{\Gr}{\mathbb{G}{\rm r}}
\newcommand{\Fl}{\mathbb{F}{\rm l}}

\def\appendix#1{\addtocounter{section}{1}\setcounter{equation}{0}
\renewcommand{\thesection}{\Alph{section}}
\section*{Appendix \thesection. #1}
\addcontentsline{toc}{section}{Appendix \thesection\ \ \ #1} }

\newcommand{\dslash}{\not{\hbox{\kern-2pt $\partial$}}}
\newcommand{\pslash}{\not{\hbox{\kern-2.3pt $p$}}}

 \newtoks\nslashfraction
 \nslashfraction={.13}
 \newcommand{\nslash}[1]{\setbox0\hbox{$ #1 $}
   \setbox0\hbox to \the\nslashfraction\wd0{\hss \box0}/\box0 }

\def\ii{{\,{\rm i}\,}}

\newtheorem{theorem}[equation]{Theorem}
\newtheorem{lemma}[equation]{Lemma}
\newtheorem{cor}[equation]{Corollary}
\newtheorem{proposition}[equation]{Proposition}
\newtheorem{definition}[equation]{Definition}

\newtheorem{remark}[equation]{Remark}

\numberwithin{equation}{section}

\begin{document}

\maketitle

\begin{abstract}
We continue our study of the noncommutative algebraic and differential
geometry of a particular class of deformations of toric varieties,
focusing on aspects pertinent to the construction and
enumeration of noncommutative instantons on these varieties. We
develop a noncommutative version of twistor theory, which introduces
a new example of a noncommutative four-sphere. We
develop a braided version of the ADHM construction and show that it
parametrizes a certain moduli space of framed torsion free
sheaves on a noncommutative projective plane. We use these
constructions to explicitly build instanton gauge bundles with
canonical connections on the noncommutative four-sphere that satisfy appropriate
anti-selfduality equations. We construct projective moduli spaces for
the torsion free sheaves and demonstrate that they are smooth. We
define equivariant partition functions of these
moduli spaces, finding that they coincide with the usual
instanton partition functions for supersymmetric gauge theories on~$\complex^2$.
\end{abstract}

\tableofcontents

\parskip 1ex

\renewcommand{\thefootnote}{\arabic{footnote}}
\setcounter{footnote}{0}

\newsection*{Introduction}

This paper is the second part of a series of articles devoted to the
construction and study of new noncommutative deformations of toric
varieties. In the first part~\cite{CLSI} the general theory was developed. In the
present work we elaborate on and extend some of these developments, and in particular derive a theory of instantons on the noncommutative projective
planes $\CP_\theta^2$ constructed in~\cite{CLSI} in several different
contexts.

In the commutative
situation, moduli spaces of framed sheaves on the complex projective plane have
been studied intensively due to their connection with moduli spaces of framed instantons on the four-sphere; they are the basis for \emph{instanton counting}. Generally, the Hitchin--Kobayashi correspondence establishes
an identification between the moduli space of anti-selfdual
irreducible connections on a hermitean vector bundle $E$ over a
K\"ahler surface $X$ and the set of equivalence classes of stable
holomorphic bundles over $X$ which are topologically equivalent to
$E$. Instanton counting consists in computing BPS invariants of
supersymmetric gauge theories in terms of ``integrals'' over
equivariant cohomology classes of the moduli spaces. Equivariant
cohomology groups of moduli spaces of framed sheaves on generic toric
surfaces are much less well-understood in general than those of $\CP^2$. 

One of the main results of this paper is a detailed description of the moduli
space of torsion free sheaves on the noncommutative projective planes 
$\CP_\theta^2$ with a trivialization on a noncommutative line ``at
infinity''. We will identify points in this moduli space with classes
of instantons on $\CP_\theta^2$. We shall establish a bijective correspondence between
torsion free sheaves on $\CP_\theta^2$ and certain sets of
braided ADHM data, which correspond to stable representations of the
ADHM quiver with a certain $q$-deformation of the usual relation. We will study some details of the
corresponding moduli spaces, which have good properties analogous to
those of the commutative case making them tractable for uses in
instanton counting problems, provided that one considers them as
members of flat families in an appropriate sense~\cite{NS}. This is in
the same spirit as our general noncommutative deformations which
occur as members of flat families of quantizations of toric varieties, and it
agrees with general expectations~\cite{LPRvS,BL,BL1} that instanton moduli 
spaces in the $\theta$-deformed case produce ``families'' of
instantons. Our results contain, in particular, an explicit
realization of the construction of instantons on a $q$-deformed
euclidean space $\real_q^4$ 
{similar to the one} 
sketched in~\cite[\S9]{KKO},
our $q$-deformation {being}
somewhat different.

Along the way we encounter some new constructions. A new noncommutative
twistor theory is developed. In particular, we construct a new noncommutative
four-sphere $S_\theta^4$ and describe some of its properties. Our construction of noncommutative instantons is partly inpired by the gluing construction of Frenkel and Jardim~\cite{FJ}, although our noncommutative spaces and instanton gauge fields are rather different. Our analysis thus extends the existing examples of noncommutative instantons and spaces. Although our instanton moduli spaces are generically (commutative) deformations of those in the classical case, we will find that the equivariant counting problems, which define instanton partition functions of supersymmetric gauge theories, coincide with those of the classical limit in this case.

It is hoped that the constructions of noncommutative instantons presented here can be extended to more complicated toric varieties in four dimensions, such as ALE spaces or Hirzebruch surfaces, and ultimately to generalized instantons of six-dimensional noncommutative toric geometries pertinent to a noncommutative version of Donaldson--Thomas theory. The development of such generalizations would further extend the uses of instanton counting in enumerative geometry, and could possibly lead to new classes of enumerative invariants. They may also lead to new examples of BPS states in supersymmetric gauge theories and string theory. See~\cite{CLSI} and~\cite{Szabo} for further motivation and background
behind our constructions.

\subsection*{Outline of paper}

In \S\ref{sec:braidedsym} we review and extend the quantization of toric varieties introduced in~\cite{CLSI}. Our treatment follows the formalism of cocycle twist quantization and noncommutative geometry in braided monoidal categories, see e.g.~\cite{BL1}.

In \S\ref{sec:ncproj} we introduce noncommutative projective varieties following~\cite{CLSI}, focusing on the noncommutative projective spaces $\CP_\theta^n$. We study coherent sheaves on these varieties and their invariants, and develop a theory of noncommutative monad complexes.

In \S\ref{sec:nctwistor} we develop a new noncommutative version of twistor theory in four dimensions following~\cite{CLSI}. We construct a new example of a noncommutative sphere $S_\theta^4$ and corresponding twistor transforms, correspondences and fibrations. Though not developed here, it would be extremely interesting to investigate further the geometry of this noncommutative sphere, such as its cyclic cohomology and how it fits as the base of a noncommutative Hopf fibration whose total space is a seven-sphere.

In \S\ref{Instcounting} we establish a bijection between framed torsion free sheaves on $\CP_\theta^2$ and a certain deformation of the usual ADHM construction. Our construction mimicks that of the commutative case and relies on many results of~\cite{KKO} and~\cite{BGK} which were obtained in an analogous but rather different setting.

In \S\ref{Instconn} we explicitly construct instanton gauge bundles and canonical connections on $S_\theta^4$ using our noncommutative twistor correspondence and ADHM constructions. We verify that these connections satisfy appropriate anti-selfduality equations with respect to a natural metric, thus justifying them as bonafide noncommutative instantons. We do not develop a suitable (Yang--Mills) gauge theory for these instantons in this paper.

In \S\ref{Instmodsp} we use results of Nevins and Stafford~\cite{NS}
to construct corresponding moduli spaces of noncommutative instantons
and demonstrate that they are smooth. We work out the corresponding
deformation theory and explicitly compute the tangent spaces in terms
of our braided ADHM construction. We work through many explicit
examples which illustrate in what sense these moduli spaces are deformations of their classical counterparts, including a (commutative) deformation of the Hilbert scheme
of points.

Finally, in \S\ref{sec:gaugetheory} we use our moduli space constructions to compute standard equivariant counting functions for our noncommutative instantons with respect to a natural torus action on the instanton moduli space. We find a combinatorial classification of the torus fixed points in moduli space and show that it coincides with that of the classical limit. We comment on how it may be possible to construct instanton partition functions which capture more deeply the deformation of the moduli spaces.

\subsection*{Conventions and notation}

Unless otherwise indicated, all tensor products are taken over the
base field of complex numbers $\complex$. All varieties considered are
reduced separated schemes of finite type over $\complex$. All algebras
are associative over $\complex$.

\subsection*{Acknowledgments}

We thank Roger Bielawski, Simon Brain, Ken Brown, Alastair Craw, Iain Gordon,
Harald Grosse, Derek Harland, Chiara Pagani, Tony Pantev, Christian S\"amann and David Treumann for helpful
discussions and correspondence. The work of GL was partially supported by the Italian
Project ``Cofin08--Noncommutative Geometry, Quantum Groups and
Applications''. The work of RJS was supported in part by grant ST/G000514/1 ``String Theory
Scotland'' from the UK Science and Technology Facilities Council. 

\newsection{Braided symmetries of noncommutative toric varieties\label{sec:braidedsym}}

\subsection{Cocycle deformations of the algebraic torus}\label{se:cdal}

Let $L$ be a lattice of rank $n$, and let
$T=L\otimes_\zed\complex^\times$ be the associated algebraic torus of
dimension $n$ over $\complex$. The Pontrjagin dual group
$\widehat{T}=\Hom_\complex(T,\complex^\times)$ is the group
of characters $\{\chi_p\}_{p\in L^*}$ parametrized by elements of the
dual lattice $L^*=\Hom_\zed(L,\zed)$. The dual pairing $L^*\times
L\to\zed$ between
lattices is denoted $(p,v)\mapsto p\cdot v$. Upon fixing a $\zed$-basis $e_1,\dots,e_n$ for
$L$, with corresponding dual basis $e_1^*,\dots,e_n^*$ for $L^*$, one has
$T\cong(\complex^\times)^n$ and $\widehat{T}\cong\zed^n$. For $p=\sum_i\,p_i\,e^*_i\in L^*$ and
$t=\sum_i\,e_i\otimes t_i\in T$, the characters are given by
\beq
\chi_p(t)=t^p:=t_1^{p_1}\cdots t_n^{p_n} \ . 
\label{Tchar}\eeq

The unital algebra $\hil=\alg(T)$ of coordinate functions on the torus $T$ is the Laurent
polynomial algebra
$$
\hil:=\complex(t_1,\dots,t_n)
$$
generated by elements $t_i$, $i=1,\dots,n$. It is equipped
with the Hopf algebra structure
$$
\Delta(t^p)=t^p\otimes t^p \ , \qquad \epsilon(t^p)=1 \ , \qquad
S(t^p)=t^{-p}
$$
for $p\in L^*$, with the coproduct and the counit respectively extended as algebra
morphisms $\Delta:\hil\to \hil\otimes \hil$ and $\epsilon:\hil\to\complex$, and the
antipode as an anti-algebra morphism $S:\hil\to \hil$. The canonical right
action of $T$ on itself by group multiplication dualizes to give a
left {$\hil$-}coaction
\beq
&&
\Delta_L\,:\, \alg(T)~\longrightarrow~ \hil\otimes \alg(T) \ , \qquad
\Delta_L(u_i)= t_i\otimes u_i \ , \quad \Delta_L\big(u_i^{-1} \big)=
t_i^{-1}\otimes u_i^{-1} \ ,
\label{DeltaLT}\eeq
where we write $u_i$, $u_i^{-1}$, $i=1,\dots,n$ for the generators of $\alg(T)$
viewed as a left comodule algebra over itself, 
when distinguishing the coordinate algebra $\alg(T)$ from the Hopf algebra $\hil$. 
This coaction is
equivalent to a grading of the algebra $\alg(T)$ by the dual lattice
$L^*$, for which the homogeneous elements are the characters
(\ref{Tchar}).

A two-cocycle $F_\theta:\hil\otimes \hil\to\complex$ on $\hil$ is defined by choosing a
complex skew-symmetric $n\times n$ matrix $\theta=(\theta^{ij})$,
regarded as a homomorphism $\theta:L^*\to L\otimes_\zed\complex$,
setting
$$
F_\theta(t^p,t^q)=\exp\big(\mbox{$\frac\ii2$}\,
p\cdot\theta q\big)
$$
on characters $t^p,t^q\in \hil$, and extending by linearity. The map $F_\theta$
is a convolution-invertible Hopf bicharacter and obeys
$$
F_\theta\circ(S\otimes \Id)=F_\theta^{-1}= F_\theta\circ(\Id\otimes S) \ , \qquad
F_\theta\circ(S\otimes S)=F_\theta \ .
$$
It follows that $F_\theta$ is completely determined by its values on
generators
$$
F_\theta(t_i,t_j)= \exp\big(\mbox{$\frac\ii2$}\,\theta^{ij}\big) =: q_{ij}
$$
for $i,j=1,\dots,n$.

Given such a map, one constructs the cotwisted Hopf
algebra $\hil_\theta$ which as a coalgebra is the same as $\hil$ but which
generally has a modified product and antipode. In the present case one
easily finds that the product and antipode are in fact undeformed by
$F_\theta$, so $\hil=\hil_\theta$ as a Hopf algebra. On the other hand, $\hil$ and
$\hil_\theta$ differ as coquasitriangular Hopf algebras. While the Hopf algebra $\hil$ naturally carries the
trivial coquasitriangular structure
$\Rcal=\epsilon\otimes\epsilon:\hil\otimes\hil\to \complex$, the cotwisted Hopf algebra $\hil_\theta$ has twisted coquasitriangular
structure given by the convolution-invertible Hopf bicharacter
$\Rcal_\theta:\hil_\theta\otimes\hil_\theta\to\complex$ defined 
as $\Rcal_\theta=F_\theta^{-2}$; on generators one has explicitly
$$
\Rcal_\theta(t_i,t_j)= F_\theta(t_j,t_i)\, F_\theta^{-1}(t_i,t_j) =
F_\theta^{-2}(t_i,t_j) =q_{ij}^{-2} \ .
$$
On the other hand, the algebra structure of $\alg(T)$ gets deformed to
a noncommutative product; this is
a particular instance of the quantization functor we describe in \S\ref{toriccat}.

\subsection{Toric symmetries in braided monoidal categories\label{toriccat}}

A left $\hil$-comodule structure on a vector space $V$ will be denoted
as in (\ref{DeltaLT}) by $\Delta_L:V\to \hil\otimes V$, together with a Sweedler
notation $\Delta_L(v)=v^{(-1)}\otimes v^{(0)}$ for $v\in V$ and with
implicit summation. Let ${}^\hil\Module$ denote the additive category of left
$\hil$-comodules. A map $V\xrightarrow{\ \sigma\ }W$ is a morphism of this
category if and only if it is $\hil$-coequivariant, i.e. it sits in
the commutative diagram
$$
\xymatrix{
V~\ar[r]^\sigma \ar[d]_{\Delta_L} & ~ W
\ar[d]^{\Delta_L} & \\
\hil\otimes V~\ar[r]_{\Id\otimes\sigma} & ~ \hil\otimes W
}
$$
or more explicitly $v^{(-1)}\otimes
\sigma(v^{(0)})=\sigma(v)^{(-1)}\otimes\sigma(v)^{(0)}$ for all $v\in V$. The category ${}^\hil\Module$ has a natural monoidal structure given by the
tensor product coaction
$$
\Delta_{V\otimes W}(v\otimes w) = v^{(-1)}\, w^{(-1)}\otimes\big(v^{(0)}\otimes w^{(0)}\big)
$$
for $v\in V$, $w\in W$.
With the trivial coquasitriangular structure
$\Rcal=\epsilon\otimes\epsilon$ on the Hopf algebra $\hil$, the
category ${}^\hil\Module$ 
{of left $\hil$-comodules} is
{(trivially)}
braided by the collection $\Psi=\{\Psi_{V,W}:V\otimes W\to W\otimes
V\}$ of functorial flip isomorphisms
$\Psi_{V,W}(v\otimes w)=w\otimes v$ for each pair of objects $V,W$ of
${}^\hil\Module$, and for all $v\in V$ and $w\in W$.

With the deformed coquasitriangular
structure $\Rcal_\theta=F_\theta^{-2}$ on the cotwisted Hopf algebra
$\hil_\theta$, the monoidal category ${}^{\hil_\theta}\Module$ of
left $\hil_\theta$-comodules is braided by the collection of
functorial isomorphisms $\Psi^\theta=\{\Psi^\theta_{V,W}:V\otimes W\to W\otimes V\}$, where
\beq
\Psi^\theta_{V,W}(v\otimes w)=F^{-2}_\theta\big(w^{(-1)}\,,\,v^{(-1)}\big)\,
w^{(0)}\otimes v^{(0)}
\label{PsithetaVW}\eeq
for each pair $V,W$ of left $\hil_\theta$-comodules with $v\in V$ and
$w\in W$. Since the functor $\Psi^\theta$ is an involution,
i.e. $\Psi^\theta\circ\Psi^\theta$ is isomorphic to the identity functor of the
category ${}^{\hil_\theta}\Module$, the braiding is symmetric and
makes $ {}^{\hil_\theta}\Module$ into a tensor category. In particular, if $A$ and $B$ are left
$\hil_\theta$-comodule algebras with product maps $\mu_A:A\otimes
A\to A$ and $\mu_B:B\otimes B\to B$, then one can define the braided
tensor product algebra $A\,{\otimes}_\theta\,B$ to be the
vector space $A\otimes B$ with the
product map
$\mu_{A\,{\otimes}_\theta\,B}:(A\,{\otimes}_\theta\,B)\otimes(
A\,{\otimes}_\theta\,B) \to A\,{\otimes}_\theta\,B$ given by
$$
\mu_{A\,{\otimes}_\theta\,B}=\big(\mu_A\otimes\mu_B\big)\circ \big(\Id_A\otimes
\Psi^\theta_{B,A} \otimes \Id_B\big) \ .
$$
The resulting algebra is an object of the category ${}^{\hil_\theta}\Module$ by the tensor
product coaction.

The deformation in passing from $\hil$ to $\hil_\theta$
takes the form of a functorial isomorphism $\mathscr{F}_\theta:
{}^{\hil}\Module\to {}^{\hil_\theta}\Module$ of braided monoidal
  categories. The functor $\mathscr{F}_\theta$ acts as the identity
  on objects and morphisms of ${}^{\hil}\Module$, but defines a new
    monoidal structure on the category ${}^{\hil_\theta}\Module$ by
$$
\lambda_\theta\,:\, \mathscr{F}_\theta(V)\otimes \mathscr{F}_\theta(W) ~\longrightarrow~
\mathscr{F}_\theta(V\otimes W) \ , \qquad \lambda_\theta(v\otimes w) =
F_\theta\big(v^{(-1)}\,,\, w^{(-1)}\big)\, v^{(0)}\otimes w^{(0)} \ .
$$
This makes $\mathscr{F}_\theta$ into a monoidal functor which
intertwines the braidings in ${}^{\hil}\Module$ and
  ${}^{\hil_\theta}\Module$, given respectively by the flip functor
  $\Psi$ and the functor $\Psi^\theta$ defined in (\ref{PsithetaVW}).

The quantization functor $\mathscr{F}_\theta$ simultaneously deforms
all $\hil$-coequivariant constructions to corresponding versions which are
coequivariant under $\hil_\theta$. For example, if $A$ is an algebra in
the category ${}^{\hil}\Module$, then the functor $\mathscr{F}_\theta$
  takes its product map $\mu_A:A\otimes A\to A$ to a map
  $\mathscr{F}_\theta(\mu_A):\mathscr{F}_\theta(A\otimes A)\to A_\theta:=
  \mathscr{F}_\theta(A)$. Composing this morphism with $\lambda_\theta$ gives
  rise to a new product map
$$
\mu_{A_\theta}:= \mathscr{F}_\theta(\mu_A)\circ \lambda_\theta\,:\,
A_\theta\otimes A_\theta~\longrightarrow~ A_\theta
$$
with
\beq
a\star_\theta b:=\mu_{A_\theta}(a\otimes b)=F_\theta\big(a^{(-1)}\,,\,
b^{(-1)}\big)~ \mu_A\big(a^{(0)}\otimes b^{(0)}\big) \ ,
\label{startheta}\eeq
which automatically makes $A_\theta$ into an $\hil_\theta$-comodule
algebra. If $A$ is a commutative algebra, then the algebra $A_\theta$
is no longer commutative in general but only 
\emph{braided commutative},
i.e. $\mu_{A_\theta}=\mu_{A_\theta}\circ
\Psi^\theta_{A_\theta,A_\theta}$.

Another standard construction in braided monoidal categories yields
deformations of exterior algebras. If $V$ is a finite-dimensional object of the category
${}^{\hil_\theta}\Module$ of left $\hil_\theta$-comodules, then the
exterior algebra of $V$ in degree $d$ is given by~\cite{CLSI}
\begin{equation}
\label{qest}
\mbox{$\bigwedge^d_{\theta}$}\,V := V^{\otimes d }\,\big/\,\big\langle
v_1\otimes v_2 + \Psi_{V,V}^\theta(v_1\otimes v_2) \big\rangle_{v_1,v_2 \in
  V} \ .
\end{equation}
For $\theta=0$ we recover the usual exterior algebra $\bigwedge^dV$ of
the vector space $V$, while for
$\theta\neq 0$ we obtain a braided skew-commutative algebra
$\bigwedge^d_{\theta}\,V$. We also write
\beq
\mbox{$\bigwedge^\bullet_{\theta}$}\,V := T(V)\,\big/\,\big\langle
v_1\otimes v_2 + \Psi_{V,V}^\theta(v_1\otimes v_2) \big\rangle_{v_1,v_2 \in V}
\label{braidedtensoralg}\eeq
where
{$T(V)=\bigoplus_{n\geq0}\,V^{\otimes n}$ }
is the free
tensor algebra of the complex vector space $V$ with
${V^0}:=\complex$.

\subsection{Quantization of toric varieties\label{subsec:Toricquant}}

As already mentioned at the end of \S\ref{se:cdal},
for $\alg=\alg(T)$ regarded as a left comodule
algebra over itself, the functorial quantization constructed in \S\ref{toriccat}
twists the {algebra} multiplication into a new product given by
combining (\ref{DeltaLT}) and (\ref{startheta}) to get
$$
u_i\star_\theta u_j=F_\theta(t_i,t_j)~ u_i\,u_j =q_{ij}~ u_i\,u_j \ .
$$
Let $\alg(T_\theta)$ be the Laurent polynomial algebra generated by $u_i$ with
this product. It has relations
$$
u_i\star_\theta u_j=q_{ij}^2~ u_j\star_\theta u_i \ , \qquad
u_i^{-1}\star_\theta u_j=q_{ij}^{-2} ~u_j\star_\theta u_i^{-1}
$$
for each $i,j=1,\dots,n$. This quantizes the torus $T$ into the
noncommutative algebraic torus
$T_\theta=(\complex_\theta^\times)^n$ {dual to the algebra $\alg(T_\theta)$}. In the sequel we drop the
star-product symbols $\star_\theta$ from the notation for simplicity.

A toric variety $X$ of dimension $n$ is a complex algebraic variety with an
algebraic action of the torus $T=(\complex^\times)^n$ and a $T$-equivariant injection
$T\hookrightarrow X$ with dense image in the Zariski topology, where $T$ acts on itself by group
multiplication. In~\cite{CLSI} we constructed a natural flat family of
quantizations $X\to X_\theta$, with dual algebras $\alg(X_\theta)$, 
over the coordinate algebra $\alg(\bigwedge^2T)=\complex(q_{ij} \, , \, 1\leq
i<j\leq n)$ of the algebraic torus
$\bigwedge^2T:=\Hom_\zed(\bigwedge^2L^*,\complex^\times)$ of
dimension $\frac12\,n\,(n-1)$. The original commutative toric variety
$X=X_{\theta=0}$ is the fibre over the identity element of
$\alg(\bigwedge^2T)$. Since the Zariski tangent space to the family at the
identity is naturally isomorphic to $\Hom_\zed(\bigwedge^2L^*,\complex)$, the family is universal for torus coinvariant
noncommutative deformations in the following sense. Let ${\sf Alg}$ be
the category of commutative unital noetherian $\complex$-algebras and
${\sf Set}$ the category of sets. Let $\mathscr{F}_X:{\sf Alg}\to{\sf
  Set}$ be the covariant functor which sends an algebra $A$ to the set
$\mathscr{F}_X(A)=\alg(X_A)$ of algebras dual to a flat family of
$T$-coinvariant deformations $X\to X_A$ parametrized by $A$. Then
there is a unique algebra homomorphism $\alpha:\alg(\bigwedge^2T)\to
A$ such that $\mathscr{F}_X(A)\cong
\mathscr{F}_X(\alpha)(\alg(X_\theta))$. Whence the pair
$\big(\alg(\bigwedge^2T)\,,\, \alg(X_\theta) \big)$ is a universal object representing the functor $\mathscr{F}_X$, and the
toric variety $X$ thus has a fine moduli space $\bigwedge^2T$ of toric noncommutative
deformations of dimension $\frac12\,n\,(n-1)$
(see~\cite[\S2.2]{ingalls}).

Noncommutative affine toric varieties
correspond to finitely-generated
$\hil_\theta$-comodule subalgebras of the algebra $\alg(T_\theta)$ of the
noncommutative torus. Using the combinatorics of fans, one can glue
these subalgebras together via algebra automorphisms in the category
${}^{\hil_\theta}\Module$ to form global noncommutative toric
varieties; see~\cite[\S3]{CLSI} for explicit constructions. The cones of the fan define the subcategory $\Open(X_\theta)$
of toric open sets of the noncommutative variety $X_\theta$. In
particular, the association of cones with $\hil_\theta$-comodule
subalgebras of $\alg(T_\theta)$ defines the structure sheaf
$\sheaf_{X_\theta}$ of noncommutative $\complex$-algebras on
$\Open(X_\theta)$. The corresponding category of coherent sheaves of
right $\sheaf_{X_\theta}$-modules on $\Open(X_\theta)$ is denoted
$\coh(X_\theta)$. While $\coh(X_\theta)\ncong\coh(X)$ in general, the coequivariant
homological algebra of the noncommutative toric variety $X_\theta$ always
coincides with that of the commutative fibre~$X$ (see~\cite[\S7]{ingalls}).

The data defining a noncommutative deformation of a toric variety also
defines a circle bundle over the (maximally) compact dual torus
$T_\real^*=(L^*\otimes_\zed\real)/L^*\cong{\rm U}(1)^n$, as both are given by pairings on
the character lattice $L^*$. A basic operation in the theory of
constructible sheaves yields ``twisted'' sheaves on $T_\real^*$
associated to this circle bundle, i.e. sheaves on the total space of
the bundle whose monodromy around each fibre is given by a parameter
$\lambda\in\complex^\times$. A version of the coherent-constructible
correspondence~\cite{FTLZ} then takes a coherent sheaf on the
noncommutative toric variety $X_\theta$ to a complex of twisted
constructible sheaves on the dual torus $T_\real^*$. These twisted
complexes are equivalent to data in an analogue of the Fukaya
category of lagrangian submanifolds of the cotangent
bundle $\Tcal^*$ of $T_\real^*$~\cite{NZ}; in this sense the correspondence
is a version of Kontsevich's homological mirror symmetry
equivalence. In particular, this mirror correspondence relates the enumerative geometry
of ideal sheaves (or instantons) on $X_\theta$ to that of lagrangians
in the mirror manifold $\Tcal^*$. In topological string theory, it provides an
equivalence between the category of noncommutative B-branes on $X_\theta$ and a
certain category of twisted lagrangian A-branes on the mirror $\Tcal^*$.

In this paper we work exclusively with
noncommutative projective varieties, as defined in~\S\ref{NCCPn}
{below}. The
ensuing simplification is that we can work for the most part directly at the level of
the homogeneous coordinate algebras, without resorting to the local
picture provided by the combinatorial fan data of the toric
variety. An alternative version of the homological mirror symmetry
correspondence for $\CP_\theta^2$, and more generally for toric
noncommutative weighted projective planes, is described
in~\cite{AKO}; in this case the deformation parameter
$\theta:=\theta^{12}\in\complex$ parametrizes a non-exact
deformation of the complexified
K\"ahler class of the mirror Landau--Ginzburg model.

\newsection{Coherent sheaves on noncommutative projective varieties\label{sec:ncproj}}

\subsection{Projective spaces
  $\complex\P_\theta^n$\label{NCCPn}}

The homogeneous coordinate algebra
$\alg(\complex\P_\theta^n)$ of the noncommutative toric variety
$\complex\P_\theta^n$ is the graded polynomial algebra in $n+1$ generators $w_i$, $i=1,\dots,n+1$ of degree~one with the
quadratic relations
\bea
w_{n+1}\,w_i&=&w_i\,w_{n+1} \ , \qquad i=1,\dots,n \ , 
\nonumber \\[4pt]
w_i\,w_j&=&q_{ij}^2~w_j\,w_i \ , \qquad i,j=1,\dots,n \ .
\label{wquadrels}\eea
This algebra is naturally an $\hil_\theta$-comodule algebra with left
coaction
$$
\Delta_L\,:\, \alg(\complex\P_\theta^n)~ \longrightarrow~
\hil_\theta\otimes\alg(\complex\P_\theta^n)
$$
given on generators by
\begin{eqnarray}
\Delta_L(w_i)&=& t_i\otimes w_i \ , \qquad i=1,\dots,n \ , 
\nonumber \\[4pt]
\Delta_L(w_{n+1}) &=& 1\otimes w_{n+1} \ ,
\label{DeltaLCPn}\end{eqnarray}
and extended as an algebra morphism. {As before,} we denote the
coaction on an arbitrary element $f\in\alg(\complex\P_\theta^n)$ by
$\Delta_L(f)=f^{(-1)}\otimes f^{(0)}$.
The {algebra $\alg=\alg(\complex\P_\theta^n)$ is quadratic and graded by the usual 
polynomial degree} as
$$
\alg(\complex\P_\theta^n)=\bigoplus_{k=0}^\infty\,\alg_k \ ,
$$
with $\alg_0=\complex$ and $\alg_k=\bigoplus_{i_1+\cdots+
  i_{n+1}=k}\,\complex w_1^{i_1}\cdots w_{n+1}^{i_{n+1}}$ for $k>0$. 
  
Each monomial $w_i$ generates a
left denominator set in $\alg(\complex\P_\theta^n)$, and the degree~zero
subalgebra of the left Ore localization of $\alg(\complex\P_\theta^n)$ with respect to
$w_i$ is naturally isomorphic to the noncommutative coordinate algebra
of the $i$-th maximal cone in the fan of $\complex\P_\theta^n$, for each
$i=1,\dots,n+1$~\cite[Thm.~5.4]{CLSI}.

If $\Ical \subset\alg(\complex\P_\theta^n)$ is a graded 
two-sided ideal generated by a set of homogeneous polynomials
$f_1,\dots,f_m$, then the quotient algebra
$\alg(\complex\P_\theta^n)/\Ical$ is identified as the graded homogeneous coordinate
algebra of a {noncommutative projective variety} $X_\theta(\Ical)$. It has
generators $w_1,\dots,w_{n+1}$ subject to the
relations (\ref{wquadrels}) and $f_1=\cdots=f_m=0$. 
We work only with varieties $X_\theta(\Ical)$ whose coordinate algebras $\alg(X_\theta(\Ical))=\alg(\CP_\theta^n)/\Ical$ are regular. Noncommutative
projective varieties inherit many of the properties of
$\complex\P_\theta^n$ that we describe in the following. 

The Koszul dual $\alg^!=\bigoplus_{k\geq0}\,\alg_k^!$ of the quadratic algebra 
{$\alg=\alg(\CP_\theta^n)$} was worked out in~\cite[Prop.~6.1]{CLSI}
and it is a
deformation of the exterior algebra of $\alg^*$, graded again by
polynomial degree, where throughout $(-)^*$ denotes the
$\complex$-dual $\Hom_\complex(-,\complex)$. In the category ${}^{\hil_{\theta}}\Module$ of left
$\hil_{\theta}$-comodules, it is given by
$$
\alg_k^!=\mbox{$\bigwedge_{\theta}^k$}\,
\alg^*_1 =\big(\alg_1^*\big)^{\otimes k}\,\big/\,\big\langle
a_1\otimes a_2 + \Psi_{\alg_1^*,\alg_1^*}^\theta(a_1\otimes a_2) \big\rangle_{a_1,a_2 \in \alg_1^*} \ ,
$$
and hence the dual algebra $\alg^!$ is generated by degree one elements
$\check w_i\in\alg_1^!$, $i=1,\dots,n+1$, with the relations
\begin{eqnarray}
\check w_i^2&=&0 \ , \qquad i=1,\dots,n+1 \ , \nonumber \\[4pt]
\check w_i\,\check w_{n+1}+\check w_{n+1}\,\check w_i&=&0 \ , \qquad
i=1,\dots,n \ , \nonumber \\[4pt] 
\check w_i\,\check w_j+q_{ij}^2~\check w_j\,\check w_i&=&0 \ , \qquad 
i,j=1,\dots,n \ .
\label{Koszuldualrels}\end{eqnarray}
The coaction $\Delta_L:\alg^!\to \hil_\theta\otimes\alg^!$ is dual to the
coaction (\ref{DeltaLCPn}) and is given by
\begin{eqnarray}
\Delta_L(\check w_i)&=& t_i^{-1}\otimes \check w_i \ , \qquad i=1,\dots,n \ , 
\nonumber \\[4pt]
\Delta_L(\check w_{n+1}) &=& 1\otimes \check w_{n+1} \ .
\label{DeltaLdual}\end{eqnarray}
By using the associated Koszul resolution of the trivial right $\alg$-module $\alg_0=\complex$~\cite[\S6.1]{CLSI}, together with the fact that $\alg=\alg(\CP_\theta^n)$ is an Ore extension of a commutative polynomial algebra~\cite[\S2.3]{AKO}, one shows that the algebra $\alg$ is a noetherian regular algebra of homological dimension $n+1$ (see~\cite[Prop.~2.6]{AKO} and \cite[Cor.~6.5]{CLSI}).

\subsection{Coherent sheaves\label{CPSheaves}}

With $\alg=\alg(\CP_\theta^n)$, 
let $\gr(\alg)$ be the abelian category of finitely-generated graded right
$\alg$-modules $M=\bigoplus_{k\geq0}\,M_k$ with morphisms given by
module homomorphisms of degree zero. Let $\tor(\alg)$ be the full subcategory of $\gr(\alg)$
consisting of graded torsion $\alg$-modules $M$ which have finite
dimension over $\complex$,
i.e. $M_k=0$ for $k\gg0$. In~\cite{CLSI} it was shown that one can identify the
category $\coh(\CP_\theta^n)$ of coherent sheaves of right
$\sheaf_{\CP_\theta^n}$-modules on $\Open(\CP_\theta^n)$ with the abelian
quotient category $\gr(\alg)/\tor(\alg)$. We denote by $\pi:\gr(\alg)\to\coh(\CP_\theta^n)$ the
canonical projection functor. Under this correspondence, the structure
sheaf $\sheaf_{\CP_\theta^n}$ of noncommutative $\complex$-algebras on
$\Open(\CP_\theta^n)$ is the image $\pi(\alg)$ of the
homogeneous coordinate algebra itself, regarded as a free right
$\alg$-module of rank one. Throughout
we abbreviate 
$$
\Hom(E,F):=\Hom_{\coh(\CP_\theta^n)}(E,F)
$$ 
for $E,F\in\coh(\CP_\theta^n)$. By~\cite[Cor.~2.19]{AKO}, if $n\times
n$ complex skew-symmetric matrices $\theta$ and $\theta'$ are related
by $\theta'\,^{ij}=\theta^{ij}+\vartheta^i-\vartheta^j$ for some
$\vartheta^1,\dots,\vartheta^n\in \complex$, then the abelian
categories $\coh(\CP_\theta^n)$ and $\coh(\CP_{\theta'}^n)$ are equivalent.

The abelian category $\gr(\alg)$ is equipped with a shift functor which is the autoequivalence sending a
graded module $M=\bigoplus_{k\geq0}\,M_k$ to the shifted module $M(l)$
defined by $M(l)_k=M_{l+k}$. The induced shift functor on the quotient
category $\coh(\CP_\theta^n)$ sends a sheaf $E=\pi(M)$ to
$E(k):=\pi(M(k))$. Since $\alg$ is a noetherian regular algebra, the correspondence which sends a sheaf
$E\in\coh(\CP_\theta^n)$ to the graded module 
$$
\Gamma(E):=\bigoplus_{k=0}^\infty\,
\Hom\big(\sheaf_{\CP_\theta^n}(-k)\,,\, E\big)
$$
defines a functor $\Gamma:\coh(\CP_\theta^n)\to \gr(\alg)$ such that
$\pi\circ\Gamma$ is isomorphic to the identity functor of the category $\coh(\CP_\theta^n)$~\cite[\S3--\S4]{AZ}.

Let $\gr_L(\alg)$ be the abelian category of finitely-generated graded
\emph{left} $\alg$-modules. We will denote by
$\pi_L:\gr_L(\alg)\to\qgr_L(\CP_\theta^n):=
\gr_L(\alg)\,/\,\tor_L(\alg)$ the corresponding quotient
projection, and by $\Gamma_L:\coh_L(\CP_\theta^n)\to \gr_L(\alg)$ its
right inverse such that
$\pi_L\circ\Gamma_L$ is isomorphic to the identity functor of $\coh_L(\CP_\theta^n)$. For any sheaf $E\in\coh(\CP^n_\theta)$, the graded left $\alg$-module
$$
\Homc\,\big(E\,,\,\sheaf_{\CP^n_\theta}\big)=\pi_L\Big(~
\mbox{$\bigoplus\limits_{k=0}^\infty$}\,
\Hom\big(E\,,\,\sheaf_{\CP^n_\theta}(k)\big)\,\Big)
$$
is called the \emph{dual sheaf} of $E$ and
is denoted $E^\vee\in\qgr_L(\CP^n_\theta)$. The internal Hom-functor
$\Homc\,(-,\sheaf_{\CP^n_\theta}):\coh(\CP^n_\theta)\to\qgr_L(\CP_\theta^n)$ is left exact and has corresponding right 
derived functors $\Extc^p(-,\sheaf_{\CP^n_\theta})$ given by
$$
\Extc^p\big(E\,,\,\sheaf_{\CP^n_\theta}\big)=\pi_L\Big(~
\mbox{$\bigoplus\limits_{k=0}^\infty$}\,\Ext^p\big(E\,,\,
\sheaf_{\CP^n_\theta}(k)\big)\,\Big)
$$
for $p\geq0$, where $\Ext^p(E,F)$ is the $p$-th derived functor
of the Hom-functor $\Hom(E,F)$ for $E,F\in\coh(\CP_\theta^n)$. Since $\alg$ is a noetherian regular algebra, there are isomorphisms~\cite{CLSI}
$$
\Ext^p(E,F)\cong \Ext_L^p(F^\vee,E^\vee\,):=\Ext^p_{\coh_L(\CP_\theta^n)}
(F^\vee,E^\vee\,)
$$
for any $p\geq0$ and for any pair of torsion free sheaves
$E,F\in\coh(\CP_\theta^n)$.  
The sheaves $E=\sheaf_{\CP^n_\theta}(k)$, $k\in\zed$ are locally free, i.e. $\Extc^p(E,\sheaf_{\CP^n_\theta})=0$ for all $p>0$,
with $\Homc\,(\sheaf_{\CP^n_\theta}(k),\sheaf_{\CP^n_\theta}(l))=
\sheaf_{\CP^n_\theta}(l-k)$ as sheaves of bimodules.

Under the conditions spelled out in~\cite[Prop.~6.4]{CLSI},
a noncommutative version of the Beilinson spectral sequence can be
developed following~\cite{KKO} by using a double Koszul bicomplex of the algebra
$\alg$. 
For this, split the left Koszul complex
$\comp^\bullet(\alg)\cong\alg\otimes(\alg^!)^*$ of
$\alg$--$\alg^!$-bimodules into
finite-dimensional subcomplexes $\comp^\bullet_{(p)}(\alg)$ for the
total degree $p$. Then for any sheaf $F\in\coh(\CP_\theta^n)$, there
is a spectral sequence with first term
\beq
&& \qquad
E_1^{p,q}=\Ext^q(\Qcal^p,F)\otimes\sheaf_{\CP^n_\theta}(-p) \qquad
\Longrightarrow \qquad E_\infty^i=\left\{\begin{array}{rl}
F \quad , & \quad i=p+q=0 \ , \\ 0 \quad , & \quad \mbox{otherwise} \
, \end{array} \right.
\label{Beilinsonseq}\eeq
where $p=0,1,\dots,n$, and $\Qcal^p=\pi_L\big(\comp_{(0)}^p(\alg)\big)^\vee$ is
the sheaf on $\Open(\CP_\theta^n)$ corresponding to the cohomology of
the truncated left Koszul complex for total degree zero given by
\beq
\comp_{(0)}^p(\alg)=\ker\Big(\alg(-p)\otimes\big(\alg_p^!\big)^*~
\longrightarrow~\mbox{$\bigoplus\limits_{k=1}^p$}\,
\alg(k-p)\otimes\big(\alg_{p-k}^!\big)^*\Big) \ .
\label{truncKoszul}\eeq
For $p=1$, it is shown in~\cite[Ex.~6.10]{CLSI} that the cohomology module
$\comp_1(\alg)$ of the left Koszul complex of $\alg$ truncated at the
first term can be naturally identified with the coherent sheaf
$$
\Omega_{\CP_\theta^n}^1=\ker\big(\mu_\alg:(\alg_1^!)^*\otimes\alg\to
\alg\big)
$$ 
of K\"ahler differentials on $\Open(\CP_\theta^n)$. Here $\mu_\alg$
denotes the product map on the algebra~$\alg$.

\subsection{Invariants of torsion free sheaves\label{Torsionfree}}

A coherent sheaf $E$ on $\Open(\CP_\theta^n)$ is torsion free if it
embeds into a locally free sheaf (a bundle), or equivalently if the right
$\alg$-module $M=\Gamma(E)$ contains no finite-dimensional submodules. They have natural isomorphism
invariants associated to them. The (Goldie) rank of $E$ is the maximal
number of nonzero direct summands of $E$, regarded as
$\alg$-submodules. This agrees with the notion of rank given
in~\cite[Def.~4.7]{CLSI} and is denoted ${\rm rank}(E)$. There is also a well-defined Euler
characteristic
$$
\chi(E)=\sum_{p\geq0}\,(-1)^p\,\dim_\complex\big(H^p(\CP^n_\theta,E)\big)
\ ,
$$
where $H^p(\CP^n_\theta,E):=\Ext^p(\sheaf_{\CP_\theta^n},E)$ are
finite-dimensional vector spaces over $\complex$ by the
$\chi$-condition of~\cite[Prop.~6.7]{CLSI}, together with the Hilbert polynomial
$$
h_E(s)=\chi\big(E(s)\big) ~\in~ \rat[s] \ .
$$

The first Chern class $c_1$ is defined by the requirement of
additivity on short exact sequences together with
\beq
c_1\big(\sheaf_{\CP_\theta^n}(k)\big)=k \ .
\label{c1Ok}\eeq
This uniquely determines $c_1(E)$ upon using the ampleness property of~\cite[Prop.~6.7]{CLSI} to construct a resolution of $E$ by shifts
of the structure sheaf $\sheaf_{\CP_\theta^n}$, and then applying
additivity. With this definition, one has~\cite{NS}
\beq
c_1\big(E(k)\big)=c_1(E)+k~{\rm rank}(E) \ .
\label{c1Ek}\eeq
In particular, for any ideal sheaf $I\in\coh(\CP_\theta^n)$, i.e. a torsion free sheaf of
rank one on $\Open(\CP_\theta^n)$, there is a unique shift $I(k)$
of $I$ which has $c_1=0$.
\begin{proposition}
If $E\in\coh(\CP_\theta^n)$ is a torsion free sheaf, then
$\Extc^{n}(E,\sheaf_{\CP_\theta^n})=0$.
\label{torsionfreeprop}\end{proposition}
\Proof{
This is a special case of~\cite[Prop.~2.0.6]{BGK}.
}

In our constructions of instanton moduli spaces, we shall need appropriate notions of stability.
\begin{definition}
A torsion free sheaf $E\in\coh(\CP_\theta^n)$ is said to be \emph{$\mu$-stable} {(resp. \emph{$\mu$-semistable})} if for every proper non-trivial subsheaf $F\subset E$, one has
$$
\frac{c_1(F)}{\rank(F)}\ <\ \frac{c_1(E)}{\rank(E)}
$$
(resp. $\leq$).
\label{slopestability}\end{definition}

\subsection{Monads\label{Monads}}

We shall now describe a general construction of coherent sheaves,
which will be instrumental in our analysis of instanton moduli spaces 
later on. 
\begin{definition}
A \emph{monad} on a (regular) noncommutative projective variety
$X_\theta=X_\theta(\Ical)$ is a complex
$$
\underline{\calg}\,_\bullet\,:\, 0~\longrightarrow~
\calg_{-1}~\xrightarrow{\sigma_{\underline{\calg}\,_\bullet}}~
\calg_0~\xrightarrow{\tau_{\underline{\calg}\,_\bullet}}~\calg_1~
\longrightarrow~0
$$
of locally free sheaves on $\Open(X_\theta)$ which is exact at the
first and last terms. The coherent sheaf
$\calg=H^0(\,\underline{\calg}\,_\bullet\,)=
\ker(\tau_{\underline{\calg}\,_\bullet})/{\rm
  im}(\sigma_{\underline{\calg}\,_\bullet})$ on $\Open(X_\theta)$ is
called the \emph{cohomology of the monad
  $\underline{\calg}\,_\bullet$}. A 
\emph{morphism of monads} is a homomorphism of complexes.
\label{monaddef}\end{definition}

In this paper we are primarily interested in \emph{linear monads} on
the noncommutative projective spaces
$\CP_\theta^n$, which are complexes of sheaves of free right
$\alg$-modules of the form
\beq
\qquad 
\underline{\calg}\,_\bullet\,:\, 
0~\longrightarrow~V_{-1}\otimes\sheaf_{\CP_\theta^n}(-1)~
\xrightarrow{\sigma_w}~V_0\otimes\sheaf_{\CP_\theta^n}~
\xrightarrow{\tau_w}~V_1\otimes\sheaf_{\CP_\theta^n}(1)~
\longrightarrow~0
\label{linmonadCPn}\eeq
for finite-dimensional complex vector spaces $V_{-1}$, $V_0$ and
$V_1$, where $\sigma_w\in V_{-1}^*\otimes V_0\otimes\alg_1$
(resp. $\tau_w\in V_0^*\otimes V_1\otimes\alg_1$) is an injective
(resp. surjective) $\alg$-module
homomorphism such that $\tau_w\circ\sigma_w=0$. Here $\alg_1$ is the
degree one component of the graded coordinate algebra
$\alg=\alg(\CP_\theta^n)$, i.e. the vector space spanned by the
generators $w_1,\dots,w_{n+1}$. Note that $\alg_1\cong
H^0(\CP_\theta^n,\sheaf_{\CP_\theta^n}(1))$ by~\cite[Prop.~6.8]{CLSI}. This definition is also well-posed on 
any regular noncommutative projective variety $X_\theta(\Ical)$.
\begin{proposition}
If $E\in\coh(\CP_\theta^n)$ is the cohomology sheaf of a linear monad complex
(\ref{linmonadCPn}), then it has invariants
\begin{eqnarray*}
{\rm
  rank}(E)&=&\dim_\complex(V_0)-\dim_\complex(V_{-1})-\dim_\complex(V_1)
\ , 
\\[4pt]
c_1(E)&=&\dim_\complex(V_{-1})-\dim_\complex(V_1) \ , \\[4pt]
\chi(E)&=&\dim_\complex(V_0)-(n+1)\,\dim_\complex(V_1) \ .
\end{eqnarray*}
\label{monadinvsprop}\end{proposition}
\Proof{
From (\ref{linmonadCPn}) and~\cite[Prop.~2.0.4~(1)]{BGK} it follows
that the kernel sheaf $\ker(\tau_w)$ is locally free, and there are
short exact sequences of sheaves of $\alg$-modules
\beq
0~\longrightarrow~\ker(\tau_w)~\longrightarrow~
V_0\otimes\sheaf_{\CP_\theta^n}~
\xrightarrow{\tau_w}~V_1\otimes\sheaf_{\CP_\theta^n}(1)~
\longrightarrow~0
\label{shortkerV0}\eeq
and
\beq
0~\longrightarrow~V_{-1}\otimes\sheaf_{\CP_\theta^n}(-1)~
\xrightarrow{\sigma_w}~\ker(\tau_w)~\longrightarrow~E~
\longrightarrow~0 \ .
\label{shortV1ker}\eeq
We apply ${\rm rank}$, $c_1$ and $\chi$ to these sequences using the
fact that they are all additive in exact sequences~\cite{NS}. The
formula for the rank then follows immediately, while the formula for
the first Chern class follows from (\ref{c1Ok}). The expression for
the Euler characteristic follows by~\cite[Prop.~6.8]{CLSI} which
gives $\chi(\sheaf_{\CP_\theta^n})= \dim_\complex(\alg_0)=1$,
$\chi(\sheaf_{\CP_\theta^n}(1))= \dim_\complex(\alg_1)=n+1$, and
$\chi(\sheaf_{\CP_\theta^n}(-1))=0$.
}
\begin{cor}
A linear monad complex (\ref{linmonadCPn}) on $\CP_\theta^n$ exists only when
$$
\dim_\complex(V_0)\ \geq\ \dim_\complex(V_{-1})+\dim_\complex(V_1) \ .
$$
\label{monadinvscor}\end{cor}
A fruitful feature of such sheaves $E$ coming from linear monads
is that the cohomology groups 
of the shifts $E(k)$ are qualitatively similiar to those of the
structure sheaf, in the sense that there is at most one non-trivial
cohomology group, and only for degree shifts in the ranges $k\geq0$ and
$k\leq-n-1$, as in~\cite[Prop.~6.8]{CLSI}.
\begin{proposition}
If $E\in\coh(\CP_\theta^n)$ is the cohomology of a linear monad complex
(\ref{linmonadCPn}), then it has the following sheaf cohomology
groups:
\begin{itemize}
\item[(1)] $H^p\big(\CP_\theta^n\,,\,E(k)\big)=0$ for all pairs of
  integers $(p,k)=(0,k<0)$, $(1,k<-1)$, $(2\leq p\leq n-2,k\in\zed)$,
  $(n-1,k>-n)$ and $(n,k\geq -n)$; \\
\item[(2)] $H^1\big(\CP_\theta^n\,,\,E(-1)\big)=V_1$; and \\
\item[(3)] $\Extc^1\big(E\,,\,\sheaf_{\CP^n_\theta}\big)= {\rm
    coker}(\sigma_w^*)$, while
  $\Extc^p\big(E(k)\,,\,\sheaf_{\CP^n_\theta}\big)=0$ for all $p\geq2$
  and for all~$k\in\zed$.
\end{itemize}
\label{monadcoh0prop}\end{proposition}
\Proof{
(1) The complex (\ref{linmonadCPn}) of sheaves of free right
$\alg$-modules can be naturally extended by applying the degree $k$
shift functor for any $k\in\zed$, whose cohomology coincides with the
sheaf $E(k)$. This modifies the short exact sequences (\ref{shortkerV0})--(\ref{shortV1ker}) to 
$$
0~\longrightarrow~\big(\ker(\tau_w)\big)(k)~\longrightarrow~
V_0\otimes\sheaf_{\CP_\theta^n}(k)~
\xrightarrow{\tau_w}~V_1\otimes\sheaf_{\CP_\theta^n}(k+1)~
\longrightarrow~0
$$
and
$$
0~\longrightarrow~V_{-1}\otimes\sheaf_{\CP_\theta^n}(k-1)~
\xrightarrow{\sigma_w}~\big(\ker(\tau_w)\big)(k)~
\longrightarrow~E(k)~\longrightarrow~0 \ .
$$
They induce long exact sequences in cohomology which contain the exact
sequences
\bea
&& H^p\big(\CP_\theta^n\,,\,\big(\ker(\tau_w)\big)(k)\big)~
\longrightarrow~ V_0\otimes H^p\big(\CP_\theta^n\,,\,
\sheaf_{\CP_\theta^n}(k)\big)~\longrightarrow \nonumber\\ && \qquad
\qquad\qquad\qquad \longrightarrow~V_1\otimes 
H^p\big(\CP_\theta^n\,,\,\sheaf_{\CP_\theta^n}(k+1)\big)~
\longrightarrow~
H^{p+1}\big(\CP_\theta^n\,,\,\big(\ker(\tau_w)\big)(k)\big) \label{1stcohseq} 
\eea
and 
\bea
&& V_{-1}\otimes
H^p\big(\CP_\theta^n\,,\,\sheaf_{\CP_\theta^n}(k-1)\big) \
\longrightarrow\ H^p\big(\CP_\theta^n\,,\,\big(\ker(\tau_w)\big)(k)
\big)~\longrightarrow \nonumber\\ && \qquad \longrightarrow \ H^p\big(\CP_\theta^n\,,\,E(k)
\big)~\longrightarrow \ V_{-1}\otimes
H^{p+1} \big(\CP_\theta^n\,,\,\sheaf_{\CP_\theta^n}(k-1)\big)
\label{2ndcohseq}
\eea
for each $p\geq0$, whose first arrows are both injections for $p=0$
and $H^{p+1}(\CP_\theta^n,-)=0$ for $p=n$ by~\cite[Prop.~6.8]{CLSI}. We use these two sequences and~\cite[Prop.~6.8]{CLSI} to find the values of $(p,k)$ for which 
the isomorphisms $H^p(\CP_\theta^n,(\ker(\tau_w))(k))
\xrightarrow{\ \approx\ }H^p(\CP_\theta^n,E(k))$ and
$H^p(\CP_\theta^n,(\ker(\tau_w))(k))=0$ simultaneously hold, and the
assertions follow. \\[6pt]
(2) Set $k=-1$ in (\ref{2ndcohseq}) and use
$H^0(\CP_\theta^n,E(-1))=0$ by (1), together with
$$ H^1\big(\CP_\theta^n\,,\,\sheaf_{\CP_\theta^n}(-2)\big)=
H^2\big(\CP_\theta^n\,,\,\sheaf_{\CP_\theta^n}(-2)\big)=0$$ by~\cite[Prop.~6.8]{CLSI}, to find the isomorphism
$H^1(\CP_\theta^n,(\ker(\tau_w))(-1))
\xrightarrow{\ \approx\ }H^1(\CP_\theta^n,E(-1))$. Setting $k=-1$ in (\ref{1stcohseq}) and using
$H^0(\CP_\theta^n,\sheaf_{\CP_\theta^n}(-1))=
H^1(\CP_\theta^n,\sheaf_{\CP_\theta^n}(-1))=0$ by~\cite[Prop.~6.8]{CLSI}, there is an isomorphism $V_1\otimes 
H^0(\CP_\theta^n,\sheaf_{\CP_\theta^n})\xrightarrow{\ \approx\ }
H^{1}(\CP_\theta^n,(\ker(\tau_w))(-1))$. The result now follows by~\cite[Prop.~6.8]{CLSI} which gives
$H^0(\CP_\theta^n,\sheaf_{\CP_\theta^n})\cong
\alg_0=\complex$. \\[6pt]
(3) We use $\sheaf_{\CP_\theta^n}(l)^\vee\cong
\sheaf_{\CP_\theta^n}(-l)$ for any $l\in\zed$ (as sheaves of
bimodules) and note that, by applying the internal Hom-functor
$\Homc\,(-,\sheaf_{\CP^n_\theta})$ to the complex (\ref{linmonadCPn})
using~\cite[Prop.~6.9]{CLSI}, the cohomologies of the dual
complex 
\beq
&& \underline{\calg}\,_\bullet^\vee\,:\, 
0~\longrightarrow~\sheaf_{\CP_\theta^n}(-1)\otimes V_1^*~
\xrightarrow{\tau_w^*}~\sheaf_{\CP_\theta^n}\otimes V_0^*~
\xrightarrow{\sigma_w^*}~\sheaf_{\CP_\theta^n}(1)\otimes V_{-1}^*~ 
\longrightarrow~0
\label{dualmonad}\eeq
coincide with
$H^0(\,\underline{\calg}\,_\bullet^\vee\,)=
\Homc\,(E,\sheaf_{\CP^n_\theta})=E^\vee$ and 
$H^1(\,\underline{\calg}\,_\bullet^\vee\,)=
\Extc^1(E,\sheaf_{\CP^n_\theta})$. We now use the fact that the 
$\alg$-module $\ker(\tau_w)$ is locally free, together with the
dualizing sequence
\bea
0~\longrightarrow~E^\vee(-k) \longrightarrow
\big(\ker(\tau_w)\big)^\vee(-k) &\xrightarrow{\sigma_w^*}
& \sheaf_{\CP_\theta^n}(-k+1)\otimes V_{-1}^*~\longrightarrow
\nonumber \\ &\longrightarrow&
\Extc^1\big(E(k)\,,\,\sheaf_{\CP^n_\theta}\big)~\longrightarrow~0 \ ,
\nonumber
\eea
and the result follows.}

It is also possible to algebraically characterize those linear monads
whose cohomology sheaf is locally free or torsion free.
\begin{proposition}
Let $E\in\coh(\CP_\theta^n)$ be the cohomology of a linear monad complex
(\ref{linmonadCPn}) and let $S=\Gamma_L(\coker(\sigma_w^*))\in
\gr_L(\alg)$. Then:
\begin{itemize}
\item[(1)] $E$ is locally free if and only if $S$ is a
  finite-dimensional graded left $\alg$-module. \\
\item[(2)] $E$ is torsion free if and only if $S$ has homological
  dimension $\leq n-2$ as a graded left $\alg$-module.
\end{itemize}
\label{monadfreeprop}\end{proposition}
\Proof{
(1) By point~(3) of Proposition~\ref{monadcoh0prop} and~\cite[Prop.~6.9]{CLSI}, $E$
is locally free if and only if $\pi_L(S)=0$,
i.e. $\dim_\complex(S)<\infty$. \\[6pt]
(2) There is a spectral sequence
$$
E_2^{p,q}=\Extc^q_{\,L}\big(\,\Extc^{-p}(E,\sheaf_{\CP_\theta^n})\,,\,
\sheaf_{\CP_\theta^n}\big) \qquad
\Longrightarrow \qquad E_\infty^i=\left\{\begin{array}{rl}
E \quad , & \quad i=p+q=0 \ , \\ 0 \quad , & \quad \mbox{otherwise} \
. \end{array} \right.
$$
By point~(3) of Proposition~\ref{monadcoh0prop}, $E_2^{p,q}=0$ for all $p\leq-2$. By
Serre duality~\cite[Prop.~6.7]{CLSI}, one has
$$
\Ext_L^q\big(\,\Extc^1(E,\sheaf_{\CP_\theta^n})\,,\,
\sheaf_{\CP_\theta^n}(k)\big)\cong H_L^{n-q}\big(\CP_\theta^n\,,\,
\Extc^1(E,\sheaf_{\CP_\theta^n})(k+n+1)\big)^*
$$
for all $k\in\zed$. The right-hand side vanishes for $q=0,1$ and for
$k\gg0$, by point~(3) of Proposition~\ref{monadcoh0prop} and our hypothesis which implies
that the sheaves $\Extc^1(E,\sheaf_{\CP_\theta^n})(k)$ in
$\coh_L(\CP_\theta^n)$ have cohomological dimension $\leq n-2$ for $k$
sufficiently large. Thus $E_2^{-1,q}=0$ for $q=0,1$, and so the only
$E_\infty^{p,-p}$ term which might be non-zero is $E_\infty^{0,0}$. But
the differentials coming into $E_k^{0,0}$ are always zero, so we get a
sequence of inclusions
$$
E=E_\infty^{0,0}~\hookrightarrow~\cdots~\hookrightarrow~ E_3^{0,0}~
\hookrightarrow~E_2^{0,0} \ .
$$
The extremities imply injectivity of the canonical morphism $E\to
E^{\vee\vee}$. By iterating the proof of point~(3) of Proposition~\ref{monadcoh0prop} and
using~\cite[Prop.~6.9]{CLSI}, one shows that $E^\vee$ is locally
free. By~\cite[Prop.~2.0.4~(2)]{BGK}, the sheaf $E^{\vee\vee}$ is also
locally free. Hence $E$ is torsion free.

For the converse statement, we use~\cite[Lem.~2.0.7]{BGK} to choose an
integer $k$ large enough such that
\beq
H_L^0\big(\CP_\theta^n\,,\,\Extc^p(F,\sheaf_{\CP_\theta^n})(k-n-1)\big)
\cong H^{n-p}\big(\CP_\theta^n\,,\,F(-k)\big)^*
\label{Lem207}\eeq
for any coherent sheaf $F\in\coh(\CP_\theta^n)$ and for all
$p\geq0$. By~\cite[Prop.~6.8]{CLSI} it follows that the
homological dimension of the graded left $\alg$-module
$\Gamma_L(\,\Extc^p(F,\sheaf_{\CP_\theta^n}))$ is $\leq n-p$. If the
cohomology sheaf $E$ is torsion free, then there is an embedding
$E\hookrightarrow \bun$ into a locally free sheaf $\bun$. By applying
the functor $\Homc(-,\sheaf_{\CP_\theta^n})$ to the exact sequence
$0\to E\to\bun\to\bun/E\to 0$, one gets an isomorphism
$\Extc^1(E,\sheaf_{\CP_\theta^n}) \cong
\Extc^2(\bun/E,\sheaf_{\CP_\theta^n})$. An application of the isomorphism (\ref{Lem207}) to $F=\bun/E$ using point~(3) of Proposition~\ref{monadcoh0prop} then shows that
the homological dimension of $S$ is $\leq n-2$.
}

\subsection{Coequivariant sheaves\label{subsec:coeq-sheaves}}

We would now like to regard the sheaves which are constructed as the cohomology of a linear monad as coequivariant sheaves $E$ in the sense of~\cite[\S4.2]{CLSI}, i.e. as elements of the category ${}^{\hil_\theta}\Module$ of left
$\hil_\theta$-comodules whose coactions are compatible with the
coaction of $\hil_\theta$ on $\alg$. Generally, this occurs when the
complex $\underline{\calg}\,_\bullet$ in $\coh(X_\theta)$ of Definition~\ref{monaddef}
is also a complex in
${}^{\hil_\theta}\Module$, i.e. when the sheaves $\calg_{-1},
\calg_{0}, \calg_{1}$ are objects of ${}^{\hil_\theta}\Module$ and the maps
$\sigma_{\underline{\calg}\,_\bullet},\tau_{\underline{\calg}\,_\bullet}$
are morphisms in ${}^{\hil_\theta}\Module$; it is easy to check that
the cohomology of such a monad is a coequivariant sheaf. In contrast
to the approach of~\cite{BL1,BvS}, here we regard the set of all
$\alg$-module morphisms as a vector space, without further structure; our
construction of instanton moduli spaces later on using the larger
space of torsion free sheaves (rather than just the dense subset of
vector bundles) will naturally use
universal objects in this setting, and can be described by commutative
parameter spaces and standard geometric invariant theory quotients.

For a linear monad (\ref{linmonadCPn}), the first requirement is
automatically satisfied due to the coaction (\ref{DeltaLCPn}) which lifts to all bundles $\sheaf_{\CP_\theta^n}(k)$ for $k\in\zed$. The second requirement, on the other hand, restricts the allowed differentials. For this, we decompose the $\alg$-module homomorphisms as 
\beq
\sigma_w=\sum_{i=1}^{n+1}\, \sigma^i\otimes w_i \ , \qquad
\tau_w=\sum_{i=1}^{n+1}\, \tau^i\otimes w_i
\label{sigmataudecomp}\eeq
with $\sigma^i\in\Hom_\complex(V_{-1},V_0)$ and
$\tau^i\in\Hom_\complex(V_0,V_1)$ for $i=1,\dots,n+1$.
\begin{lemma}
The differentials $\sigma_w$ and $\tau_w$ are morphisms in the category ${}^{\hil_\theta}\Module$ if and only if the vector spaces spanned by $\sigma^i$ and $\tau^i$ for $i=1,\dots,n+1$ are objects of ${}^{\hil_\theta}\Module$ with left $\hil_\theta$-coactions given by
\begin{eqnarray*}
\Delta_L\big(\sigma^i\big)&=& t_i^{-1}\otimes \sigma^i \ , \qquad
\Delta_L\big(\tau^i\big)\ =\ t_i^{-1}\otimes \tau^i \ , \qquad i=1,\dots,n \ , \\[4pt]
\Delta_L\big(\sigma^{n+1}\big) &=& 1\otimes\sigma^{n+1} \ , \qquad 
\Delta_L\big(\tau^{n+1}\big) \ =\ 1\otimes\tau^{n+1} \ .
\end{eqnarray*}
\label{eqmapslemma}\end{lemma}
\Proof{
The requisite $\hil_\theta$-coequivariance conditions follow easily
from (\ref{DeltaLCPn}).
}

\newsection{Noncommutative twistor geometry\label{sec:nctwistor}}

\subsection{Grassmannians $\Gr_\theta(d;n)$\label{ncgrassgen}}

In~\cite[\S5.3]{CLSI} we defined noncommutative Grassmann varieties
{$\Gr_\theta(d;V)\cong\Gr_\theta(d;n)$} 
associated to an $\hil_\theta$-comodule $V$ of dimension~$n>d$. The homogeneous
coordinate algebra $\alg(\Gr_{\theta}(d;n))$ of the noncommutative grassmannian 
{$\Gr_{\theta}(d;n)$}
is defined as a quotient of the algebra of a suitable projective space
$\CP_\Theta^N\cong\mathbb{P}(\bigwedge_{\theta}^d\, V)$, with $N={n\choose d} -1$. The minors
$\Lambda^J$ which span the braided exterior algebra as in (\ref{qest}) are
labelled by ordered $d$-multi-indices $J=(j_1\cdots j_d)$, $1\leq j_\alpha\leq
n$. The noncommutativity relations
between the minors are given by~\cite{CLSI}
\begin{equation}
\label{ncrelpr}
\Lambda^{J}\,\Lambda^{K} = \Big(~\prod_{\alpha,\beta =1}^d
\,q^2_{j_{\alpha}k_{\beta}}~ \Big)~ \Lambda^{K}\,\Lambda^{J} \ .
\end{equation}
Regarding $\Lambda^J$ as homogeneous coordinates in
$\alg(\mathbb{CP}_{\Theta}^{N})$, the $N\times N$ noncommutativity matrix $\Theta$ of
the projective space containing the embedding of $\Gr_{\theta}(d;n)$
is completely determined (mod~$2\pi$) from the $n\times n$
noncommutativity matrix $\theta$ of the grassmannian as
\begin{equation}
\label{Theta}
\Theta^{JK} = \sum_{\alpha,\beta =1}^d
\,\theta^{j_{\alpha}k_{\beta}} \ .
\end{equation}
This is a necessary and sufficient
condition for the existence of an embedding of the noncommutative
grassmannian $\Gr_{\theta}(d;n)
\hookrightarrow \mathbb{CP}_{\Theta}^N$, {with $N={n\choose d} -1$}.

Given the noncommutative relations (\ref{ncrelpr}) between generators of the
projective space, the next step is to exhibit noncommutative Pl\"ucker
relations. They generate a homogeneous ideal in the homogeneous
coordinate algebra $\alg(\complex\P^N_{\Theta})$ of the projective space,
and one {defines} the noncommutative quotient algebra as the graded
homogeneous coordinate algebra $\alg(\Gr_\theta(d;n))$ of the (embedding
of the) noncommutative grassmannian. The natural noncommutative
version of Young symmetry relations takes into account the
braided antisymmetry of the minors, and read~\cite{CLSI} 
\begin{equation}
\label{ncysr}
\sum_{\gamma =1}^{d+1} \, \Big(\,  \prod_{\alpha =1}^d \,q_{i_{\gamma}i^{\gamma}_{\alpha}}\, \Big)\, \Big(\, \prod_{\beta =1}^{d-1} \, q_{i_{\gamma}j_{\beta}} \, \Big)\,(-1)^{\gamma} \; \Lambda^{I\setminus i_{\gamma}}\, \Lambda^{i_{\gamma}\cup J} = 0
\end{equation}
for all choices of $(d+1)$-multi-indices $I$
and $(d-1)$-multi-indices $J$, where $i_\alpha^\gamma\in I\setminus i_\gamma$.

With respect to the noncommutative algebraic torus $T_\theta=(\complex^\times_\theta)^n$, the coordinate algebra $\alg(\Gr_{\theta}(d;n))$ is naturally an object of the category ${}^{\hil_\theta}\Module$ by the left coaction
\beq
\Delta_L\,:\, \alg\big(\Gr_{\theta}(d;n)\big) ~\longrightarrow~ \hil_\theta\otimes \alg\big(\Gr_{\theta}(d;n)\big) \ , \qquad \Delta_L(\Lambda^J)=t_J\otimes \Lambda^J
\label{grasstoric}\eeq
where $t_J:=t_{j_1}\cdots t_{j_d}$. 

The tautological bundle $\Scal_\theta$ on $\Open(\Gr_{\theta}(d;n))$ is defined
to be the subsheaf of elements of the free module
$(f_1(\Lambda),\ldots ,f_n(\Lambda)) \in
\alg(\Gr_{\theta}(d;n))^{\oplus n}$ over the noncommutative
grassmannian, with each $f_k(\Lambda)$, $k=1,\dots,n$, a function on $\Gr_{\theta}(d;n)$, i.e. an element in $\alg(\Gr_{\theta}(d;n))$, which satisfy the equations
\begin{equation}
\label{eq2}
\sum_{\alpha=1}^{d+1}\, \Big(\, \prod_{\beta=1}^d \, q_{ j_\alpha  j_\beta^\alpha } \, \Big)\, (-1)^{\alpha} \ \Lambda^{J\setminus j_\alpha}\, f_{j_\alpha}  = 0 
\end{equation}
for every ordered $(d+1)$-multi-index $J=(j_1\cdots j_{d+1})$ with
$j_1<j_2<\cdots<j_{d+1}$, where the minors of order
$d$ obey the relations~(\ref{ncrelpr}). We can use the Pl\"ucker
map to regard the noncommutative minors $\Lambda^{{J\setminus
    j_\alpha}}$ as homogeneous coordinates in
$\mathbb{P}(\bigwedge_{\theta}^d
V)$. Then the quotient by the graded two-sided ideal generated by the
set of homogeneous relations (\ref{eq2}) defines the projection of the
free module $\mathbb{P}(\bigwedge_{\theta}^d V)\otimes
V\to\mathcal{S}_\theta$. In this case we have to consider the
restriction of (\ref{eq2}) to those elements {$\Lambda^J$} which also
satisfy the Young symmetry relations~(\ref{ncysr}). This gives the
sheaf $\Scal_\theta$ the natural structure of a graded
$\alg(\Gr_{\theta}(d;n))$-bimodule. In~\cite[\S6.4]{CLSI} it is shown that
the coherent sheaf of noncommutative K\"ahler differential forms on $\Open(\Gr_{\theta}(d;n))$ is isomorphic
to the braided tensor product
\beq
\Omega_{\Gr_\theta(d;n)}^1~\cong~
\Scal_\theta\,\otimes_\theta\,\Qcal_\theta
\label{lcinvtbraidedprod}\eeq
as a bimodule algebra over $\alg(\Gr_\theta(d;n))$, where
$\Qcal_\theta$ is the orthogonal complement of the tautological bundle
defined through the noncommutative Euler sequence 
\beq
0~\longrightarrow~\Qcal_\theta^\vee~
\longrightarrow~\alg\big(\Gr_\theta(d;n)\big)\otimes V~
\xrightarrow{\ \hat \eta\ }~\Scal_\theta~\longrightarrow~0 \ .
\label{NCGrEuler}\eeq

\begin{remark}
\textup{
For $d=1$ this construction gives an alternative description of the
homogeneous coordinate algebra of
$\CP_\theta^{n-1}\cong\Gr_\theta(1;n)^*$; in this case $N=n-1$ and
$\Theta=\theta$. The explicit mapping between the two homogeneous coordinate
algebras, with relations (\ref{wquadrels}) induced by the local fan
construction of $\CP_\theta^{n-1}$ and (\ref{ncrelpr}) induced by the
noncommutative deformation ${\rm GL}_\theta(n)$ of the general linear
group, is described in~\cite[\S5.2]{CLSI}.
}
\label{CPnremark}\end{remark}

\subsection{Klein quadric $\Gr_\theta(2;4)$\label{ncgrass}}

We explicitly work out the algebraic relations in the case $d=2$ and $n=4$. Consider the
algebra projection $\alg(\mathbb{CP}_{\Theta}^5)\to
\alg(\Gr_{\theta}(2;4))$. Give to the multi-indices labelling the
minors a lexicographic ordering, such that $1=(1\,2)$, $2=(1\,3)$,
$3=(1\,4)$, $4=(2\,3)$, $5=(2\,4)$, and $6=(3\,4)$. Then the
expression (\ref{Theta}) for the skew-symmetric noncommutativity
matrix $\Theta$ in terms of entries of $\theta$ is given by
$$
\Theta=\begin{pmatrix}
\scriptstyle{0} & \scriptstyle{-\theta^{12}+\theta^{13}+\theta^{23}} &
\scriptstyle{-\theta^{12}+\theta^{14}+\theta^{24}} &
\scriptstyle{\theta^{12}+\theta^{13}+\theta^{23}} &
\scriptstyle{\theta^{12}+\theta^{14}+\theta^{24}} &
\scriptstyle{\theta^{13}+\theta^{14}+\theta^{23}+\theta^{24}} \\
  & \scriptstyle{0}                                  &
  \scriptstyle{-\theta^{13}+\theta^{14}+\theta^{34}} &
  \scriptstyle{\theta^{12}+\theta^{13}-\theta^{23}} &
  \scriptstyle{\theta^{12}+\theta^{14}-\theta^{23}+\theta^{34}} &
\scriptstyle{\theta^{13}+\theta^{14}+\theta^{34}} \\
  &                                      & \scriptstyle{0}
  &\scriptstyle{\theta^{12}+\theta^{13}-\theta^{24}-\theta^{34}}&
    \scriptstyle{\theta^{12}+\theta^{14}-\theta^{24}} &
\scriptstyle{\theta^{13}+\theta^{14}-\theta^{34}} \\
  &                                      &
  & \scriptstyle{0} &
  \scriptstyle{-\theta^{23}+\theta^{24}+\theta^{34}} &
  \scriptstyle{\theta^{23}+\theta^{24}+\theta^{34}} \\
  &                                      &
  &   & \scriptstyle{0} &
  \scriptstyle{\theta^{23}+\theta^{24}-\theta^{34}} \\
  &                                      &
  &   &   & \scriptstyle{0}
\end{pmatrix} \ .
$$
It follows that in order to be interpreted as minors of a
noncommutative matrix, the generators of
$\alg(\mathbb{CP}^5_{\Theta})$ cannot have generic noncommutativity
relations. From the above matrix one sees that one must take for example 
$$
\Theta^{46}-\Theta^{56} = \Theta^{26}-\Theta^{36} = 2 \, \theta^{34}
\ ,
$$ 
and hence it is easy to construct a matrix $\Theta$ parametrizing a
projective space $\mathbb{CP}^5_{\Theta}$ that cannot contain any
embedding of any grassmannian $\Gr_{\theta}(2;4)$.

Let us consider the noncommutative Pl\"ucker relations
(\ref{ncysr}). Various choices for the multi-indices $I$ and $J$ yield
structure equations, i.e. the noncommutative relations among
minors. For example, if we set $I=(1\, 2\, 3)$ and  
$J=(1)$, then the equation~(\ref{ncysr}) reduces to
a two-term relation (as the additional term contains
$\Lambda^{(11)}=0$) given by 
$$ -\, q_{21}\, q_{23}\, q_{21} \ \Lambda^{(13)}\, \Lambda^{(21)} +
q_{31}\, q_{32}\ \Lambda^{(12)}\, \Lambda^{(31)} = 0 \ . $$
Using the alternating properties~\cite[eq.~(2.30)]{CLSI} $\Lambda^{(21)}=-\Lambda^{(12)}$ and
$\Lambda^{(31)}=-\Lambda^{(13)}$, this expression can be reordered as
$$\Lambda^{(12)}\, \Lambda^{(13)}=\, q_{12}^{-2}\, q_{23}^2\,
q_{13}^2\ \Lambda^{(13)}\, \Lambda^{(12)} \ , $$ which is exactly (\ref{ncrelpr}) for the two minors considered. One
also arrives at (\ref{ncrelpr}) for all other choices of multi-indices
which lead to a two-term Pl\"ucker equation. Thus from these
``trivial'' Pl\"ucker equations we can derive completely the
noncommutativity relations~(\ref{ncrelpr}).

Now let
us consider the only three-term Pl\"ucker equation, which is
a noncommutative deformation of the well-known classical equation
describing the Klein quadric $\Gr(2;4)\hookrightarrow\mathbb{CP}^5$. It comes from (\ref{ncysr}) with
$I=(1\,2\,3)$ and $J=(4)$. After rearranging all indices labelling the
minors in increasing order using antisymmetry, and the minors
themselves using (\ref{ncrelpr}), we obtain
\begin{equation}
\label{pl24}
q_{31}\, q_{32}\, q_{34}\ \Lambda^{(12)}\,\Lambda^{(34)} -
q_{21}\, q_{23}\, q_{24}~\Lambda^{(13)}\,\Lambda^{(24)} +
q_{12}\, q_{13}\,q_{14}~\Lambda^{(23)}\,\Lambda^{(14)} = 0 \ .
\end{equation}

\subsection{Twistor correspondences\label{ncflag}}

Noncommutative flag varieties associated to $\hil_\theta$ comodules $V$ of dimension $n$ are defined in generality in~\cite[\S5.4]{CLSI}. In this paper we will only need the flag varieties of $\GL(4)$; these are the ones which naturally appear in the double fibrations
underlying the most important twistor correspondences~\cite{HH}. All
noncommutative double twistor fibrations are included in the complete
noncommutative flag variety for $n=4$, according to the diagram
$$
\xymatrix{
 & \alg\big(\Fl_\theta(2,3;4)\big) \ar[d]&
 \\ \alg\big(\Gr_\theta(2;4)\big)
\ar[ur]\ar[r]\ar[d]& 
\alg\big(\Fl_\theta(1,2,3;4)\big) & \ar[l]\ar[d]\ar[ul] 
\alg\big(\Gr_\theta(3;4)\big) \\
\alg\big(\Fl_\theta(1,2;4)\big)\ar[ur] &
\alg\big(\CP_\theta^3\big) \ar[l]\ar[r] \ar[u]& 
\alg\big(\Fl_\theta(1,3;4)\big)\ar[ul]
}
$$
where all morphisms are subalgebra inclusions.

In this paper we will focus on the noncommutative correspondence diagram
\beq
\xymatrix{
 & \alg\big(\Fl_\theta(1,2;4)\big) & \\
\alg\big(\CP_\theta^3\big) \ar[ur]^{p_1} &  & 
\alg\big(\Gr_\theta(2;4)\big) \ar[ul]_{p_2}
}
\label{nccorr}\eeq
which is a noncommutative deformation of the usual Penrose twistor
correspondence, with $\alg(\CP_\theta^3)$ the ``noncommutative twistor
algebra''. This diagram, as well as all double fibrations above,
is an example of a ``noncommutative correspondence'' in the sense
of~\cite[\S5.2]{BMRS}. In the present case, the morphism
$E\mapsto p_2{}^*\,p_{1*}(E)$ gives a map from sheaves in
$\coh(\CP_\theta^3)$ to sheaves in $\coh(\Gr_\theta(2;4))$, and it
defines a noncommutative deformation of the usual Penrose--Ward
twistor transform described in more detail in \S\ref{se:tt} below.

We will describe the homogeneous coordinate algebra
of the noncommutative partial flag variety $\Fl_\theta(1,2;4)$ using
the algebra projection~\cite{CLSI}
$$
\alg\big(\CP_\theta^3\big)\,\otimes_\theta\,
\alg\big(\Gr_\theta(2;4)\big)~\longrightarrow~
\alg\big(\Fl_\theta(1,2;4)\big) \ .
$$
The braided tensor product algebra $\alg(\CP_\theta^3)\,\otimes_\theta\,
\alg(\Gr_\theta(2;4))$, as an object in the category ${}^{\hil_\theta}\Module$, is generated by the homogeneous coordinate
elements $w_i$ with relations (\ref{ncrelpr}) for
$d=1$, i.e. $w_i\, w_j=q_{ij}^2\ w_j\, w_i$ for $i,j=1,2,3,4$, 
and by the noncommutative $2\times2$ minors
$\Lambda^{(j_1j_2)}$, $1\leq j_1<j_2\leq4$, obeying (\ref{ncrelpr}) for
$d=2$ and the quadric relation (\ref{pl24}). They also obey the structure equations (relations among minors of different order, see \cite[eq.~(2.29)]{CLSI})
\beq
w_i\,\Lambda^{(j_1j_2)}=q_{12}^{-2}\,q_{i\,j_1}^2\,q_{i\,j_2}^2~
\Lambda^{(j_1j_2)}\,w_i
\label{wLambda}\eeq
for all $i=1,2,3,4$ and $1\leq j_1<j_2\leq4$. The remaining
noncommutative Pl\"ucker equations come from the Young symmetry
relations of~\cite[eq.~(5.18)]{CLSI} with $d=2$ and $d'=1$. Working through all four increasing
cyclic permutations of order~three in $S_4$ for the multi-index $I$, by
completely analogous calculations to those of \S\ref{ncgrass} one finds
the additional relations
\begin{eqnarray}
q_{12}\, q_{13}\ \Lambda^{(23)}\, w_1 - q_{21}\, q_{23}\ \Lambda^{(13)}\, w_2 + q_{31}\, q_{32}\ \Lambda^{(12)}\, w_3&=&0 \ ,
\nonumber \\[4pt]
q_{12}\, q_{14}\ \Lambda^{(24)}\, w_1 - q_{21}\, q_{24}\ \Lambda^{(14)}\, w_2 + q_{41}\, q_{42}\ \Lambda^{(12)}\, w_4 &=&0 \ ,
\nonumber \\[4pt]
q_{13}\, q_{14}\ \Lambda^{(34)}\, w_1 - q_{31}\, q_{34}\ \Lambda^{(14)}\, w_3 + q_{41}\, q_{43}\ \Lambda^{(13)}\, w_4&=&0 \ ,
\nonumber \\[4pt]
q_{23}\, q_{24}\ \Lambda^{(34)}\, w_2 - q_{32}\, q_{34}\
\Lambda^{(24)}\, w_3 + q_{42}\, q_{43}\ \Lambda^{(23)}\, w_4&=&0 \ .
\label{flagpluck}\end{eqnarray}
Other quantum deformations of these flag varieties in the context of
twistor theory can be found in~\cite{KKO,Hannabuss,FJ,MS,BM,BL}.

\subsection{Twistor transform}\label{se:tt}

As mentioned in \S\ref{ncflag}, the twistor transform of coherent sheaves on $\Open(\CP_\theta^3)$ is determined by the noncommutative correspondence diagram (\ref{nccorr}). It is the Fourier--Mukai type transform $E\mapsto
p_2{}^*\,p_{1*}(E)$ defined on $\coh(\CP_{\theta}^3)\to
\coh(\Gr_{\theta}(2;4))$ as follows. Given a graded right
$\alg(\CP_\theta^3)$-module $M$ in $\gr(\alg(\CP_\theta^3))$, the push-forward 
\beq
M'=p_{1*}(M)=M\otimes_{\alg(\CP_\theta^3)}\alg\big(\Fl_{\theta}(1,2;4)\big) 
\label{p1forward}\eeq
is a bigraded right module over $\alg(\Fl_{\theta}(1,2;4))$,
where on the right-hand side we regard the algebra
$\alg(\Fl_{\theta}(1,2;4))$ as an $\alg(\CP_\theta^3)$-bimodule
according to the action described in \S\ref{ncflag}. The diagonal
subspace of this module, in the sense of~\cite[\S8]{KKO}, induces the
push-forward functor
$p_{1*}:\coh(\CP_{\theta}^3)\to\coh(\Fl_{\theta}(1,2;4))$. Similarly,
one defines the push-forward functor
$p_{2*}:\coh(\Gr_{\theta}(2;4))
\to\coh(\Fl_{\theta}(1,2;4))$, which has a right adjoint functor denoted
$p_2{}^*:\coh(\Fl_{\theta}(1,2;4))\to\coh(\Gr_{\theta}(2;4))$.

For the noncommutative twistor transforms of elementary bundles on
$\Open(\CP_{\theta}^3)$ we have the following computation.
\begin{lemma}
The noncommutative twistor transforms of the locally free sheaves
$\sheaf_{\CP_{\theta}^3}(k)$ for $k\in\zed$ are given by 
$$
p_2{}^*\,p_{1*}\big(\sheaf_{\CP_{\theta}^3}(k)\big) = 
\left\{\begin{array}{rl}
{\rm Sym}^k_{\theta}(\Scal_{\theta}) \quad , & k\geq0 \ ,
\\ 0 \quad , & k<0 \ , \end{array} \right.
$$
where $\Scal_{\theta}$ is the tautological bundle on
$\Open(\Gr_{\theta}(2;4))$ defined in \S\ref{ncgrassgen} and
${\rm Sym}^k_{\theta}(\Scal_{\theta})$ is the bundle
associated to the graded right module
$$
\Gamma(\Scal_{\theta})^{\otimes k}\, \big/\, \big\langle s_1\otimes
s_2-\Psi_{\Gamma(\Scal_{\theta}), \Gamma(\Scal_{\theta})}^{\theta}(s_1\otimes
s_2)\big\rangle_{s_1,s_2\in\Gamma(\Scal_{\theta})}
$$
over $\alg\big(\Gr_{\theta}(2;4)\big)$. 
\label{twisttransfbundles}\end{lemma}
\Proof{
The direct image formula (\ref{p1forward}) gives
$$
p_{1*}\big(\alg(\CP_\theta^3)\big)=\alg\big(\Fl_{\theta}(1,2;4)\big)
=p_{2*}\big(\alg(\Gr_{\theta}(2;4)\big) \ ,
$$
whence
$p_2{}^*\,p_{1*}(\sheaf_{\CP_{\theta}^3})
=p_2{}^*\,p_{2*}(\sheaf_{\Gr_{\theta}(2;4)}) =
\sheaf_{\Gr_{\theta}(2;4)}$ and the result holds for $k=0$. By
the noncommutative Pl\"ucker equations (\ref{flagpluck}), the image
of the set of generators $p_2^{-1}\,(p_1(\alg(\CP_\theta^3)_1))$ consists
of elements $f_k=w_k$ satisfying the equations (\ref{eq2}) for $d=2$ and $n=4$,
which defines the tautological bundle over the noncommutative
Grassmann variety $\Gr_{\theta}(2;4)$. It follows that
$p_2{}^*\,p_{1*}(\sheaf_{\CP_{\theta}^3}(1))
=p_2{}^*\,p_{2*}(\Scal_{\theta}) =\Scal_{\theta}$. The
result for negative degree shifts is then clear, and for positive
degree shifts follows by applying the same reasoning to
$\alg(\CP_\theta^3)_1^{\otimes k}$.
}

\subsection{Sphere $S_\theta^4$\label{ncsphere}}

In order to introduce a $*$-algebra structure on the algebra $\alg(\Gr_\theta(2;4))$ and a compatible twistor construction of instanton bundles to be considered later on, 
in the remainder of this section we consider a reduction from a six complex parameter deformation to a one 
real parameter deformation. Thus, in particular, we set 
\beq
q_{12}=q_{21}^{-1}=:q \ , \qquad q_{ij}=1 \quad \mbox{otherwise} \ , 
\label{qvalues}\eeq
and in addition we assume that $q\in\real$, i.e. that the noncommutativity parameter $\theta:=\theta^{12}$ is purely imaginary.

The compatible $*$-involution is defined on the generators of $\alg(\Gr_{\theta}(2;4))$ as
\bea\label{xdelta}
&&\Lambda^{(13)}\,^\dag=q~\Lambda^{(24)} \ , \qquad
\Lambda^{(14)}\,^\dag=-q^{-1}~\Lambda^{(23)} \ , \nonumber \\ && 
\Lambda^{(12)}\,^\dag=\Lambda^{(12)} \ , \qquad
\Lambda^{(34)}\,^\dag=\Lambda^{(34)} \ , 
\label{stargrass}\eea
and then extended to the whole of $\alg(\Gr_{\theta}(2;4))$ as a
conjugate linear anti-homomorphism. When substituted in the Klein quadric equation~\eqref{pl24} we obtain
\begin{equation}
\label{realkq}
q\ \Lambda^{(12)}\,\Lambda^{(34)} -
q^{-1}~\Lambda^{(13)}\,\Lambda^{(13)}\,^\dag -
q~\Lambda^{(14)}\,\Lambda^{(14)}\,^\dag = 0 \ . 
\end{equation}
Our choice of deformation (\ref{qvalues}) and of the
real structure (\ref{stargrass}) is distinguished by the fact that, in
the classical case, the quadratic form in (\ref{realkq}) has signature
$(5,1)$ which is the correct signature to interpret the relation (\ref{realkq}) as
the equation of a four-sphere in a real slice of $\complex\P^5$. Other choices would
lead to different signature and hence to other real embedded subvarieties,
see e.g.~\cite[Chap.~III.1]{Atiyahbook}.

We interpret the corresponding $*$-algebra as the coordinate algebra
$\alg(S_\theta^4)$ of a noncommutative four-sphere $S_\theta^4$ with a
non-central ``radius''. It is rather different from the spheres considered
e.g. in~\cite{cl} {or even in~\cite{lpr,BL11}}. To see this, let us redefine the
generators by writing $\frac q2\, (\Lambda^{(12)}- \Lambda^{(34)}) =: X$ and
$\frac q2\,(\Lambda^{(12)}+\Lambda^{(34)})=: R$, where both elements $R,X$ are hermitean
and commute with each other but not with the remaining minors. Simple algebra then transforms the relation \eqref{realkq} into
\begin{equation}
\label{realkq-bis}
\Lambda^{(13)}\,\Lambda^{(13)}\,^\dag +
q^{2}~\Lambda^{(14)}\,\Lambda^{(14)}\,^\dag + X^2 = R^2 \ . 
\end{equation}
This is a deformation of the equation of a four-sphere in homogeneous
coordinates on a real slice of $\complex\P^5$, with a 
{non-central radius~$R$. Due to this, one is not allowed to fix it to some real number (typically $R=1$ in the classical case). Furthermore, in the present noncommutative setting the homogeneous element $R$ does not generate 
a right or left denominator set, i.e. powers of the corresponding fraction element would not close to an algebra.  
Thus it is not possible to consider Ore localization with respect to
$R$, an operation that classically would correspond to ``rescaling''
the homogeneous coordinates to get a description of the sphere in
affine coordinates of $\real^5$.
Thus, for our noncommutative sphere there is no hope for any global
affine description. However, we will now see that one can still use localization to go to local patches for the sphere.}

Using the two hermitean generators $\Lambda^{(12)}$ and $\Lambda^{(34)}$ we can define two
localizations to affine subvarieties of the grassmannian
$\Gr_{\theta}(2;4)$, whose ``real slices'' are interpreted as local
patches of the sphere $S_\theta^4$. The two generators
$\Lambda^{(12)}$ and $\Lambda^{(34)}$ are rather different in nature, and
hence so are the corresponding localizations, as $\Lambda^{(34)}$ is central
while $\Lambda^{(12)}$ is not. From the commutation relations
(\ref{ncrelpr}) with $d=2$ and $n=4$,
together with the $q$-values (\ref{qvalues}), one finds indeed that $\Lambda^{(34)}$ is a central element of the
homogeneous coordinate algebra
$\alg(\Gr_{\theta}(2;4))$.

\begin{proposition}
The central noncommutative minor $\Lambda^{(34)}$ generates a right
denominator set in $\alg(\Gr_{\theta}(2;4))$, and the degree~zero
subalgebra $\alg(\Gr_{\theta}(2;4))[\Lambda^{(34)}\,^{-1}]_0$ of
the right Ore localization of $\alg(\Gr_{\theta}(2;4))$ with
respect to $\Lambda^{(34)}$ is isomorphic to the $\complex$-algebra generated by
elements $\xi_i,\bar\xi_i$, $i=1,2$ with the relations
\begin{eqnarray*}
\xi_1\,\bar\xi_1=q^2~\bar\xi_1\,\xi_1 \qquad &,& \qquad
\xi_2\,\bar\xi_2=q^{-2}~\bar\xi_2\,\xi_2 \ , \\[4pt]
\xi_1\,\xi_2=q^2~\xi_2\,\xi_1 \qquad &,& \qquad
\bar\xi_1\,\bar\xi_2=q^{-2}~\bar\xi_2\,\bar\xi_1 \ , \\[4pt]
\xi_1\,\bar\xi_2=\bar\xi_2\,\xi_1 \qquad &,& \qquad
\xi_2\,\bar\xi_1=\bar\xi_1\,\xi_2 \ .
\end{eqnarray*}
\label{R4prop}
\end{proposition}
\Proof{
As already noted, the minor $\Lambda^{(34)}$ is a central element of
$\alg(\Gr_{\theta}(2;4))$, whence its non-negative powers forms a right (and
left) denominator set. The degree~zero subalgebra of
$\alg(\Gr_{\theta}(2;4))[\Lambda^{(34)}\,^{-1}]$ is generated by
the elements
\begin{eqnarray*}
\xi_1=-\Lambda^{(14)}\,\Lambda^{(34)}\,^{-1} \qquad &,& \qquad
\xi_2= -\Lambda^{(24)}\,\Lambda^{(34)}\,^{-1} \ , \\[4pt]
\bar\xi_1=\Lambda^{(23)}\,\Lambda^{(34)}\,^{-1} \qquad &,& \qquad
\bar\xi_2=\Lambda^{(13)}\,\Lambda^{(34)}\,^{-1} \ ,
\end{eqnarray*}
together with
$$
\rho=q\ \Lambda^{(12)}\,\Lambda^{(34)}\,^{-1} \ .
$$
Since $\alg(\Gr_{\theta}(2;4))[\Lambda^{(34)}\,^{-1}]$ is a commutative localization, a straightforward calculation using (\ref{ncrelpr}) 
establishes the commutation relations among the generators $\xi_i,\bar\xi_i$, $i=1,2$,
while the Pl\"ucker relation (\ref{pl24}) becomes
$$
\rho=\xi_1\,\bar\xi_1-\bar\xi_2\,\xi_2 \ ,
$$
showing that the generator $\rho$ in the algebra
$\alg(\Gr_{\theta}(2;4))[\Lambda^{(34)}\,^{-1}]_0$ is redundant.
}

The $*$-involution of \eqref{xdelta} gives on the generators $\xi_i,\bar\xi_i$, $i=1,2$ the relations
\beq
\xi_1^\dag=q^{-1}~ \bar\xi_1 \ , \qquad \xi_2^\dag=-q^{-1}~ \bar\xi_2 \ ,  
\label{starxi}\eeq
from which it also follows that $\rho= q\ \xi_1\, \xi_1^\dag + q^3\ \xi_2\, \xi_2^\dag$. The corresponding 
$*$-algebra is dual to
a noncommutative real variety denoted $\real_\theta^4$. This
noncommutative space differs from the $q$-deformed euclidean space $\real_{q,\hbar=0}^4$
considered in~\cite[\S9]{KKO}; it is somewhat analogous to the quantum Minkowski
space constructed in~\cite{FJ}, although its origin is very
different. The present deformation orginates through the general categorical prescription of \S\ref{toriccat} via the natural action of the torus $T=(\complex^\times)^2$ on $\real^4$ described by the left coaction 
$$
\Delta_L\,:\, \alg\big(\real^4\big)~ \longrightarrow~ \hil\otimes \alg\big(\real^4\big)
$$ 
with
\begin{eqnarray}
\Delta_L(\xi_1)=t_1\otimes\xi_1 \quad &,& \quad \Delta_L(\xi_2)=t_2\otimes\xi_2 \ , \nonumber \\[4pt]
\Delta_L(\bar\xi_1)=t_2\otimes\bar\xi_1 \quad &,& \quad \Delta_L(\bar\xi_2)=t_1\otimes\bar\xi_2 \ ,
\label{DeltaLR4}\end{eqnarray}
and extended as an algebra map. These relations can be directly
obtained from the torus action on $\Gr(2;4)$ described in (\ref{grasstoric}) for our choice of deformation (\ref{qvalues}), which amount to acting 
non-trivially only on the indices $1$ and $2$. 

Geometrically, the space $\real_\theta^4$ is regarded
as an open affine subvariety of the noncommutative four-sphere $S_\theta^4$ defined before. 
A second open affine subvariety of $S_\theta^4$ is obtained by the
localization onto another affine subvariety of the grassmannian $\Gr_{\theta}(2;4)$ via the non-central hermitean minor 
$\Lambda^{(12)}$.
\begin{proposition}
The non-central noncommutative minor $\Lambda^{(12)}$ generates a left
denominator set in $\alg(\Gr_{\theta}(2;4))$, and the degree~zero
subalgebra ${}_0[\Lambda^{(12)}\,^{-1}]\alg(\Gr_{\theta}(2;4))$ of
the left Ore localization of $\alg(\Gr_{\theta}(2;4))$ with
respect to $\Lambda^{(12)}$ is isomorphic to the $\complex$-algebra generated by
elements $\zeta_i,\bar{\zeta_i}$, $i=1,2$ with the relations
\begin{eqnarray*}
\zeta_1\,\bar\zeta_1=q^{-2}~\bar\zeta_1\,\zeta_1 \qquad &,& \qquad
\zeta_2\,\bar\zeta_2=q^{2}~\bar\zeta_2\,\zeta_2 \ , \\[4pt]
\zeta_1\,\zeta_2=q^{-2}~\zeta_2\,\zeta_1 \qquad &,& \qquad
\bar\zeta_1\,\bar\zeta_2=q^{2}~\bar\zeta_2\,\bar\zeta_1 \ , \\[4pt]
\zeta_1\,\bar\zeta_2=\bar\zeta_2\,\zeta_1 \qquad &,& \qquad
\zeta_2\,\bar\zeta_1=\bar\zeta_1\,\zeta_2 \ .
\end{eqnarray*} 
\label{tildeR4prop}
\end{proposition}
\Proof{
From the commutation relations (\ref{ncrelpr}) with $d=2$, $n=4$,
together with the $q$-values (\ref{qvalues}), one has
$$
\big(\Lambda^{(12)}\big)\,\alg\big(\Gr_{\theta}(2;4)\big)=
\alg\big(\Gr_{\theta}(2;4)\big)\,\big(\Lambda^{(12)}\big)
$$ 
as sets. Whence the
element $\Lambda^{(12)}$ is left (and right) permutable in
$\alg(\Gr_{\theta}(2;4))$, so the set of non-negative powers of
$\Lambda^{(12)}$ is a left (and
right) denominator set. The degree~zero subalgebra of
$[\Lambda^{(12)}\,^{-1}]\alg(\Gr_{\theta}(2;4))$ is generated by
the elements
\begin{eqnarray*}
\zeta_1=-\Lambda^{(12)}\,^{-1}\,\Lambda^{(14)} \qquad &,& \qquad
\zeta_2=-\Lambda^{(12)}\,^{-1}\,\Lambda^{(24)} \ , \\[4pt]
\bar{\zeta_1}=\Lambda^{(12)}\,^{-1}\,\Lambda^{(23)} \qquad &,& \qquad
\bar{\zeta_2}=\Lambda^{(12)}\,^{-1}\,\Lambda^{(13)} \ ,
\end{eqnarray*}
together with
$$
\tilde\rho=q\ \Lambda^{(12)}\,^{-1}\,\Lambda^{(34)} \ .
$$ 
A straightforward calculation using (\ref{ncrelpr}) together with the rules for noncommutative Ore localization establishes the
commutation relations among the generators $\zeta_i,\bar \zeta_i$, $i=1,2$,
while the Pl\"ucker relation (\ref{pl24}) becomes
$$
\tilde\rho=q^2~\zeta_1\,\bar\zeta_1-q^{2}~\bar\zeta_2\,\zeta_2  = \bar\zeta_1 ~\zeta_1 - \zeta_2~\bar\zeta_2 
$$
similarly to the previous case.
}

The $*$-involution on the algebra generated by
$\zeta_i,\bar\zeta_i$, $i=1,2$ induced from the real structure (\ref{stargrass}) reads 
\beq
\zeta_1^\dag=q~\bar \zeta_1 \ , \qquad 
\zeta_2^\dag=-q^{-3}~\bar \zeta_2 \ , 
\label{starzeta}\eeq
from which it also follows that $\tilde\rho= q\ \zeta_1 \,
\zeta_1^\dag + q^3\ \zeta_2 \, \zeta_2^\dag$. The corresponding noncommutative real
variety is denoted $\widetilde\real_\theta^4$. The counterpart of the coaction \eqref{DeltaLR4} is now given 
by the dual left coaction 
$$
\Delta_L\,:\, \alg\big(\widetilde\real_\theta^4\big)~ \longrightarrow~ \hil_\theta \otimes \alg\big(\widetilde\real_\theta^4\big)
$$ 
with
\begin{eqnarray}
\Delta_L(\zeta_1)=t_2^{-1}\otimes\zeta_1 \quad &,& \quad \Delta_L(\zeta_2)=t_1^{-1}\otimes\zeta_2 \ , \nonumber \\[4pt]
\Delta_L(\bar\zeta_1)=t_1^{-1}\otimes \bar\zeta_1 \quad &,& \quad \Delta_L(\bar\zeta_2)=t_2^{-1}\otimes \bar\zeta_2 \ ,
\label{DeltaLR4tilde}\end{eqnarray}
and again extended as an algebra map.

The intersection of the two
open affine subvarieties $\real_\theta^4$ and $\widetilde\real_\theta^4$ is
described by adjoining the element $\tilde\rho$ to $\alg(\real_\theta^4)$
and $\rho$ to $\alg(\widetilde\real_\theta^4)$, and computing the
gluing automorphism between the two resulting algebras. This also defines
the noncommutative real variety $S_\theta^4$ in an analogous manner as our
generic noncommutative toric varieties. 
\begin{proposition}
The algebras $\alg(\real_\theta^4)[\tilde\rho\, ]$ and
$[\rho]\alg(\widetilde\real_\theta^4)$ are
isomorphic as $*$-algebras in the category ${}^{\hil_{\theta}}\Module$ of left
$\hil_{\theta}$-comodules.
\label{R4glueprop}\end{proposition}
\Proof{
Define an algebra morphism
$$
G\,:\, \alg(\real_\theta^4)[\tilde\rho\, ] ~\longrightarrow~
[\rho]\alg(\widetilde\real_\theta^4)
$$
on generators by
$$
\big(\xi_j\,,\,\bar\xi_j \, \big)~\longmapsto~ \big(\rho ~\zeta_j\,,
\rho ~\bar\zeta_j\,\big) \qquad \mathrm{for} \quad j=1,2 \ . 
$$
The inverse map $G^{-1}$ is then given by
$$
\big(\zeta_j\,,\,\bar\zeta_j \, \big)~\longmapsto~ \big(\tilde\rho ~\xi_j\,, \tilde\rho ~\bar\xi_j\,\big)
\qquad \mathrm{for} \quad j=1,2 \ ,
$$
and one can check that $G$ is an algebra isomorphism. Moreover, one has $G(\xi_i^\dag\,) = G(\xi_i)^\dag$, and hence $G(a^\dag\,)=G(a)^\dag$ for all 
$a \in \alg(\real_\theta^4)[\tilde\rho\, ]$. Finally, using the
coactions (\ref{DeltaLR4}) and (\ref{DeltaLR4tilde}) we compute
$$ \Delta_L(\rho)=t_1\,t_2\otimes\rho \ , \qquad
\Delta_L(\tilde\rho\, )=(t_1\,t_2)^{-1}\otimes\tilde\rho \ , $$
from which
one easily shows that the map $G$ is coequivariant,
i.e. $\Delta_L\circ G=(\Id\otimes G)\circ\Delta_L$, and hence
is a $*$-isomorphism in the category ${}^{\hil_{\theta}}\Module$.
}

\begin{remark}
\textup{
The ``geometric'' interpretation of the map $G$ in the proof of
Proposition~\ref{R4glueprop} is as follows. In the overlap of the two
patches there are two sets of generators to describe ``points'', and $G$
describes how to pass from the affine coordinates $\xi_j$ to the affine
coordinates $\zeta_j$. It is indeed the identity map in terms of the
homogeneous coordinates $\Lambda^J$ on the grassmannian, i.e. we do not ``move'' points, we just describe how the coordinates of the two patches are related. 
}
\label{Ginterpret}\end{remark}

\subsection{Twistor fibration\label{Twistorfib}}
{As a particular case of the noncommutative projective spaces in \S\ref{NCCPn}, one has 
the \emph{noncommutative twistor algebra} $\alg^{\rm tw}=\alg(\CP_{\theta}^3)$, } the
homogeneous coordinate algebra generated by $w_i$,
$i=1,2,3,4$ and the relations
\bea
w_i\,w_k&=&w_k\,w_i \ , \qquad i=1,2,3,4~,~k=3,4 \ , \nonumber\\[4pt]
w_1\,w_2&=&q^2~w_2\,w_1 \ .
\label{CP3rels}\eea
It is dual to the (complex) twistor space of the noncommutative sphere~$S_\theta^4$.  
{When} $q\in\real$,
there is a natural real structure on $\alg^{\rm tw}$ such that
\beq
w_1^\dag=w_2 \ , \qquad w_2^\dag=w_1 \ , \qquad w_3^\dag=w_4 \ ,
\qquad w_4^\dag=w_3 \ .
\label{starCP3}\eeq

The restriction functor induced by Proposition~\ref{R4prop} is denoted
$$
j^\bullet\,:\, \coh\big(\Gr_{\theta}(2;4)\big) ~\longrightarrow~
\coh\big(\real_\theta^4\big) \ . 
$$
At the level of the noncommutative
correspondence algebra $\alg(\Fl_{\theta}(1,2;4))$, regarded as
an $\alg(\Gr_{\theta}(2;4))$-bimodule, the Pl\"ucker
equations (\ref{flagpluck}) in the localized coordinate algebra
$\alg(\Fl_{\theta}(1,2;4))[\Lambda^{(34)}\,^{-1}]_0$ now read
\beq
w_1=-w_3\,\xi_1-w_4\,\bar\xi_2 \ , \qquad
w_2=-w_3\,\xi_2-w_4\,\bar\xi_1 \ .
\label{w1w2triv}\eeq
Using the commutation relations (\ref{wLambda}) together with the
multiplication rule of noncommutative Ore
localization~\cite[\S1.3]{CLSI}, one easily checks that the generators
$w_3,w_4$ commute not only among themselves but also with
$\xi_i,\bar\xi_i$, $i=1,2$, and it follows that
\beq
\alg\big(\Fl_{\theta}(1,2;4)\big)
\big[\Lambda^{(34)}\,^{-1}\big]_0
\cong\alg\big(\real_\theta^4\big)\otimes \alg\big(\CP^1\big) \ ,
\label{flagloc34}\eeq
with $\alg(\CP^1)=\complex[w_3,w_4]$ the homogeneous coordinate algebra of a commutative projective line $\CP^1$. This isomorphism implies that, in the image of the functor
$j^\bullet$, the tautological bundle $\Scal_{\theta}$ obtained
through the twistor transform via Lemma~\ref{twisttransfbundles} restricts to the free right
$\alg(\real_\theta^4)$-module of rank~two, spanned by $w_3$ and $w_4$.

The situation is somewhat different for the noncommutative localization described by
Proposition~\ref{tildeR4prop}. The Pl\"ucker
equations (\ref{flagpluck}) in
${}_0[\Lambda^{(12)}\,^{-1}]\alg(\Fl_{\theta}(1,2;4))$ are
\beq
w_3=-q~ \bar\zeta_1\, w_1 + q^{-1}\ \bar\zeta_2\,w_2 \ , \qquad
w_4=q~ \zeta_2\,w_1 - q^{-1}\ \zeta_1\,w_2 \ .
\label{w3w4triv}\eeq
Now, however, the generators $w_1,w_2$ do not commute with
$\zeta_i,\bar\zeta_i$, $i=1,2$ in general; one finds
\begin{eqnarray*}
 w_1\, \zeta_1=q^{-2}~ \zeta_1\, w_1 \qquad &,& \qquad w_1\,\bar\zeta_1= \bar\zeta_1\, w_1 \ , \\[4pt]
 w_1\, \zeta_2= \zeta_2\, w_1 \qquad &,& \qquad w_1\,\bar\zeta_2=q^{-2}~\bar\zeta_2\, w_1 \ , \\[4pt]
 w_2\, \zeta_1= \zeta_1\, w_2 \qquad &,& \qquad w_2\,\bar\zeta_1= q^{2}~ \bar\zeta_1\, w_2 \ , \\[4pt]
 w_2\, \zeta_2=q^{2}~  \zeta_2\, w_2 \qquad &,& \qquad w_2\,\bar\zeta_2= \bar\zeta_2\, w_2 \ .
\end{eqnarray*} 
As a consequence, the localized coordinate algebra
has the structure of the braided tensor product algebra
$$
{}_0\big[\Lambda^{(12)}\,^{-1}\big]\alg\big(\Fl_{\theta}(1,2;4)\big)
\cong
\alg\big(\CP_\theta^1\big)\otimes_\theta\,
\alg\big(\widetilde\real_\theta^4\big) \ ,
$$
where the noncommutative projective line $\CP_\theta^1$ has
homogeneous coordinates $w_1,w_2$ subject to the relations
(\ref{CP3rels}). We will show in Proposition~\ref{CP1equivprop} below that coherent sheaves on $\CP_\theta^1$
can be functorially identified with sheaves on a commutative line
$\CP^1$, 
and hence, in the image of the restriction functor $\tilde j\,^\bullet:\coh(\Gr_{\theta}(2;4)) \to
\coh(\widetilde\real_\theta^4)$ induced by Proposition~\ref{tildeR4prop},
the tautological bundle $\Scal_\theta$ restricts to the free right
$\alg(\widetilde\real_\theta^4)$-module of rank~two. The free modules
$\alg(\real_\theta^4)\otimes\complex^2$ and
$\complex^2\otimes\alg(\widetilde\real_\theta^4)$ carry natural
$*$-involutions induced by (\ref{starxi}), (\ref{starzeta}) and
(\ref{starCP3}), and by Proposition~\ref{R4prop},
Proposition~\ref{tildeR4prop} and Proposition~\ref{R4glueprop} there is naturally an isomorphism
$$
G_2\,:\, \alg(\real_\theta^4)[\tilde\rho\,]\otimes\complex^2
~\longrightarrow~ \complex^2\otimes[\rho]\alg(\widetilde\real_\theta^4)
$$
in the category ${}^{\hil_\theta}\Module$, which is compatible with the $*$-structures and satisfies $G_2(v\triangleleft
a)=G(a)\triangleright G_2(v)$ for all $a\in
\alg(\real_\theta^4)[\tilde\rho\,]$ and $v\in
\alg(\real_\theta^4)[\tilde\rho\,]\otimes\complex^2$. This describes
the twistor bundle over the noncommutative sphere $S_\theta^4$.

\newsection{Instanton counting on $\CP_\theta^2$\label{Instcounting}}

\subsection{Framed modules}

The noncommutative projective plane $\CP_\theta^2$ was described in
\cite[\S3.3]{CLSI} using combinatorial data encoded in
the fan of $\CP^2$,
and in \S\ref{NCCPn} above via the homogeneous coordinate algebra
$\alg=\alg(\CP_\theta^2)$ whose generators have relations 
\beq
w_1\,w_2=q^2~w_2\,w_1 \ , \qquad w_1\,w_3=w_3\,w_1 \ , \qquad
w_2\,w_3=w_3\,w_2  \ ,
\label{CP2homcoordrels}\eeq
where $q=\exp(\frac\ii2\,\theta)$ with $\theta\in\complex$. Let
$\alg_\ell:=\alg/ ( \alg\cdot w_3 )$. We identify
$\alg_\ell=\alg(\CP_\theta^1)$ as the homogeneous coordinate algebra
dual to a \emph{noncommutative projective line} $\CP_\theta^1$. The
algebra projection $p:\alg\to\alg_\ell$ is dual to a closed
embedding $\CP_\theta^1\hookrightarrow\CP_\theta^2$ of noncommutative
projective varieties.

For any pair $\theta,\theta'\in\complex\setminus \pi\,\zed$, the abelian categories
of coherent sheaves $\coh(\CP_\theta^2)$ and $\coh(\CP_{\theta'}^2)$ are equivalent, while
$\coh(\CP_\theta^2)\ncong \coh(\CP^2)$ for any $\theta\notin \pi\,\zed$. On the other
hand, the category $\coh(\CP_\theta^1)$ is independent of
$\theta\in\complex$;
the geometric structure of the algebra
$\alg_\ell$ thus agrees
with general expectations that generic deformed curves in algebraic
geometry are commutative (see e.g.~\cite[Cor.~5.3]{ingalls} and~\cite{Artin-S}).
\begin{proposition}
For any $\theta\in\complex$ there is a natural equivalence of abelian categories
$$
\coh\big(\CP_\theta^1\big)~\cong~\coh\big(\CP^1\big) \ .
$$
\label{CP1equivprop}\end{proposition}
\Proof{
By~\cite[Thm.~5.4]{CLSI}, the degree~zero subalgebra of the left Ore localization of the
noncommutative algebra $\alg(\CP_\theta^1)$ on maximal cones is given
by 
$$
\big(\alg\,\big/\,(\alg\cdot w_3)\big)\big[w_i^{-1}\big]_0
\cong\complex[y_{i+1}] \ ,
$$
where $y_{i+1}=w_i^{-1}\,w_{i+1}$ for $i=1,2$ (mod~$2$). This algebra
is the same as the corresponding localization of the commutative
homogeneous coordinate algebra $\alg(\CP^1)=\complex[w_1,w_2]$, and by~\cite[Prop.~4.6]{CLSI} the result follows.
}

This proposition will enable us to exploit the known cohomology of
sheaves on the commutative projective line $\CP^1$, since the functors 
$\Ext^p$, $\Extc^p$ and $H^p$ all commute with the functorial
equivalence. For example, a sheaf $E$ on $\Open(\CP_\theta^1)$ is
locally free if and only if it is locally free as a sheaf on $\CP^1$
under the equivalence of Proposition~\ref{CP1equivprop}, and hence by the
Birkhoff--Grothendieck theorem it is isomorphic to a finite direct sum
of rank one bundles $E\cong\sheaf_{\CP_\theta^1}(k_1)\oplus\cdots
\oplus\sheaf_{\CP_\theta^1}(k_s)$ for some integers
$k_1,\dots,k_s$. The inclusion of algebras
$i:\alg_\ell\hookrightarrow\alg$ induces a restriction functor
$i^\bullet:\coh(\CP_\theta^2)\to\coh(\CP_\theta^1)$ defined on right
$\alg$-modules $M$ by
$$
i^\bullet\big(\pi(M)\big)=\pi\big(M\,/\, (M\triangleleft w_3) \big) \ .
$$
By~\cite[Prop.~4.6]{CLSI} and~\cite[Prop.~3.3.9~(1)]{BGK}, the functor $i^\bullet$ is
exact and its right adjoint is the faithful, exact push-forward
functor $i_\bullet=p^\bullet:\coh(\CP_\theta^1)\to\coh(\CP_\theta^2)$
induced by the algebra projection $p:\alg\to\alg_\ell$, and defined on
$\alg_\ell$-modules $N$ by
$$
p^\bullet\big(\pi(N)\big)=\pi(N)
$$
with $w_3$ acting as $0$ on the right-hand side.

Let us fix (isomorphism classes of) complex vector spaces $V$ and $W$
of dimensions $k$ and $r$, respectively. The purpose of this section
is to describe the following (set-theoretic) moduli space.
\begin{definition}
\begin{itemize}
\item[(1)]
A \emph{$(W,V)$-framed sheaf} is a coherent torsion free sheaf $E$
on $\Open(\CP_\theta^2)$ such that there exists an isomorphism
$H^1(\CP_\theta^2,E(-1))\cong V$, together with an isomorphism
$i^\bullet(E)\cong W\otimes\sheaf_{\CP_\theta^1}$ called a
\emph{framing of $E$ (of type $W$) at infinity}. \\
\item[(2)]
A \emph{morphism of $(W,V)$-framed sheaves} $E$ and $E'$ is a
homomorphism of $\alg$-modules $\xi:E\to E'$ which
preserves the framing isomorphisms, i.e. there is a commutative
diagram
$$
\xymatrix{
E~\ar[rd]_{i^\bullet}\ar[r]^\xi & ~E' \ar[d]^{i^\bullet} \\
 & W\otimes\sheaf_{\CP_\theta^1} \ .
}
$$
\end{itemize}
\label{modspdef}\end{definition}

\begin{definition}
The \emph{(instanton) moduli space $\Mcal_\theta(W,V)$} is the set of isomorphism
classes $[E]$ of $(W,V)$-framed sheaves $E$. When bases have been 
fixed for the vector spaces $V\cong\complex^k$ and $W\cong\complex^r$,
we will denote this moduli space by $\Mcal_\theta(r,k)$.
\label{modspdef1}\end{definition}

\begin{proposition}
For any $(W,V)$-framed sheaf $E$ on $\Open(\CP_\theta^2)$ and for any
$k\in\zed$, there is a canonical exact sequence
\beq
0~\longrightarrow~E(k-1)~\xrightarrow{\cdot \, w_3}~E(k)~
\longrightarrow~p^\bullet\big(W\otimes\sheaf_{\CP_\theta^1}(k)\big)~
\longrightarrow~0 \ .
\label{EWcanexseq}\eeq
\label{CP21exseqprop}\end{proposition}
\Proof{
Using Proposition~\ref{CP1equivprop}, this is essentially a straightforward
adaptation of the proofs of~\cite[Lem.~6.1]{KKO}
and~\cite[Prop.~3.3.9]{BGK}.
}

The framed sheaf cohomology of $\CP_\theta^2$ can be described as
follows.
\begin{proposition}
For any $(W,V)$-framed sheaf $E$ on $\Open(\CP_\theta^2)$, one has:
\begin{itemize}
\item[(1)] $H^0\big(\CP_\theta^2\,,\,E(-1)\big) \= 0 \=
H^0\big(\CP_\theta^2\,,\,E(-2)\big) \ ; $ \\
\item[(2)] $H^2\big(\CP_\theta^2\,,\,E(-1)\big) \= 0 \=
H^2\big(\CP_\theta^2\,,\,E(-2)\big) \ ; $ and \\
\item[(3)] $H^1\big(\CP_\theta^2\,,\,E(-1)\big) \= V \=
H^1\big(\CP_\theta^2\,,\,E(-2)\big) \ . $
\end{itemize}
\label{H021CP2vanprop}\end{proposition}
\Proof{
Use Proposition~\ref{CP1equivprop} and Proposition~\ref{CP21exseqprop}, and repeat the
proofs of~\cite[Lem.~4.2.12]{BGK} and~\cite[Lem.~6.2]{KKO}. 
}
\begin{cor}
A framed sheaf $E$ with isomorphism class $[E]\in\Mcal_\theta(r,k)$
has invariants
$$
{\rm rank}(E)=r \ , \qquad c_1(E)=0 \ , \qquad \chi(E)=r-k \ .
$$
\label{framedtopinvscor}\end{cor}
\Proof{
By~\cite[Lem.~6.1]{NS}, the Hilbert polynomial of $E$ can be expressed
as
\beq
h_E(s)=\mbox{$\frac12$}~\rank(E)\,s\,(s+1)+\big(c_1(E)+\rank(E)
\big)\,s+\chi(E) \ .
\label{hEsCP2}\eeq
Using (\ref{hEsCP2}), Proposition~\ref{CP21exseqprop} 
and~\cite[Prop.~3.3.9]{BGK}, one has
$h_{i^\bullet(E)}(s)=h_E(s)-h_E(s-1)$ and hence
$$
\rank(E)=\rank\big(i^\bullet(E)\big)=\rank\big(W\otimes
\sheaf_{\CP_\theta^1}\big)=r \ .
$$
Using Proposition~\ref{H021CP2vanprop}, and setting $s=-1$ and $s=-2$ in
(\ref{hEsCP2}), we arrive at the respective equations
$$
-k=-r-c_1(E)+\chi(E) \ , \qquad -k=-r-2c_1(E)+\chi(E) \ ,
$$
which together yield $c_1(E)=0$ and $\chi(E)=r-k$.
}

In~\S\ref{Instmodsp} we will indeed demonstrate that the set $\Mcal_\theta(r,k)$ is a (coarse) moduli space by constructing universal modules (co)representing the moduli functor of framed torsion free objects of $\coh(\CP_\theta^2)$. The construction will rely crucially on the following basic property.
\begin{lemma}
Every $(W,V)$-framed sheaf on $\Open(\CP_\theta^2)$ is $\mu$-semistable.
\label{framedstablelemma}\end{lemma} 
\Proof{
Let $[E]\in\Mcal_\theta(W,V)$ and suppose that $0\neq F\subsetneq E$
is a proper subsheaf of $E$. Without loss of generality we may assume
that $E/F$ is torsion free. The restriction functor
$i^\bullet:\coh(\CP_\theta^2)\to \coh(\CP_\theta^1)$ is exact, so
$i^\bullet(F)\subset i^\bullet(E)$. Then by additivity of the first Chern class and Corollary~\ref{framedtopinvscor} it follows that
$$
c_1(F)=c_1\big(i^\bullet(F)\big) \leq c_1\big(i^\bullet(E)\big) =
c_1(E)=0 \ ,
$$
and thus $F$ cannot de-semistabilize $E$.
}

\subsection{Noncommutative ADHM construction\label{NCADHM}}

We will now reduce the study of moduli of $(W,V)$-framed sheaves on
$\Open(\CP_\theta^2)$ to a problem of \emph{linear} algebra, which
defines a noncommutative deformation of the standard ADHM data. For
this, we consider triples 
\bea
(B,I,J)&\in& \Xcal(W,V) \nonumber\\ && := \Hom_\complex\big(V\,,\,
V\otimes(\alg_\ell^!)_1\big)~\oplus~ \Hom_\complex\big(W\,,\,
V\otimes(\alg_\ell^!)_2\big)~ 
\oplus~\Hom_\complex(V,W) \ ;
\label{BIJsubvar}\eea
here $\alg_\ell^!=\alg^!/ ( \alg^!\cdot \check w_3 ) $ is the Koszul dual
of the homogeneous coordinate algebra of the noncommutative projective
line $\CP_\theta^1$ (see (\ref{Koszuldualrels})), with
generators $\check w_1,\check w_2$ satisfying relations
\beq
\check w_1^2=0=\check w_2^2 \ , \qquad
\check w_1\,\check w_2=-q^2~\check w_2\,\check w_1 \ ,
\label{CP1Koszulrels}\eeq
while for its components one has isomorphisms $(\alg_\ell^!)_1\cong (\alg_\ell)_1^*$ and $(\alg_\ell^!)_2\cong
\bigwedge_\theta^2\,(\alg_\ell)_1^*$ as vector spaces in the category ${}^{\hil_{\theta}}\Module$ of left
$\hil_{\theta}$-comodules. Given $B$ as above, define a vector
space morphism $B\wedge_\theta B:V\to V\otimes(\alg_\ell^!)_2$ as the
composition of maps
\beq
B\wedge_\theta B=\big(\Id_V\otimes\mu_{\alg_\ell^!}\big)\circ
\big(B\otimes\Id_{\alg_\ell^!}\big)\circ B \ ,
\label{BwedgeB}\eeq
where $\mu_{\alg_\ell^!}:\alg_\ell^!\otimes \alg_\ell^!\to
\alg_\ell^!$ is the multiplication map of the algebra
$\alg_\ell^!$.

Since $(\alg_\ell^!)_1\cong \complex\check
w_1\oplus\complex \check w_2$, we can write the map $B:V\to
V\otimes(\alg_\ell^!)_1$ in the form
\beq
B=B_1\otimes\check w_1+B_2\otimes\check w_2
\label{Bvexpl}\eeq
with $B_1,B_2\in\End_\complex(V)$. Using (\ref{CP1Koszulrels}) we can
then evaluate the morphism (\ref{BwedgeB}) as
$$
B\wedge_\theta B=[B_1,B_2]_\theta\otimes\check w_1\,\check w_2
$$
where $[B_1,B_2]_{-\theta}\in\bigwedge^\bullet_\theta\,\End_\complex(V)$
is the \emph{braided commutator}
\bea
&& \label{thetacomm} \\
&& [B_1,B_2]_{-\theta} :=\mu_{\End_\complex(V)}\big(B_1\otimes
B_2-\Psi_{\End_\complex(V),\End_\complex(V)}^\theta(B_1\otimes B_2)\big)=
B_1\,B_2-q^{2}~B_2\,B_1 \ . \nonumber
\eea
The braiding here arises when we consider $\hil_\theta$-coequivariant
sheaves, as in \S\ref{subsec:coeq-sheaves}, using the natural coactions of the cotwisted Hopf algebra $\hil_\theta$
induced by the torus $T=(\complex^\times)^2$ on the moduli
spaces. Then by point~(1) of Definition~\ref{modspdef} the
finite-dimensional vector spaces $V$ and $W$ are
$\hil_\theta$-comodules, i.e. objects of the category
${}^{\hil_{\theta}}\Module$, and
hence so is the vector space $\End_\complex(V)\cong V^*\otimes
V\cong V\otimes V$. The explicit coaction on the triples
(\ref{BIJsubvar}) can be computed from the
coaction (\ref{DeltaLdual}), Lemma~\ref{eqmapslemma}, and the monadic
parametrization (\ref{sigmaBIJ})--(\ref{tauBIJ}) of framed sheaves
below. In this way one finds that the triples are
$\hil_\theta$-coinvariants, i.e.
\beq
\Delta_L(B,I,J)=(1\otimes B,1\otimes I, 1\otimes J) \ ,
\label{DeltaLBIJ}\eeq
which using (\ref{Bvexpl}) implies
$$
\Delta_L(B_i)=t_i\otimes B_i \qquad \mbox{for} \quad i=1,2 \ .
$$
As mentioned in \S\ref{subsec:coeq-sheaves}, here we only regard the
parameter subspace of $\End_\complex(V)$ as a vector space object of
${}^{\hil_{\theta}}\Module$, hence we do not deform the product on the
endomorphism algebra in (\ref{thetacomm}); the deformation in
(\ref{thetacomm}) follows from (\ref{PsithetaVW}).

\begin{definition}
The variety $\Mcal_\theta^{\rm ADHM}(W,V)$ of \emph{noncommutative
  complex ADHM data} is the locally closed subvariety of triples
(\ref{BIJsubvar}) subject to the following two conditions: 
\begin{itemize}
\item[(1)] The \emph{noncommutative complex ADHM equation}:
\beq
B\wedge_\theta B+I\circ J=0
\label{NCADHMeqs}\eeq
in $\Hom_\complex(V,V\otimes(\alg_\ell^!)_2)$; and \\
\item[(2)] The \emph{stability condition}: \ There are no proper
  non-zero subspaces $V'$ of $V$ such that $B(V'\,)\subset V'\otimes
  (\alg_\ell^!)_1$ and $I(W)\subset V'\otimes(\alg_\ell^!)_2$. 
\end{itemize}
\label{NCADHMvar}\end{definition}

The general linear group $\GL(V)$ of automorphisms of the vector space
$V$ acts naturally on the variety $\Mcal_\theta^{\rm ADHM}(W,V)$ as 
\beq
g\triangleright(B,I,J)=\big(\tilde g\circ B\circ g^{-1}\,,\,\tilde
g\circ I\,,\,J\circ g^{-1}\big) \ , 
\label{GLVaction}\eeq
where $g\in \GL(V)$ and $\tilde g:=g\otimes\Id_{\alg_\ell^!}$.
\begin{lemma}
The natural action of $\GL(V)$ on $\Mcal_\theta^{\rm ADHM}(W,V)$ is
free and proper.
\label{stablem}\end{lemma}
\Proof{
Suppose that the triple $(B,I,J)\in\Mcal_\theta^{\rm ADHM}(W,V)$ is
fixed by $g\in \GL(V)$. Then $\tilde g\circ B\,\circ g^{-1}=B$ and
$\tilde g\circ I=I$, which respectively imply that $V'=\ker(g-\Id_V)$
is $B$-coinvariant and that $I(W)\subset V'\otimes(\alg_\ell^!)_2$. By
the stability condition of Definition~\ref{NCADHMvar}, it follows
that~$g=\Id_V$. Thus $\GL(V)$ acts freely on $\Mcal_\theta^{\rm
  ADHM}(W,V)$.

Suppose now that the $\GL(V)$-orbit of some triple
$(B,I,J)\in\Mcal_\theta^{\rm ADHM}(W,V)$ is not closed. Then there is
a non-trivial one-parameter subgroup $\lambda:\complex^\times\to \GL(V)$
such that the limit
$(B_0,I_0,J_0)=\lim_{t\to0}\,\lambda(t)\triangleright (B,I,J)$ exists
but does not belong to the orbit $\GL(V)\triangleright (B,I,J)$. Let
$V=\bigoplus_{m\in\zed}\,V_m$ be the weight decomposition of the
vector space $V$ with respect to the subgroup
$\lambda(\complex^\times)$. The existence of the limit $(B_0,I_0,J_0)$ implies that
$B(V_m)\subset \big(\bigoplus_{k\geq
  m}\,V_k\big)\otimes(\alg_\ell^!)_1$ and $I(W)\subset
\big(\bigoplus_{k\geq0}\,V_k\big)\otimes(\alg_\ell^!)_2$. Set
$V'=\bigoplus_{k\geq0}\,V_k$. Then $B(V'\,)\subset
V'\otimes(\alg_\ell^!)_1$ and $I(W)\subset
V'\otimes(\alg_\ell^!)_2$. Since
$(B_0,I_0,J_0)$ does not belong to the orbit $\GL(V)\triangleright
(B,I,J)$, we must have that 
$\det(\lambda(t))=t^N$ for some $N<0$. This implies that $V'$ is a proper subspace of $V$, which
contradicts the stability condition of Definition~\ref{NCADHMvar}. Thus
$\GL(V)$ acts properly on $\Mcal_\theta^{\rm ADHM}(W,V)$.
}

It follows from Lemma~\ref{stablem} that the quasi-projective variety of
closed $\GL(V)$-orbits on the space $\Mcal_\theta^{\rm ADHM}(W,V)$ is
given by the geometric invariant theory quotient
\beq
\widehat\Mcal_\theta^{\rm ADHM}(W,V):=\Mcal_\theta^{\rm ADHM}(W,V)\,
\big/\,\GL(V) \ .
\label{MhatADHM}\eeq
Our first characterization of
the moduli space of Definition~\ref{modspdef1} is
then as follows.
\begin{theorem}
There is a natural (set-theoretic) bijection between the moduli space
$\widehat\Mcal_\theta^{\rm ADHM}(W,V)$ of braided linear
algebraic ADHM data and the moduli space $\Mcal_\theta(W,V)$
of framed sheaves on $\Open(\CP_\theta^2)$. 
\label{ADHMbij}\end{theorem}

{This theorem is proved in~\S\ref{Beilinson} below.}
In~\S\ref{Instmodsp} we will analyse to what extent this bijection establishes an
isomorphism of algebraic varieties; in particular, we will show that
the parametrization (\ref{MhatADHM}) is a fine moduli space for
$(W,V)$-framed sheaves on $\Open(\CP_\theta^2)$ and that this
isomorphism induces the bijection of Theorem~\ref{ADHMbij}. It shows that framed torsion free
sheaves on $\Open(\CP_\theta^2)$ are in a bijective correspondence
with stable framed representations of the ADHM
quiver 
$$
\xymatrix{
v \ \bullet \ \ar@(ur,ul)|{\, b_1 \,} \ar@(dl,dr)|{\, b_2 \,} \ar@/^/[rr]|{\, j \,} &&
\ \bullet \ w \ar@/^/[ll]|{\, i \,}
}
$$
in the category of complex vector spaces, with a deformation of the
usual relation specified by the $\complex$-linear combination of paths
\beq
b_1\,b_2 - q^{-2}\ b_2\, b_1 +i\,j \ .
\label{ADHMquiverrels}\eeq
In the
$\hil_\theta$-coequivariant case, the bijection gives a $T$-equivariant
isomorphism of algebraic spaces and this category can be replaced with
the braided monoidal category ${}^{\hil_{\theta}}\Module$ of left $\hil_\theta$-comodules. 

\subsection{Beilinson monads for framed sheaves\label{Beilinson}}

We will now prove Theorem~\ref{ADHMbij}. For this, we will mimick the
classical approach. Thus we will exploit the Beilinson spectral
sequence of \S\ref{CPSheaves} to obtain a monadic description of
$(W,V)$-framed sheaves on $\Open(\CP_\theta^2)$, i.e. in terms of the 
cohomology of linear monads on $\CP_\theta^2$ as defined in
\S\ref{Monads}. We will need the following vanishing lemma, analogous 
to Proposition~\ref{H021CP2vanprop}.
\begin{lemma}\label{Kahlervanlem}
For any $(W,V)$-framed sheaf $E$ on $\Open(\CP_\theta^2)$, one has
$$
\Hom_L\big(E^\vee(1)\,,\,\Omega^1_{\CP_\theta^2}(1)\big) \= 0 \=
\Ext_L^2\big(E^\vee(1)\,,\,\Omega^1_{\CP_\theta^2}(1)\big) \ .
$$
\end{lemma}
\Proof{
By (\ref{truncKoszul}) and~\cite[Ex.~6.10]{CLSI}, the sheaf of K\"ahler 
differentials $\Omega^1_{\CP_\theta^2}$ is a bundle of bimodules which
can be included in the noncommutative Euler sequence
$$
0~\longrightarrow~\Omega^1_{\CP_\theta^2}~\longrightarrow~
\sheaf_{\CP_\theta^2}(-1)\otimes\alg_1~
\longrightarrow~\sheaf_{\CP_\theta^2}~ \longrightarrow~0 \ .
$$
Applying the functor $\Hom_L(E^\vee(1),-)$ to the degree~one shift
autoequivalence of this sequence in the category
$\qgr_L(\CP_\theta^2)$ of sheaves of 
left $\alg$-modules, one induces a long exact sequence of Ext$_L$-modules
which begins with the exact sequence
$$
0~\longrightarrow~\Hom_L\big(E^\vee(1)\,,\,
\Omega^1_{\CP_\theta^2}(1)\big)~\longrightarrow~H^0
\big(\CP_\theta^2\,,\,E(-1)\big)\otimes \alg_1 \ ;
$$
here we have used $\Ext^p_L(E^\vee(k),\sheaf_{\CP_\theta^2}(l))\cong
H^p(\CP_\theta^2,E(l-k))$ for $p\geq0$ and $k,l\in\zed$. Whence by
point~(1) of Proposition~\ref{H021CP2vanprop}, one has
$\Hom_L\big(E^\vee(1)\,,\,\Omega^1_{\CP_\theta^2}(1)\big)= 0$. By the construction
of the Koszul complex given in~\cite[\S6.1]{CLSI} and \S\ref{CPSheaves} above,
the sheaf $\Omega^1_{\CP_\theta^2}(1)$ can also be included in the
exact sequence
$$
0~\longrightarrow~\sheaf_{\CP_\theta^2}(-2)~\longrightarrow~
\sheaf_{\CP_\theta^2}(-1)\otimes\big(\alg_2^!\big)^*~
\longrightarrow~\Omega^1_{\CP_\theta^2}(1)~\longrightarrow~0 \ ,
$$
with $(\alg_2^!)^*$ the subspace of $\alg_1\otimes \alg_1$ spanned by
the quadratic relations (\ref{CP2homcoordrels}). By point~(2) of~\cite[Prop.~6.8]{CLSI}, one has
$H^3(\CP_\theta^2,F)=0$ for all $F\in\coh(\CP_\theta^2)$, and so by
applying $\Hom_L(E^\vee(1),-)$ one induces a long exact sequence which
terminates at the exact sequence
$$
H^2\big(\CP_\theta^2\,,\,E(-2)\big)\otimes\big(\alg_2^!\big)^* ~
\longrightarrow~\Ext^2_L\big(E^\vee(1)\,,\,
\Omega^1_{\CP_\theta^2}(1)\big)~\longrightarrow~0 \ .
$$
By point~(2) of Proposition~\ref{H021CP2vanprop}, one thus finds
$\Ext^2_L\big(E^\vee(1)\,,\,\Omega^1_{\CP_\theta^2}(1)\big)=0$.
}

\begin{theorem}
Up to isomorphism, 
any $(W,V)$-framed sheaf on $\Open(\CP_\theta^2)$ is the middle
cohomology of a monad complex
$$
\underline{\calg}\,_\bullet(W,V)\,:\, 0~\longrightarrow~
V\otimes\sheaf_{\CP_\theta^2}(-1)~
\xrightarrow{\sigma_w}~\begin{matrix}
  V\otimes\big(\alg_\ell^!\big)_1\otimes\sheaf_{\CP_\theta^2}
  \\[4pt] \oplus \\[4pt] W\otimes\sheaf_{\CP_\theta^2} \end{matrix}
~ \xrightarrow{\tau_w}~V\otimes\sheaf_{\CP_\theta^2}(1)~
\longrightarrow~0 \, .
$$
Conversely, any linear monad
$\underline{\calg}\,_\bullet(W,V)$ on $\CP_\theta^2$ of this form,
such that $S=\pi_L(\coker(\sigma_w^*))$ is a graded left artinian
$\alg$-module, defines an isomorphism class in $\Mcal_\theta(W,V)$.
\label{sheafmonadthm}\end{theorem}
\Proof{
We use the Beilinson spectral sequence of \S\ref{CPSheaves}. In the
present case, the only non-vanishing sheaves $\Qcal^p$ appearing in
(\ref{Beilinsonseq}) are given by
\begin{eqnarray*}
\Qcal^0&=&\pi(\alg)^\vee\=\sheaf_{\CP_\theta^2} \ , \\[4pt]
\Qcal^1&=&\pi\big(\ker(\alg(-1)\otimes\alg_1\to\alg)\big)^\vee \=
\Omega_{\CP_\theta^2}^1(1)^\vee \ , \\[4pt]
\Qcal^2&=&\pi\big(\alg(-1)\big)^\vee \= \sheaf_{\CP_\theta^2}(1) \ .
\end{eqnarray*}
The spectral sequence converges to $F\in\coh(\CP_\theta^2)$
concentrated in degree~zero. We apply this sequence to the sheaf
$F=E(-1)$ where $[E]\in\Mcal_\theta(W,V)$. One has the sheaf cohomology
groups 
\begin{eqnarray*}
\Ext^q\big(\sheaf_{\CP_\theta^2}(1)\,,\,E(-1)\big) &=&
H^q\big(\CP_\theta^2\,,\,E(-2)\big) \ , \\[4pt]
\Ext^q\big(\sheaf_{\CP_\theta^2}\,,\,E(-1)\big) &=&
H^q\big(\CP_\theta^2\,,\,E(-1)\big)
\end{eqnarray*}
which both vanish for $q\neq1$ by points~(1)
and~(2) of Proposition~\ref{H021CP2vanprop}. Likewise, one has
$$
\Ext^q\big(\Omega_{\CP_\theta^2}^1(1)^\vee\,,\,
E(-1)\big)=\Ext^q_L\big(E^\vee(1)\,,\,\Omega_{\CP_\theta^2}^1(1)\big)
$$
which is also
trivial for $q\neq1$ by Lemma~\ref{Kahlervanlem}. It follows that
$E_1^{p,q}=0$ unless $p=0,1,2$ and $q=1$, and hence the spectral
sequence (\ref{Beilinsonseq}) is simply the three-term complex
\bea
0~\longrightarrow~H^1\big(\CP_\theta^2\,,\,E(-2)\big)\otimes
\sheaf_{\CP_\theta^2}(-2) &\longrightarrow&\Ext^1_L\big(E^\vee(1)\,,\,
\Omega_{\CP_\theta^2}^1(1)\big)\otimes\sheaf_{\CP_\theta^2}(-1)~ \longrightarrow \nonumber\\
&\longrightarrow& H^1\big(\CP_\theta^2\,,\,E(-1)\big)\otimes
\sheaf_{\CP_\theta^2}~\longrightarrow~0 \ .
\label{specseq3term}\eea
Apply the degree~one shift autoequivalence to write this complex in
the generic monadic form of (\ref{linmonadCPn}), i.e. with
finite-dimensional vector spaces given by
$V_{-1}=H^1(\CP_\theta^2,E(-2))$, 
$V_0=\Ext_L^1(E^\vee(1),\Omega_{\CP_\theta^2}(1))$, and
$V_1=H^1(\CP_\theta^2,E(-1))$. By Proposition~\ref{monadinvsprop} with $n=2$
and Corollary~\ref{framedtopinvscor}, one has $\dim_\complex(V_{\pm\,1})=k$
and $\dim_\complex(V_0)=2k+r$, and by point~(3) of Proposition~\ref{H021CP2vanprop} there
are vector space isomorphisms $V_{\pm\,1}\xrightarrow{\ \approx\ }V$.

It remains to identify the $\complex$-vector space $V_0$ with
$(V\otimes(\alg_\ell^!)_1)\oplus W$. For this, we apply the exact
restriction functor $i^\bullet$ to the short exact sequences (\ref{shortkerV0})--(\ref{shortV1ker}). Since the
cohomological dimension of the category $\coh(\CP_\theta^1)$ is~one~\cite[Prop.~6.8~(1)]{CLSI},
the corresponding induced long exact cohomology sequences (\ref{1stcohseq})--(\ref{2ndcohseq}) are
non-trivial only for $p=0,1$. Since by~\cite[Prop.~6.8~(1)]{CLSI} one has
$$
H^0\big(\CP_\theta^1\,,\,\sheaf_{\CP_\theta^1}(-1)\big)=
H^1\big(\CP_\theta^1\,,\,\sheaf_{\CP_\theta^1}(-1)\big)= 0 \ , \qquad
H^0\big(\CP_\theta^1\,,\,\sheaf_{\CP_\theta^1}\big)=
(\alg_\ell)_0=\complex \ ,
$$
and $i^\bullet(E)\cong
W\otimes\sheaf_{\CP_\theta^1}$, the sequence (\ref{2ndcohseq}) gives
isomorphisms
\begin{eqnarray*}
H^0\big(\CP_\theta^1\,,\,i^\bullet(\ker(\tau_w))\big)&
\xrightarrow{\ \approx\ }& H^0\big(\CP_\theta^1\,,\,i^\bullet(E)\big)~
\cong~ W \ , \\[4pt]
H^1\big(\CP_\theta^1\,,\,i^\bullet(\ker(\tau_w))\big)&
\xrightarrow{\ \approx\ }& H^1\big(\CP_\theta^1\,,\,i^\bullet(E)\big)
\=0 \ .
\end{eqnarray*}
Using~\cite[Prop.~6.8~(1)]{CLSI} we identify the finite-dimensional
complex vector space
$$
H^0\big(\CP_\theta^1\,,\,\sheaf_{\CP_\theta^1}(1)\big)= (\alg_\ell)_1
$$ 
with its
Koszul dual $(\alg_\ell^!)_1\cong (\alg_\ell)_1^*$. Then the
exact sequence (\ref{1stcohseq}) truncates to the short exact sequence
\beq
0~\longrightarrow~ W~\longrightarrow~ V_0~\longrightarrow~ 
V\otimes\big(\alg_\ell^!\big)_1~\longrightarrow~0 \ .
\label{WV0exactseq}\eeq
Since $i^\bullet(E)$ is locally free and its dual coincides with
$H^0(i^\bullet(\,\underline{\calg}\,_\bullet(W,V)^\vee\,))$, we can
apply the same argument to the dual monad (\ref{dualmonad}) to get the exact sequence
$$
0~\longrightarrow~ H^0\big(\CP_\theta^1\,,\,
i^\bullet(\ker(\sigma_w^*))\big)~ \longrightarrow~ V_0^*~
\xrightarrow{\sigma^*}~ (\alg_\ell)_1\otimes V^*~ \longrightarrow~0 \
,
$$
which implies that the map $\sigma:V\otimes(\alg_\ell^!)_1\to V_0$ is
injective. Thus the sequence (\ref{WV0exactseq}) splits, and we get the
identification $V_0\cong (V\otimes(\alg_\ell^!)_1)\oplus W$ as desired. By
Proposition~\ref{monadcoh0prop}, the vector space isomorphisms
$V_{\pm\,1}\xrightarrow{\ \approx\ } V$ and $V_0\xrightarrow{\ \approx\ }
(V\otimes(\alg_\ell^!)_1)\oplus W$ can be uniquely extended to an
isomorphism between the monad (\ref{specseq3term}) and one of the form
$\underline{\calg}\,_\bullet(W,V)$ which is compatible with the
framing isomorphisms (see~\cite[Lem.~4.2.15]{BGK}
and~\cite[Thm.~6.7]{KKO}).

Conversely, if $\calg$ is the middle cohomology sheaf of the monad
$\underline{\calg}\,_\bullet(W,V)$, then one has
$H^1(\CP_\theta^2,\calg(-1))\cong V$ by
point~(2) of Proposition~\ref{monadcoh0prop}. Moreover, by Proposition~\ref{CP1equivprop} the
restriction complex $i^\bullet(\,\underline{\calg}\,_\bullet(W,V))$
is a complex of sheaves on $\Open(\CP_\theta^1)$ which is canonically
quasi-isomorphic to $W\otimes \sheaf_{\CP_\theta^1}$. The artinian
condition on the module $S$ implies that $\coker(\sigma_w^*)$ is an
Artin sheaf in the sense of~\cite[Def.~2.0.8]{BGK}. By
point~(2) of Proposition~\ref{monadfreeprop} with $n=2$, it follows that $\calg$ is a
torsion free sheaf on $\Open(\CP_\theta^2)$.

It remains to prove that the construction above is independent of the
choices of representatives for the respective isomorphism
classes. By~\cite[Lem.~4.2.15]{BGK}, every isomorphism $\xi:E\to E'$
of $(W,V)$-framed sheaves on $\Open(\CP_\theta^2)$ extends uniquely to
an isomorphism between the corresponding monads
$\underline{\calg}\,_\bullet(W,V)\to
\underline{\calg}\,_\bullet'(W,V)$. The fact that the isomorphism
$\xi$ preserves the framing isomorphisms then forces the isomorphism
between monads to be given by a commutative diagram
\bea\label{monadisog}
&& \quad 
\xymatrix{
\underline{\calg}\,_\bullet(W,V)\,:\, 0~\ar[r] &  
~V\otimes\sheaf_{\CP_\theta^2}(-1)~\ar[d]^{g\otimes\Id}
\ar[r]^{\ \ \sigma_w} & ~V_0\otimes\sheaf_{\CP_\theta^2}~
\ar[r]^{\!\!\tau_w}\ar[d]^{(\tilde g\oplus\Id_W)\otimes\Id} & 
~V\otimes\sheaf_{\CP_\theta^2}(1)~ \ar[r]\ar[d]^{g\otimes\Id} & ~0 \\
\underline{\calg}\,_\bullet'(W,V)\,:\, 0~\ar[r] &
~V\otimes\sheaf_{\CP_\theta^2}(-1)~\ar[r]_{\ \ \sigma_w'} &
~V_0\otimes\sheaf_{\CP_\theta^2}~ \ar[r]_{\!\!\tau_w'} &
~V\otimes\sheaf_{\CP_\theta^2}(1)~ \ar[r] & ~0
}
\eea
for some $g\in \GL(V)$. Conversely, any isomorphism of monads of the
form (\ref{monadisog}) induces an isomorphism between the
corresponding cohomology
sheaves $E=H^0(\,\underline{\calg}\,_\bullet(W,V))$ and
$E'=H^0(\,\underline{\calg}\,_\bullet'(W,V))$ which preserves the
framing isomorphisms; the automorphism $(\tilde g\oplus\Id_W)\otimes\Id$ maps $E$ onto $E'$, regarded as submodules of $V_0\otimes\sheaf_{\CP_\theta^2}$.
}

The proof of Theorem~\ref{ADHMbij} is now completed by demonstrating that
the ADHM moduli space (\ref{MhatADHM}) has the same monadic description as
that in Theorem~\ref{sheafmonadthm}. Given a triple of ADHM data $(B,I,J)\in \Mcal_\theta^{\rm
  ADHM}(W,V)$, we define a canonical sheaf morphism
\bea
\sigma_{(B,I,J)}& =& \begin{pmatrix}
B\otimes w_3+\Id_V\otimes\big(q\,\check w_1\otimes w_1+q^{-1}\,
\check w_2\otimes w_2\big) \\[4pt] J\otimes w_3 \end{pmatrix} \nonumber \\[4pt]
&:& V\otimes\sheaf_{\CP_\theta^2}(-1)~
\longrightarrow~ \big((V\otimes(\alg_\ell^!)_1)\oplus W\big)
\otimes\sheaf_{\CP_\theta^2} \ ,
\label{sigmaBIJ}\eea
where the homogeneous coordinates $\check w_i$ and $w_j$ act by
multiplication in $\alg_\ell^!$ and $\alg$, respectively. Similarly,
define
\begin{align}\label{tauBIJ}
\tau_{(B,I,J)} &= \begin{pmatrix} \big(\Id_V\otimes
\mu_{\alg_\ell^!}\big)\circ\big(B\otimes\Id_{\alg_\ell^!}\big)\otimes
w_3+ \Id_V\otimes\big(q^{-1}\,\check w_1\otimes w_1+q\,
\check w_2\otimes w_2\big) & I\otimes w_3 \end{pmatrix} \nonumber \\[4pt]
&\: : \big((V\otimes(\alg_\ell^!)_1)\oplus W\big)
\otimes\sheaf_{\CP_\theta^2}~
\longrightarrow~ \big(V\otimes(\alg_\ell^!)_2\big)
\otimes\sheaf_{\CP_\theta^2}(1) \ .
\end{align}
Henceforth we will use the natural vector space isomorphism
$(\alg_\ell^!)_2\cong\complex$. Then the maps (\ref{sigmaBIJ})--(\ref{tauBIJ}) determine a chain of morphisms of coherent sheaves on
$\Open(\CP_\theta^2)$ given by
\bea && \label{BIJchain} \\ \nonumber &&
\underline{\calg}\,_\bullet(B,I,J)\,:\,
V\otimes\sheaf_{\CP_\theta^2}(-1)~
\xrightarrow{\sigma_{(B,I,J)}}~\begin{matrix}
  V\otimes\big(\alg_\ell^!\big)_1\otimes\sheaf_{\CP_\theta^2}
  \\[4pt] \oplus \\[4pt] W\otimes\sheaf_{\CP_\theta^2} \end{matrix}
~ \xrightarrow{\tau_{(B,I,J)}}~V\otimes\sheaf_{\CP_\theta^2}(1) \ .
\eea
We will show that the sequence (\ref{BIJchain}) is a monad of the type considered
in Theorem~\ref{sheafmonadthm}, and that every $(W,V)$-framed torsion free
sheaf arises from this construction.

\begin{theorem}
For any triple $(B,I,J)\in\Mcal^{\rm ADHM}_\theta(W,V)$, the chain
$\underline{\calg}\,_\bullet(B,I,J)$ of morphisms is a monad complex which
defines an isomorphism class in $\Mcal_\theta(W,V)$. Conversely, any
linear monad complex $\underline{\calg}\,_\bullet(W,V)$ of the form given in
Theorem~\ref{sheafmonadthm} defines an isomorphism class in the moduli space 
$\widehat\Mcal^{\rm ADHM}_\theta(W,V)$.
\label{BIJmonadthm}\end{theorem}
\Proof{
Apply the exact restriction functor $i^\bullet$ to (\ref{BIJchain})
with
\begin{eqnarray*}
i^\bullet(\sigma_{(B,I,J)})&=&\begin{pmatrix}
\Id_V\otimes\big(q\,\check w_1\otimes w_1+q^{-1}\,
\check w_2\otimes w_2\big) \\[4pt] 0 \end{pmatrix} \ , \\[4pt]
i^\bullet(\tau_{(B,I,J)}) &=& \begin{pmatrix} 
\Id_V\otimes\big(q^{-1}\,\check w_1\otimes w_1+q\,
\check w_2\otimes w_2\big) & 0 \end{pmatrix} \ .
\end{eqnarray*}
It follows that the morphism $i^\bullet(\sigma_{(B,I,J)})$ is
injective, $i^\bullet(\tau_{(B,I,J)})$ is surjective, while using
(\ref{CP2homcoordrels}) and (\ref{CP1Koszulrels}) one easily finds
$i^\bullet(\tau_{(B,I,J)})\circ i^\bullet(\sigma_{(B,I,J)})=0$. It
follows that $i^\bullet(\underline{\calg}\,_\bullet(B,I,J))$ is a
monad on $\CP_\theta^1$ whose cohomology is naturally isomorphic to
$W\otimes\sheaf_{\CP_\theta^1}$. More generally, an elementary
calculation using (\ref{CP2homcoordrels}) and
(\ref{CP1Koszulrels})--(\ref{Bvexpl}) shows 
$$
\tau_{(B,I,J)}\circ\sigma_{(B,I,J)}=(B\wedge_\theta B+I\circ J)
\otimes w_3^2 \ ,
$$
from which it follows that (\ref{BIJchain}) is a complex if and only
if the triple $(B,I,J)$ satisfies the noncommutative ADHM equations
(\ref{NCADHMeqs}).

Arguing as in Proposition~\ref{CP21exseqprop} using
$i^\bullet(\ker(\sigma_{(B,I,J)}))=0$ from above, multiplication by
$w_3$ yields isomorphisms of sheaves $\ker(\sigma_{(B,I,J)})(k-1)\cong
\ker(\sigma_{(B,I,J)})(k)$ for all
$k\in\zed$. By~\cite[Cor.~5.3]{KKO}, one has the trivial sheaf cohomology
$H^p(\CP_\theta^2,\ker(\sigma_{(B,I,J)})(k))=0$ for $k\gg0$ and for all
$p\geq1$. It follows that $\ker(\sigma_{(B,I,J)})$ is an Artin sheaf,
which is locally free by~\cite[Prop.~2.0.4]{BGK}. Hence
$\ker(\sigma_{(B,I,J)})=0$ by~\cite[Prop.~2.0.9~(4)]{BGK}, and the map
$\sigma_{(B,I,J)}$ is injective. The same argument shows that
$\coker(\sigma_{(B,I,J)}^*)$ is an Artin sheaf, and hence that the graded
left $\alg$-module $S=\pi_L(\coker(\sigma_{(B,I,J)}^*))$ has
homological dimension~zero. Analogously, $\coker(\tau_{(B,I,J)})$ is an
Artin sheaf such that there are isomorphisms of sheaves $\coker(\tau_{(B,I,J)})(k-1)\cong
\coker(\tau_{(B,I,J)})(k)$ for all $k\in\zed$, and we can mimick the
proof of~\cite[Lem.~4.1.9]{BGK}. For this, we note
that by point~(1) of~\cite[Prop.~6.8]{CLSI} we can write 
$$
\tau_{(B,I,J)}=
\big(\Id_V\otimes\mu_\alg\big)\circ \big(H^0(\tau_{(B,I,J)})\otimes\Id\big) \ ,
$$
where
$H^0(\tau_{(B,I,J)}):V_0\to V\otimes\alg_1$ is the induced map on
cohomology. Similarly, the canonical projection
$$
\phi\,:\, V\otimes\sheaf_{\CP_\theta^2}(1)~\longrightarrow~
\coker(\tau_{(B,I,J)})
$$
induces a map $H^0(\phi)(-1):V\to
H^0(\CP_\theta^2,\coker(\tau_{(B,I,J)}))$. Let
$V'=\ker\big(H^0(\phi)(-1)\big)$. Then (\ref{tauBIJ}) and the
composition $H^0(\phi\circ \tau_{(B,I,J)})=0$ imply $B(V'\,)\subset
V'\otimes(\alg_\ell^!)_1$ and $I(W)\subset
V'\otimes(\alg_\ell^!)_2$. By the stability condition~(2) of
Definition~\ref{NCADHMvar}, one has $V'=V$. Since $\phi$ is surjective, it
follows that $\coker(\tau_{(B,I,J)})=0$ and hence $\tau_{(B,I,J)}$ is
surjective. Putting everything together, the chain (\ref{BIJchain}) is
a monad whose cohomology sheaf is torsion free by
point ~(2) of Proposition~\ref{monadfreeprop}. 

Conversely, given a monad $\underline{\calg}\,_\bullet(W,V)$, we
mimick the proof of~\cite[Thm.~6.7]{KKO}. For this, we use (\ref{sigmataudecomp}) for $n=2$ to decompose the
maps $\sigma_w$ and $\tau_w$ into constant linear maps $\sigma^i:V\to V\otimes(\alg_\ell^!)_1\oplus
W$ and $\tau^i:V\otimes(\alg_\ell^!)_1\oplus W\to
V\otimes(\alg_\ell^!)_2$ for $i=1,2,3$. Then the monadic condition
$\tau_w\circ\sigma_w=0$ is equivalent to the system of vanishing
morphism compositions
\bea
\tau^i\,\sigma^i&=&0 \ , \qquad i=1,2,3 \ , \nonumber\\[4pt]
\tau^2\,\sigma^1+q^2~\tau^1\,\sigma^2&=&0 \ , \nonumber \\[4pt]
\tau^i\,\sigma^3+\tau^3\,\sigma^i&=&0 \ , \qquad i=1,2 \ .
\label{tausigma0comps}\eea
Since $i^\bullet(\sigma_w)=\sigma^1\otimes w_1+\sigma^2\otimes w_2$
and $i^\bullet(\tau_w)=\tau^1\otimes w_1+\tau^2\otimes w_2$ with
cohomology $W\otimes\sheaf_{\CP_\theta^1}\cong
\ker(i^\bullet(\tau_w))/{\rm im}(i^\bullet(\sigma_w))$, from the proof
of Theorem~\ref{sheafmonadthm} it follows that, as in~\cite[Thm.~6.7]{KKO},
the composition $\tau^1\,\sigma^2=-\tau^2\,\sigma^1$ is an isomorphism
$V\xrightarrow{\ \approx\ }V\otimes(\alg_\ell^!)_2$. With a suitable
choice of basis for the vector space $V$ we can set
$\tau^1\,\sigma^2=q^{-2}~\Id_V\otimes\check w_1\,\check w_2$. We can
then choose a basis for $W$ such that the first set of equations for
$i=1,2$ and the second equation in (\ref{tausigma0comps}) are
solved by
\begin{eqnarray*}
\sigma^1=\begin{pmatrix} q~\Id_V\otimes\check w_1 \\[4pt] 0 \end{pmatrix}
\qquad &,& \qquad \sigma^2=\begin{pmatrix} q^{-1}~\Id_V\otimes\check
  w_2 \\[4pt] 0 \end{pmatrix} \ , \\[4pt]
\tau^1=\begin{pmatrix}q^{-1}~\Id_V\otimes\check w_1 & 0 \end{pmatrix}
\qquad &,& \qquad \tau^2=\begin{pmatrix}q~\Id_V\otimes\check w_2 &
  0 \end{pmatrix} \ .
\end{eqnarray*}
Then the third set of equations in (\ref{tausigma0comps}) are solved
by
$$
\sigma^3=\begin{pmatrix} B_1\otimes\check w_1+B_2\otimes\check w_2 \\[4pt]
  J \end{pmatrix} \ , \qquad \tau^3=\begin{pmatrix} B_1\otimes\check
  w_1+B_2\otimes\check w_2 & I \end{pmatrix}
$$
with arbitrary maps $B_1,B_2\in\End_\complex(V)$,
$J\in\Hom_\complex(V,W)$, and
$I\in\Hom_\complex(W,V\otimes(\alg_\ell^!)_2)$. Finally, the first
equation in (\ref{tausigma0comps}) with $i=3$ is equivalent to
$$
\big(B_1\,B_2-q^{-2}~B_2\,B_1\big)\otimes\check w_1\,\check w_2
+I\circ J=0 \ ,
$$
which is just the ADHM equation~(\ref{NCADHMeqs}). The morphisms
(\ref{sigmataudecomp}) then take the forms given in (\ref{sigmaBIJ})--(\ref{tauBIJ}).

It remains to check the stability condition~(2) of
Definition~\ref{NCADHMvar}. Suppose that there exists a non-trivial proper
subspace $V'\subset V$ such that
$$
B_i(V'\,)\subseteq V' \ , \qquad {\rm im}(I)\subseteq V'\otimes
\big(\alg_\ell^!\big)_2
$$
for $i=1,2$. Let $B_i'=B_i\big|_{V'}$, $i=1,2$, and let $I':W\to V'$
be the factorization of $I$ through $V'\subset V$. Then there are
natural inclusions
$$
{\rm im}(I'\,)\hookrightarrow V' \ , \qquad
{\rm im}(B_i'\,)\hookrightarrow V'
$$
for $i=1,2$. Set $V_\perp':=V/V'$, and consider the induced dual
linear map on cohomology $H^0(\tau_w^*)(1):V^*\to\alg_1^*\otimes
V_0^*$. Since $(V_\perp'\,)^*\subset\ker(I'\,^*)$, if $V_\perp'\neq0$
then by (\ref{tauBIJ}) the induced map $H^0(\tau_w^*)(1)$ is not
injective and hence $\tau_w$ cannot be surjective as a morphism of
sheaves. This contradicts the assumption that
$\underline{\calg}\,_\bullet(W,V)$ is a monad. Therefore, a
destabilizing proper subspace $V'\subset V$ as above cannot
exist. Finally, any isomorphism of monads preserving the choices of
bases for $V$ and $W$ made above is of the form (\ref{monadisog}), and
coincides with the natural action of the group $\GL(V)$ on~$\Mcal^{\rm
  ADHM}_\theta(W,V)$.
}

\begin{remark}
\textup{
By point~(3) of Proposition~\ref{monadcoh0prop} and Theorem~\ref{sheafmonadthm} it follows that
a $(W,V)$-framed sheaf is locally free if and only if the dual map
$\sigma_w^*$ is surjective. From the proof of Theorem~\ref{BIJmonadthm} this
is true if and only if the corresponding dual set of ADHM data
$(B^*,J^*,I^*)$ satisfies the stability condition in
$\Mcal_\theta^{\rm ADHM}(V^*,W^*)$. The set of all such
triples coincides with the inverse image under the affine map
$\sigma_w^*$ of the open subset of surjective vector space morphisms
contained in $\alg_1^*\otimes\Hom_\complex(V^*,V_0^*)$, and hence is
itself an open subset of $\Mcal^{\rm ADHM}_\theta(W,V)$ in the Zariski
topology. It follows that the moduli space of $(W,V)$-framed \emph{vector bundles} form a
dense open subset in $\Mcal_\theta(W,V)$, and hence the space
$\Mcal_\theta(W,V)$ can be regarded as its (partial) compactification. 
}
\label{ADHMbundlerem}\end{remark}

\subsection{Self-conjugate monads on the twistor variety\label{scmonad}}

We will now provide another characterization of the moduli space
$\Mcal_\theta(W,V)$ in terms of ``real'' linear algebraic data.
In the sequel we shall assume that $q_{12}=q\in\real$, i.e. that the
noncommutativity parameter $\theta$ is purely imaginary. There is a
natural $*$-structure on the algebra $\alg_\ell^!$ given by the
$\complex$-conjugate linear anti-algebra anti-involution
$\Jscr:\alg_\ell^!\to (\alg_\ell^!)^{\rm op}$ defined on generators by
\beq
\Jscr(\check w_1,\check w_2)=(\check w_2,-\check w_1) \ .
\label{Jscrcheckw}\eeq
We will extend $\Jscr$ as a morphism in the category ${}^{\hil_\theta}\Module$ of left
$\hil_\theta$-comodules; in particular, it is coequivariant
$$
(\Id\otimes\Jscr)\circ \Delta_L=\Delta_L\circ\Jscr \ .
$$
As in \S\ref{ncsphere}, this real structure restricts the toric action,
in this case to a coaction of the diagonal subgroup
$\complex^\times\subset T$ with $t_1=t_2$.

Let us fix hermitean inner products on the complex vector spaces $V$
and $W$. Then the space of complex ADHM data (\ref{BIJsubvar}) also
acquires a natural anti-involution
$$
\Jscr\,:\,\Mcal^{\rm ADHM}_\theta(W,V)~\longrightarrow~ \Mcal^{\rm
  ADHM}_\theta(W,V) \ , \qquad
\Jscr(B,I,J)=\big(B^\dag\,,\,-J^\dag\,,\,I^\dag\,\big) \ ,
$$
where we implicitly always use the vector space isomorphisms
$(\alg_\ell^!)_2\cong\complex$ and $(\alg_\ell^!)^{\rm
  op}\cong\alg_\ell^!$, and we set 
\beq
B^\dag:=\Jscr(B)=-B_2^\dag\otimes\check w_1+B_1^\dag\otimes\check w_2
\label{BdagJscr}\eeq
with respect to the decomposition (\ref{Bvexpl}). With these
definitions one has
$$
\Jscr(B\wedge_\theta B)=B^\dag\wedge_\theta B^\dag \ .
$$

Given any pair of linear maps
$B,B'\in\Hom_\complex(V,V\otimes(\alg_\ell^!)_1)$, we generalize the
definition (\ref{BwedgeB}) to the morphism
$$
B\wedge_\theta B'=\big(\Id_V\otimes\mu_{\alg_\ell^!}\big)\circ
\big(B\otimes\Id_{\alg_\ell^!}\big)\circ B'
$$
in $\Hom_\complex(V,V\otimes(\alg_\ell^!)_2)$. In general, this cannot
be represented as in (\ref{thetacomm}), but
an explicit expression in terms of braided commutators is obtained for
the sum
$$
B\wedge_\theta B'+B'\wedge_\theta B=\big([B_1,B_2'\,]_\theta-q^{-2}~
[B_2,B_1'\,]_{-\theta}\big)\otimes\check w_1\,\check w_2 \ .
$$
We will demonstrate how to relate certain framed torsion free sheaves on
$\CP_\theta^2$ to the following quotient of the space of complex ADHM
data.
\begin{definition}
The variety $\Mcal_\theta^{\rm tw}(W,V)$ of \emph{noncommutative real
  ADHM data} is the subspace of $\Mcal_\theta^{\rm ADHM}(W,V)$
consisting of triples (\ref{BIJsubvar}) which satisfy, in addition to
conditions~(1) and~(2) of Definition~\ref{NCADHMvar}, the \emph{noncommutative real ADHM
  equation}
\beq
B\wedge_\theta B^\dag+B^\dag\wedge_\theta B+I\circ I^\dag-J^\dag\circ
J=0
\label{realADHMeq}\eeq
in $\End_\complex(V\otimes(\alg_\ell^!)_2)$.
\label{realADHMdata}\end{definition}

The natural $\GL(V)$-action on $\Mcal_\theta^{\rm ADHM}(W,V)$ reduces 
on $\Mcal_\theta^{\rm tw}(W,V)$ to an action of the group $\U(V)$ of
unitary automorphisms of the vector space $V$. The corresponding
space of stable orbits is denoted $\widehat\Mcal_\theta^{\rm
  tw}(W,V)$. We will now provide a monadic description of this moduli
space. 

Consider the natural embedding of the noncommutative projective plane
$\CP_\theta^2$ into the noncommutative twistor space
$\CP_{\theta}^3$ with homogeneous coordinate algebra {$\alg^{\rm tw}=\alg(\CP_{\theta}^3)$} described in \S\ref{Twistorfib}.
The homogeneous coordinate algebra of the original space
$\CP_\theta^2$ is then recovered through $\alg\cong\alg^{\rm tw}/\langle
w_4\rangle$. Let $\iota:\alg\hookrightarrow\alg^{\rm tw}$ be the
natural algebra inclusion. We will again denote by
$i:\alg_\ell\hookrightarrow\alg^{\rm tw}$ the algebra inclusion of the
noncommutative projective line $\CP_\theta^1$ with
$\alg_\ell\cong \alg^{\rm tw}/\langle w_3,w_4\rangle$. Define a non-degenerate conjugate linear
anti-involution $\Jscr:\alg^{\rm tw} \to (\alg^{\rm tw})^{\rm op}$
acting on generators as
\beq
\Jscr(w_1,w_2,w_3,w_4)=(w_2,-w_1,w_4,-w_3) \ .
\label{Jscrw}\eeq

Consider a linear monad on $\CP_{\theta}^3$ of the form
$$
\underline{\calg}\,^{\rm tw}_\bullet(W,V)\,:\, 0~\longrightarrow~
V\otimes\sheaf_{\CP_{\theta}^3}(-1)~
\xrightarrow{\sigma_w}~V_0\otimes\sheaf_{\CP_{\theta}^3}~
\xrightarrow{\tau_w}~V\otimes\sheaf_{\CP_{\theta}^3}(1)~
\longrightarrow~0
$$
with $V_0=V\otimes(\alg_\ell^!)_1\oplus W$. Its restriction
$i^\bullet(\,\underline{\calg}\,^{\rm tw}_\bullet(W,V))$ is again a monad on
$\CP_\theta^1$ which is quasi-isomorphic to
$W\otimes\sheaf_{\CP_\theta^1}$. The anti-homomorphism $\Jscr$ induces
a functor between the categories of sheaves
$\Jscr^\bullet:\coh(\CP_{\theta}^3)\to 
\coh_L(\CP_{\theta}^3)$. Composing this functor with the
dualizing functor $\Homc\,(-,\sheaf_{\CP_{\theta}^3})$ gives a
functor $\coh(\CP_{\theta}^3)\to \coh(\CP_{\theta}^3)$,
which we denote by $E\mapsto E^\dag:=\Jscr^\bullet(E)^\vee$. This
functor can be extended to the derived category of
$\coh(\CP_{\theta}^3)$ and applied to a monad
$\underline{\calg}\,^{\rm tw}_\bullet(W,V)$ to give a monad
$\underline{\calg}\,^{\rm tw}_\bullet(W,V)^\dag:=
\Jscr^\bullet(\,\underline{\calg}\,^{\rm tw}_\bullet(W,V)^\vee\,)$, with
$$
\underline{\calg}\,^{\rm tw}_\bullet(W,V)^\dag\,:\, 0~\longrightarrow~
\overline{V}\,^*\otimes\sheaf_{\CP_{\theta}^3}(-1)~
\xrightarrow{\sigma_w^\dag}~\overline{V_0}\,^*
\otimes\sheaf_{\CP_{\theta}^3}~
\xrightarrow{\tau_w^\dag}~\overline{V}\,^*
\otimes\sheaf_{\CP_{\theta}^3}(1)~\longrightarrow~0 \, .
$$
Here the bars denote complex conjugation while
$\sigma_w^\dag:=\Jscr^\bullet(\sigma_w)^*\in
\overline{V_0}\,^*\otimes\overline{V}\otimes (\alg_1^{\rm tw})^{\rm
  op}$ and $\tau_w^\dag:=\Jscr^\bullet(\tau_w)^*\in
\overline{V}\,^*\otimes\overline{V_0}\otimes (\alg_1^{\rm tw})^{\rm
  op}$. We will say that a monad of the form $\underline{\calg}\,^{\rm
  tw}_\bullet(W,V)$ is \emph{self-conjugate} if there is an isomorphism
$\underline{\calg}\,^{\rm tw}_\bullet(W,V)^\dag \cong
\underline{\calg}\,^{\rm tw}_\bullet(W,V)$ of complexes.

\begin{theorem}
There is a natural (set-theoretic) bijective correspondence between isomorphism
classes of self-conjugate linear monad complexes $\underline{\calg}\,^{\rm
  tw}_\bullet(W,V)$ on $\CP_{\theta}^3$ and isomorphism
classes in the moduli space~$\widehat\Mcal_\theta^{\rm tw}(W,V)$ of
braided real ADHM data.
\label{selfconjtwistorthm}\end{theorem}
\Proof{
We argue exactly as we did in \S\ref{Beilinson}. Decompose the differentials as in (\ref{sigmataudecomp}), with constant linear maps $\sigma^i:V\to V_0$ and $\tau^i:V_0\to V$
for $i=1,2,3,4$. By suitable choices of bases for the vector spaces
$V$ and $W$, we can put these maps into the forms
\begin{eqnarray*}
\sigma^1=\begin{pmatrix} q~\Id_V\otimes\check w_1 \\[4pt] 0 \end{pmatrix}
\qquad &,& \qquad \sigma^2=\begin{pmatrix} q^{-1}~\Id_V\otimes\check
  w_2 \\[4pt] 0 \end{pmatrix} \ , \\[4pt]
\sigma^3=\begin{pmatrix} B_1\otimes\check w_1+B_2\otimes\check w_2 \\[4pt]
  J \end{pmatrix} \qquad &,& \qquad
\sigma^4=\begin{pmatrix} B_1'\otimes\check w_1+B_2'\otimes\check w_2 \\[4pt]
  J' \end{pmatrix}
\end{eqnarray*}
and
\begin{eqnarray*}
\tau^1=\begin{pmatrix}q^{-1}~\Id_V\otimes\check w_1 & 0 \end{pmatrix}
\qquad &,& \qquad \tau^2=\begin{pmatrix}q~\Id_V\otimes\check w_2 &
  0 \end{pmatrix} \ , \\[4pt]
\tau^3=\begin{pmatrix} B_1\otimes\check
  w_1+B_2\otimes\check w_2 & I \end{pmatrix} \qquad &,& \qquad
\tau^4=\begin{pmatrix} B_1'\otimes\check
  w_1+B_2'\otimes\check w_2 & I'\, \end{pmatrix} \ ,
\end{eqnarray*}
with $B_1,B_2,B_1',B_2'\in\End_\complex(V)$,
$J,J'\in\Hom_\complex(V,W)$, and
$I,I'\in\Hom_\complex(W,V\otimes(\alg_\ell^!)_2)$.

Self-conjugacy
$\underline{\calg}\,^{\rm tw}_\bullet(W,V)^\dag \cong
\underline{\calg}\,^{\rm tw}_\bullet(W,V)$ is equivalent to the
conditions $\sigma_w^\dag=-\tau_w$ and $\tau_w^\dag=\sigma_w$, or
equivalently that $\sum_i\,\sigma^i\otimes
w_i=\sum_i\,\tau^i\,^\dag\otimes \Jscr(w_i)$ where
$\tau^i\,^\dag:=(\overline{\tau}\,^i)^*$ are the adjoint linear maps
with respect to the induced hermitean metrics. Using (\ref{Jscrw}),
equating coefficients of the generators of $\alg^{\rm tw}$ yields the
relations
$$
\sigma^1=-\tau^2\,^\dag \ , \qquad \sigma^2=\tau^1\,^\dag \ , \qquad
\sigma^3=-\tau^4\,^\dag \ , \qquad \sigma^4=\tau^3\,^\dag \ ,
$$
which using (\ref{BdagJscr}) implies $(B_1',B_2',I',J'\,)=
(-B_2^\dag,B_1^\dag,-J^\dag,I^\dag\,)$, i.e. $$(B',I',J'\,
)=\Jscr(B,I,J) \ . $$ 
By using (\ref{Bvexpl}) the differentials may thus 
be expressed as
\beq
\sigma_w = \begin{pmatrix}
B\otimes w_3+B^\dag\otimes w_4+\Id_V\otimes\big(q\,\check w_1\otimes
w_1+q^{-1}\,\check w_2\otimes w_2\big) \\[6pt] J\otimes w_3 +I^\dag\otimes
w_4\end{pmatrix}
\label{sigmawreal}\eeq
and
\begin{multline}
\tau_w = 
\Big( (\Id_V\otimes \mu_{\alg_\ell^!})\circ(B\otimes\Id_{\alg_\ell^!})\otimes w_3+ (\Id_V\otimes\mu_{\alg_\ell^!})\circ(B^\dag \otimes\Id_{\alg_\ell^!})\otimes w_4  \\ + \Id_V\otimes(q^{-1}\,\check w_1\otimes w_1+q\, 
\check w_2\otimes w_2)  \\  
 I\otimes w_3 -J^\dag\otimes w_4 \Big)  \ .
\label{tauwreal}
\end{multline}
By (\ref{CP1Koszulrels}) and (\ref{CP3rels}), we
compute the composition of the maps (\ref{sigmawreal})--(\ref{tauwreal}) to get
\begin{eqnarray*} 
\tau_w\circ\sigma_w&=&(B\wedge_\theta B+I\circ J)\otimes w_3^2+
\big(B^\dag\wedge_\theta B^\dag-J^\dag\circ I^\dag\,\big)\otimes w_4^2
\\ && +\,\big(B\wedge_\theta B^\dag+B^\dag\wedge_\theta B+I\circ
I^\dag-J^\dag\circ J\big)\otimes w_3\,w_4 \ .
\end{eqnarray*}
The monadic condition $\tau_w\circ\sigma_w=0$ is thus equivalent,
respectively, to the complex ADHM equation~(\ref{NCADHMeqs}), its
image under the anti-involution $\Jscr$, and the real ADHM equation~(\ref{realADHMeq}). 
Stability follows similarly to the proof of
Theorem~\ref{BIJmonadthm}, and is again equivalent to surjectivity of the map (\ref{tauwreal}). Naturality also follows as before.
}

By repeating the arguments of \S\ref{Beilinson}, the cohomology of a
self-conjugate monad complex of the form $\underline{\calg}\,^{\rm
  tw}_\bullet(W,V)$, under the correspondence of
Theorem~\ref{selfconjtwistorthm}, is a torsion free $(W,V)$-framed sheaf $E$ 
on $\Open(\CP_{\theta}^3)$ obeying the reality condition
$E^\dag\cong E$, and with vanishing cohomology $H^1(\CP_{\theta}^3,E(-2))=0$ by
point~(1) of Proposition~\ref{monadcoh0prop}. Conversely, by adapting the proof of
Theorem~\ref{sheafmonadthm} using the Beilinson spectral sequence of
\S\ref{CPSheaves}, applied to $\CP_{\theta}^3$, any such sheaf
$E$ is the cohomology of a self-conjugate linear monad
$\underline{\calg}\,^{\rm tw}_\bullet(W,V)$ on $\CP_{\theta}^3$,
with $V=V_{-1}\cong
\Ext_L^1\big(E^\vee(1)\,,\,\Omega_{\CP_{\theta}^3}^{1}(2)\big)$ and
the remaining monadic vector spaces analogous to those in the proof of
Theorem~\ref{sheafmonadthm}. The restriction
$\iota^\bullet(\,\underline{\calg}\,^{\rm tw}_\bullet(W,V))$ of a
self-conjugate monad on $\CP_{\theta}^3$ is a monad on
$\CP_\theta^2$ which by Theorem~\ref{BIJmonadthm} defines an isomorphism
class in $\Mcal_\theta(W,V)$. This gives a map of moduli spaces
$\widehat\Mcal_\theta^{\rm tw}(W,V)\to \Mcal_\theta(W,V)$. At the
level of noncommutative ADHM data, this map is just the natural
inclusion of varieties $\Mcal_\theta^{\rm tw}(W,V)\hookrightarrow
\Mcal_\theta^{\rm ADHM}(W,V)$. The virtue of this restriction is that
this smaller class of torsion free sheaves on $\Open(\CP_\theta^2)$
corresponds directly to a class of anti-selfdual connections on a
canonically associated ``instanton bundle'', as we
demonstrate in \S\ref{Instconn}. We do not know if this construction is complete.

\newsection{Construction of noncommutative instantons\label{Instconn}}

\subsection{Instanton bundles on $S_\theta^4$\label{subsec:instbun}}

In this section we will use the monadic description of noncommutative
real ADHM data on the twistor space $\CP_{\theta}^3$ to construct
canonical bundles, called ``instanton bundles'', on the
noncommutative sphere $S_\theta^4$ obtained in~\S\ref{ncsphere}. This
will be achieved via the twistor transform of framed torsion free sheaves $E$ on $\Open(\CP_{\theta}^3)$ which satisfy the reality
condition $E^\dag\cong E$ of \S\ref{scmonad}.
It is determined by the
noncommutative correspondence diagram
$$
\xymatrix{
 & \alg\big(\Fl_{\theta}(1,2;4)\big) & \\
{\alg^{\rm tw}=\alg\big(\CP_{\theta}^3\big)} \ar[ur]^{p_1} &  & 
\alg\big(\Gr_{\theta}(2;4)\big) \ar[ul]_{p_2} \ .
}
$$
The generators $w_i$, $i=1,2,3,4$ and $\Lambda^{(ij)}$, $1\leq
i<j\leq4$ of the homogeneous coordinate algebra of the noncommutative
flag variety $\Fl_{\theta}(1,2;4)$ obey relations given by
(\ref{CP3rels}) and by (\ref{ncrelpr}) for $d=2,n=4$, the Pl\"ucker
equations (\ref{pl24}), and the relations
(\ref{wLambda}) {and} (\ref{flagpluck}), with the $q$-values
(\ref{qvalues}) for $q\in\real$. 

Using Lemma~\ref{twisttransfbundles} we can apply the derived functor
of the twistor transform $p_2{}^*\,p_{1*}$ to a self-conjugate
linear monad complex $\underline{\calg}\,^{\rm tw}_\bullet(W,V)$ on
$\CP_{\theta}^3$ to get
\beq \qquad
p_2{}^*\,p_{1*}\big(\,\underline{\calg}\,^{\rm tw}_\bullet(W,V)\big)
\,:\, 0~\longrightarrow~V_0\otimes \sheaf_{\Gr_{\theta}(2;4)}~
\xrightarrow{(\Id_V\otimes\hat\eta)\circ\tau_w}~V\otimes
\Scal_{\theta}~\longrightarrow~0 \ ,
\label{twisttransfmonad}\eeq
where $\hat\eta$ is the rank~two projector of (\ref{NCGrEuler})
defining the tautological bundle over the Grassmann variety as $\Scal_{\theta}\cong
\hat\eta\big(\alg(\Gr_{\theta}(2;4))\otimes\alg_1^{\rm tw} \big)${, and $\alg_1^{\rm tw}$ is the degree-one part of the algebra $\alg^{\rm tw}$}. It follows that
the image of a cohomology sheaf $E=H^0\big( \,\underline{\calg}\,^{\rm
  tw}_\bullet(W,V) \big)$ under the twistor transform is the sheaf on $\Open(\Gr_{\theta}(2;4))$
given by
$$
E'=p_2{}^*\,p_{1*}(E)=\ker\big((\Id_V\otimes\hat\eta)\circ\tau_w\big)
\ .
$$
We are interested in the restriction of this complex to the affine
subvariety $\real_\theta^4$ of the noncommutative grassmannian $\Gr_{\theta}(2;4)$
described by Proposition~\ref{R4prop} and the real structure~(\ref{starxi}). 

It follows from (\ref{flagloc34}) that the complex (\ref{twisttransfmonad}) 
restricts to the ``Dirac operator''
\bea
&& \quad 
\Dcal:=j^\bullet\,(\Id_V\otimes\hat\eta)\circ\tau_w \,:\, \big(V\otimes(\alg_\ell^!)_1\oplus
W\big)\otimes\alg\big(\real_\theta^4\big) ~\longrightarrow~
(V\oplus V)\otimes\alg\big(\real_\theta^4\big) \ ,
\label{Diracop}\eea
which can be
written explicitly using (\ref{tauwreal}) with (\ref{w1w2triv}), and the
decompositions (\ref{Bvexpl}) and (\ref{BdagJscr}) as
\begin{eqnarray}
\Dcal &=& \begin{pmatrix}
(\Id_V\otimes 
\mu_{\alg_\ell^!})\circ(B\otimes\Id_{\alg_\ell^!})\otimes 1-
\Id_V\otimes(q^{-1}\, \check w_1\otimes\xi_1+q\, \check w_2\otimes\xi_2) & I
\\[6pt] (\Id_V\otimes 
\mu_{\alg_\ell^!})\circ(B^\dag\otimes\Id_{\alg_\ell^!})\otimes 1-
\Id_V\otimes(q^{-1}\,\check w_1\otimes\bar\xi_2+q\, \check
w_2\otimes\bar\xi_1) & -J^\dag
\end{pmatrix} \nonumber \\[10pt]
&=& \begin{pmatrix}
\big(B_1-q^{-1}\, \xi_1\big)\otimes\check w_1 +
\big(B_2-q\,\xi_2\big)\otimes\check w_2 & I
\\[6pt] \big(-B_2^\dag-q^{-1}\,\bar\xi_2\big)\otimes\check w_1 + \big(B_1^\dag-q\,
\bar\xi_1\big)\otimes\check w_2 & -J^\dag
\end{pmatrix} \ .
\label{Diracopexpl}\end{eqnarray}
By construction, $\Dcal$ is a surjective morphism of free right
$\alg(\real_\theta^4)$-modules. Recall from the proof of
Theorem~\ref{BIJmonadthm} that surjectivity of the differential
$\tau_w$ is equivalent to the stability condition~(2) of
Definition~\ref{NCADHMvar}. For the same reason, and by point~(3) of
Proposition~\ref{monadcoh0prop}, the map $\tau_w^\dag$ is injective,
and hence $\Dcal^\dag$ is also injective. Here the $\dag$-involution
is the tensor product of the real structures given by (\ref{starxi}) and 
(\ref{Jscrcheckw}), and those of the chosen hermitean structures on
the vector spaces $V$ and $W$.

Next we define the ``Laplace operator''
\beq
\triangle:=\Dcal\circ\Dcal^\dag \,:\, (V\oplus
V)\otimes\alg\big(\real_\theta^4\big) ~\longrightarrow~ (V\oplus
V)\otimes\alg\big(\real_\theta^4\big) \ .
\label{Laplaceop}
\eeq

\begin{proposition} The operator
$\triangle\in\End_{\alg(\real_\theta^4)}\big((V\oplus
V)\otimes\alg(\real_\theta^4)\big)$ is an isomorphism.
\label{imkerprop}\end{proposition}
\Proof{
Since $\Dcal^\dag\in\Hom_{\alg(\real_\theta^4)}\big((V\oplus
V)\otimes\alg(\real_\theta^4)\,,\,
V_0\otimes\alg(\real_\theta^4)\big)$ is injective, it follows that
$\im(\Dcal^\dag\,)$ is a free $\alg(\real_\theta^4)$-submodule of
$V_0\otimes\alg(\real_\theta^4)$. Its orthogonal complement with
respect to the induced hermitean structures above can be naturally
identified with the kernel of $\Dcal$, and hence there is a decomposition
\beq
V_0\otimes\alg\big(\real_\theta^4\big) =\im\big(\Dcal^\dag\,\big)
\oplus \ker(\Dcal)
\label{imkerdecomp}\eeq
of $\alg(\real_\theta^4)$-modules. In particular,
$\im(\Dcal^\dag\,)\cap\ker(\Dcal)= 0$, and hence $\triangle$ is injective
since $\Dcal^\dag$ is injective.

Now let $v\in (V\oplus V)\otimes\alg(\real_\theta^4)$. Since $\Dcal$
is surjective, there exists $v_0\in V_0\otimes\alg(\real_\theta^4)$
such that $\Dcal(v_0)=v$. Using the decomposition (\ref{imkerdecomp})
one has $v_0=v_0'+v_0^{\prime\prime}$ for some
$v_0'\in\im(\Dcal^\dag\,)$ and
$v_0^{\prime\prime}\in\ker(\Dcal)$. Setting $v_0'=\Dcal^\dag(v'\,)$
for $v'\in (V\oplus V)\otimes\alg(\real_\theta^4)$, we then have
$v=\Dcal(v_0)=\Dcal(v_0'\,)= \Dcal\big(\Dcal^\dag(v'\,)\big)$ and
hence $\triangle$ is also surjective.
}

The operators $\Dcal$, $\Dcal^\dag$ and $\triangle$ are all
$\hil_\theta$-coequivariant morphisms with respect to the coactions given by
(\ref{DeltaLdual}), (\ref{DeltaLR4}) and (\ref{DeltaLBIJ}). The right $\alg(\real_\theta^4)$-module
\beq
\bun :=\ker(\Dcal) =\ker\big(
j^\bullet\,(\Id_V\otimes\hat\eta)\circ\tau_w \big)
\label{instbun}\eeq
is projective by (\ref{imkerdecomp}). It is also finitely generated,
and has rank $\dim_\complex(W)=r$ since $\Dcal$ is surjective. In
fact, the isomorphism invariants of $\bun$ are described by
Corollary~\ref{framedtopinvscor}. Using
Proposition~\ref{imkerprop} the
corresponding projection can be given as
\beq
P:= \Id_{V_0} - \Dcal^\dag\circ\triangle^{-1}\circ \Dcal\, :\,
V_0\otimes\alg\big(\real_\theta^4\big) ~\longrightarrow~ \bun \ ,
\label{instbunproj}\eeq
with $P^2=P=P^\dag$ and trace $\Tr
P=\dim_\complex(V_0)-\dim_\complex(V\oplus V)= r$. The module (\ref{instbun}) is called an
instanton bundle over $\real_\theta^4$; it defines an object of the
category ${}^{\hil_{\theta}}\Module$. Using (\ref{monadisog}), one
easily demonstrates that the isomorphism class of $\bun$ depends only
on the class of the noncommutative ADHM data $(B,I,J)$ in the moduli
variety $\Mcal_\theta^{\rm tw}(W,V)$. Moreover, the reality condition
$\bun^\dag\cong \bun$ follows from the construction of \S\ref{scmonad}.

Now we consider the restriction of the complex
(\ref{twisttransfmonad}) to the affine subvariety
$\widetilde\real_\theta^4$ of the noncommutative grassmannian
$\Gr_\theta(2;4)$ described by Proposition~\ref{tildeR4prop} and
the real structure~(\ref{starzeta}). Using (\ref{tauwreal}) with (\ref{w3w4triv}), and the
decompositions (\ref{Bvexpl}) and (\ref{BdagJscr}) we define a Dirac
operator
\begin{eqnarray}
&&
\widetilde\Dcal := \tilde j\,^\bullet\,(\Id_V\otimes\hat\eta)\circ\tau_w
= \begin{pmatrix} \begin{matrix} {
      -q\, \big(B_1\, \bar\zeta_1+B_2^\dag\,\zeta_2-q^{-2}\big)\otimes
\check w_1} \\ {-\,q\, \big(B_2\,\bar\zeta_1-B_1^\dag\,
\zeta_2\big)\otimes\check w_2} \end{matrix} & \ \
-q\,\big(I\,\bar\zeta_1 +J^\dag\, \zeta_2\big) \\~\\ \begin{matrix} {
      q^{-1}\, \big(B_1\, \bar\zeta_2+B_2^\dag\,\zeta_1\big)\otimes
\check w_1} \\ { +\,q^{-1}\, \big(B_2\,\bar\zeta_2-B_1^\dag\,
\zeta_1+ q^2\big)\otimes\check w_2} \end{matrix} & \ \
q^{-1}\,\big( I\,\bar\zeta_2+J^\dag\, \zeta_1\big)
\end{pmatrix} \nonumber \\ &&
\label{Diracoptilde}\end{eqnarray}
in $\Hom_{\alg(\widetilde\real_\theta^4)}
\big(V_0\otimes\alg(\widetilde\real_\theta^4)\,,\,(V\oplus
V)\otimes\alg(\widetilde\real_\theta^4)
\big)$. The operator $\widetilde\Dcal$ is surjective by construction, and
by repeating the arguments used above the module
$$
\tilde\bun:= \ker\big(\widetilde\Dcal\big) =\ker\big(\,\tilde
j\,^\bullet\,(\Id_V\otimes\hat\eta)\circ\tau_w\, \big)
$$
is a projective module over $\alg(\widetilde\real_\theta^4)$, called an
instanton bundle on $\widetilde\real_\theta^4$. Moreover, the
coequivariant map
\beq
\widetilde\triangle:=\widetilde\Dcal\circ \widetilde\Dcal^\dag \ \in \ \End_{\alg(\widetilde\real_\theta^4)}\big((V\oplus
V)\otimes\alg(\widetilde\real_\theta^4)
\big)
\label{Laplaceoptilde}\eeq
is an isomorphism in the category ${}^{\hil_{\theta}}\Module$, and the rank $r$ projection $\widetilde P$ 
describing $\tilde\bun$ is  
$$
\widetilde P:=\Id_{V_0}-\widetilde\Dcal^\dag \circ
\widetilde\triangle^{-1}\circ\widetilde\Dcal  \, : \,
V_0\otimes\alg\big(\widetilde\real_\theta^4\big) ~\longrightarrow~
\tilde\bun
$$
with $\widetilde P^2=\widetilde P=\widetilde P^\dag$ and $\Tr\widetilde P=r$.

It remains to determine the gluing automorphism between the two
instanton bundles $\bun$ and $\tilde\bun$. For this, we use
Proposition~\ref{R4glueprop} to write the commutative diagram
\beq
\xymatrix{
0~ \ar[r] & ~ V_0\otimes \alg\big(\real_\theta^4\big)[\tilde\rho\,] ~ \ar[r]^{\!\!\!\!\!\!\!\!\Dcal}
\ar[d]_{\Id_{V_0}\otimes G}  & ~(V\oplus
V)\otimes\alg\big(\real_\theta^4\big)[\tilde\rho\, ] ~ \ar[r]
\ar[d]^{(\galg\otimes1)\circ (\Id_{V\oplus V}\otimes G)} & ~ 0 \\ 0~ \ar[r] & ~ V_0\otimes [\rho]\alg\big(\widetilde\real_\theta^4\big) ~ \ar[r]_{\!\!\!\!\!\!\!\!\widetilde\Dcal} & ~(V\oplus
V)\otimes[\rho]\alg\big(\widetilde\real_\theta^4\big) ~ \ar[r] & ~ 0
}
\label{commdiaginst}\eeq
in the category ${}^{\hil_{\theta}}\Module$, where the linear map
$\galg:V\oplus V\to(V\oplus V)\otimes\alg(\widetilde\real_\theta^4)$ is
defined by
$$
\begin{pmatrix} v_1 \\[6pt] v_2 \end{pmatrix} \ \longmapsto \
\begin{pmatrix} -q\,\big(v_1\, \bar\zeta_1-v_2\,\zeta_2\big) \\[6pt]
  q^{-1}\,\big(v_1 \, \bar \zeta_2-v_2\, \zeta_1\big) \end{pmatrix} \ .
$$
The cohomology of the first row is
$\bun[\tilde\rho\,]:=\bun\otimes_{\alg(\real_\theta^4)}
\alg(\real_\theta^4)[\tilde\rho\,]$, while the cohomology of the
second row is $[\rho]\tilde\bun:=
[\rho]\alg(\widetilde\real_\theta^4)\otimes_{\alg(\widetilde\real_\theta^4)}
\tilde\bun$. In this way the isomorphism of
Proposition~\ref{R4glueprop} induces the desired isomorphism of modules
$G_\bun:\bun[\tilde\rho\,] \to[\rho]\tilde\bun$ which is compatible
with the $\dag$-involutions and obeys $G_\bun(\sigma\triangleleft
a)=G(a)\triangleright G_\bun(\sigma)$ for
$\sigma\in\bun[\tilde\rho\,]$,
$a\in\alg(\real_\theta^4)[\tilde\rho\,]$, thus defining an $\hil_\theta$-coequivariant instanton
bundle on the noncommutative sphere $S_\theta^4$.

\subsection{Instanton gauge fields\label{sec:instconn}}

We will now construct canonical connections on the instanton bundles
introduced in \S\ref{subsec:instbun}. For this, we first need to write down a canonical
exterior differential algebra over the sphere $S_\theta^4$ as a deformation of its classical counterpart, following the general construction of K\"ahler differentials given in~\cite[\S4.4]{CLSI}. We begin by constructing differential forms on the open subvariety $\real_\theta^4$.

Starting with the real affine variety $\real^4$, let
$\Omega^\bullet_{\real^4}=\bigwedge^\bullet\Omega_{\real^4}^1$ be the
usual classical differential calculus on the coordinate
algebra $\alg(\real^4)$, generated as a differential graded
algebra by degree zero elements $\xi_i,\bar\xi_i$ and degree one
elements $\dd\xi_i,\dd\bar\xi_i$ satisfying the skew-commutation relations
$$
\dd\xi_i\wedge\dd\xi_j=-\dd\xi_j\wedge\dd\xi_i \ , \qquad
\dd\xi_i\wedge\dd\bar\xi_j=-\dd\bar\xi_j\wedge\dd\xi_i \ , \qquad
\dd\bar\xi_i\wedge\dd\bar\xi_j=-\dd\bar\xi_j\wedge\dd\bar\xi_i
$$
and the symmetric $\alg(\real^4)$-bimodule structure
$$
\xi_i~\dd\xi_j=\dd\xi_j~\xi_i \ , \qquad
\xi_i~\dd\bar\xi_j=\dd\bar\xi_j~\xi_i \ , \qquad
\bar\xi_i~\dd\bar\xi_j=\dd\bar\xi_j~\bar\xi_i
$$
for $i,j=1,2$. The differential
$\dd:\Omega^0_{\real^4}:=\alg(\real^4)\to\Omega^1_{\real^4}$ is
defined by $\xi_i\mapsto\dd\xi_i$, $\bar\xi_i\mapsto\dd\bar\xi_i$. It
is extended uniquely to a map of degree one, $\dd:\Omega^n_{\real^4}\to
\Omega^{n+1}_{\real^4}$, using $\complex$-linearity and the graded
Leibniz rule
$$
\dd(\omega\wedge\omega'\,)=\dd\omega\wedge\omega'+(-1)^{{\rm
    deg}(\omega)}\, \omega\wedge\dd\omega'
$$
with $\dd^2=0$, 
where the product is taken over $\alg(\real^4)$ using the
bimodule structure of $\Omega^\bullet_{\real^4}$. We will demand that the differential calculus on
$\alg(\real^4)$ is \emph{coequivariant} for the torus coaction
$\Delta_L$ given in (\ref{DeltaLR4}). Then we can extend this coaction
to a left coaction $\Delta_L:\Omega^\bullet_{\real^4}\to
\hil\otimes\Omega^\bullet_{\real^4}$ such that $\dd$ is a left
$\hil$-comodule morphism and $\Delta_L$ is an $\alg(\real^4)$-bimodule morphism.

Following the prescription of \S\ref{toriccat}, we can now use the
comodule cotwist to deform the differential structure in the same way
as for the algebra itself in Proposition~\ref{R4prop}. One finds
\begin{eqnarray*}
\xi_i\blacktriangleright_\theta\dd\xi_j= F_\theta(t_i,t_j)\,\xi_i~\dd\xi_j \quad &,& \quad \bar\xi_i\blacktriangleright_\theta\dd\bar\xi_j= F_\theta(t_{i+1},t_{j+1})\,\bar\xi_i~\dd\bar\xi_j \ , \\[4pt]
\xi_i\blacktriangleright_\theta\dd\bar\xi_j= F_\theta(t_i,t_{j+1})\,\xi_i~\dd\bar\xi_j \quad &,& \quad \bar\xi_i\blacktriangleright_\theta\dd\xi_j= F_\theta(t_{i+1},t_j)\,\bar\xi_i~\dd\xi_j \ , \\[4pt]
\dd\xi_i\wedge_\theta\dd\xi_j= F_\theta(t_i,t_j)~ \dd\xi_i\wedge \dd\xi_j \quad &,& \quad \dd\bar\xi_i\wedge_\theta\dd\bar\xi_j= F_\theta(t_{i+1},t_{j+1})~ \dd\bar\xi_i\wedge \dd\bar\xi_j \ , \\[4pt]
\dd\xi_i\wedge_\theta\dd\bar\xi_j= F_\theta(t_i,t_{j+1})~ \dd\xi_i\wedge \dd\bar\xi_j \quad &,& \quad \dd\bar\xi_i\wedge_\theta\dd\xi_j= F_\theta(t_{i+1},t_j)\,\dd\bar\xi_i\wedge \dd\xi_j \ ,
\end{eqnarray*}
with indices read modulo~$2$. The braided exterior product
$\wedge_\theta$ describes how the tensor product acts on the quotient of the
tensor algebra
$$
T_\theta\big(\Omega_{\real^4}^1\big)=\alg\big(\real_\theta^4\big) \ \oplus \ \bigoplus_{n\geq1}\,
\big(\Omega_{\real^4}^1\big)^{\otimes_{\alg(\real_\theta^4)}\, n}
$$ 
by the
ideal generated by the braided skew-commutation relations (see (\ref{braidedtensoralg})). 
The undeformed differential $\dd$ is still a
derivation (of degree one) of the deformed product $\wedge_\theta$, as
follows from general results of twist deformation theory~\cite{MajidPlanck} or
simply by direct computation
\begin{eqnarray*}
\dd(\omega_1\wedge_\theta\omega_2) &=& F_\theta\big(\omega_1^{(-1)}\,,\,
\omega_2^{(-1)}\big)~ \dd\big(\omega_1^{(0)} \wedge\omega_2^{(0)}
\big) \\[4pt] &=& F_\theta\big(\omega_1^{(-1)}\,,\,
\omega_2^{(-1)}\big)~ \big(\dd\omega_1^{(0)} \wedge\omega_2^{(0)}
+(-1)^{{\rm deg}(\omega_1^{(0)})}\, \omega_1^{(0)} \wedge\dd \omega_2^{(0)}
\big) \\[4pt] &=& \dd\omega_1 \wedge_\theta \omega_2
+(-1)^{{\rm deg}(\omega_1)}\, \omega_1\wedge_\theta \dd \omega_2
\end{eqnarray*}
for homogeneous forms $\omega_1$, $\omega_2$, with the usual Sweedler notation $\Delta_L(\omega)=\omega^{(-1)}\otimes \omega^{(0)}$.

The above construction defines a canonical differential graded algebra
$\Omega_{\real_\theta^4}^\bullet=
\bigwedge^\bullet_\theta\Omega_{\real^4}^1$ for $\alg(\real_\theta^4)$
with the same generators and the same differential $\dd$, but now
subject to the braided skew-commutation relations
\begin{eqnarray}
\dd\xi_i\wedge \dd\xi_j\= -q_{ij}^2~\dd\xi_j\wedge\dd\xi_i \quad &,& \quad
\dd\bar\xi_i\wedge \dd\bar\xi_j\= -q_{i+1\,j+1}^2~\dd\bar\xi_j\wedge\dd\bar\xi_i \ , \nonumber \\[4pt] 
\dd\xi_i\wedge \dd\bar\xi_j &=& -q_{i\,j+1}^2~\dd\bar\xi_j\wedge\dd\xi_i
\label{R4skewcomm}\end{eqnarray}
and the braided symmetric $\alg(\real_\theta^4)$-bimodule structure
\begin{eqnarray*}
\xi_i~\dd\xi_j\= q_{ij}^2~\dd\xi_j~\xi_i \quad &,& \quad
\bar\xi_i~\dd\bar\xi_j\= q_{i+1\,j+1}^2~\dd\bar\xi_j~\bar\xi_i \ , \\[4pt]
\xi_i~\dd\bar\xi_j\= q_{i\,j+1}^2~\dd\bar\xi_j~\xi_i \quad &,& \quad
\bar\xi_i~\dd\xi_j\= q_{i+1\,j}^2~\dd\xi_j~\bar\xi_i \ .
\end{eqnarray*}
Again we drop the explicit deformation symbols from the products. This
also identifies $\Omega_{\real_\theta^4}^\bullet$ via the restriction
$$
j^\bullet\Omega_{\Gr_\theta(2;4)}^1\cong \Omega_{\real_\theta^4}^1
$$
of the bundle of noncommutative K\"ahler differentials
(\ref{lcinvtbraidedprod}). The real structure (\ref{starxi}) extends
to $\Omega_{\real_\theta^4}^\bullet$ by graded extension of the morphism $\xi_i\mapsto\xi_i^\dag$.

We are now ready to construct a canonical connection on the right $\alg(\real_\theta^4)$-module $\bun$, i.e. a $\complex$-linear map
$$
\nabla\,:\, \bun ~\longrightarrow~ \bun\otimes_{\alg(\real_\theta^4)} \Omega^1_{\real_\theta^4}
$$
satisfying the Leibniz rule
$$
\nabla(\sigma\triangleleft f)=(\nabla\sigma)\triangleleft f+ \sigma\otimes \dd f
$$
for $f\in\alg(\real_\theta^4)$ and $\sigma\in\bun$. The connection $\nabla$ extends to one-forms as a $\complex$-linear map
$$
\nabla\,:\, \bun\otimes_{\alg(\real_\theta^4)} \Omega^1_{\real_\theta^4} ~\longrightarrow~ \bun\otimes_{\alg(\real_\theta^4)} \Omega^2_{\real_\theta^4}
$$
satisfying
$$
\nabla(\sigma\otimes\omega)=(\nabla\sigma)\otimes \omega+ \sigma\otimes\dd\omega
$$
for $\omega\in\Omega^1_{\real_\theta^4}$ and $\sigma\in\bun$. Two
connections $\nabla$ and $\nabla'$ are \emph{gauge equivalent} if
there exists an automorphism $g\in\Aut_{\alg(\real_\theta^4)}(\bun)$ such that $\nabla=\tilde g\circ\nabla'\circ g^{-1}$, where $\tilde g=g\otimes\Id$. The curvature $F_\nabla=\nabla\circ\nabla$ is defined by the composition
$$
\bun~ \xrightarrow{\ \nabla\ }~\bun\otimes_{\alg(\real_\theta^4)} \Omega^1_{\real_\theta^4} ~\xrightarrow{\ \nabla\ }~ \bun\otimes_{\alg(\real_\theta^4)} \Omega^2_{\real_\theta^4} \ .
$$
Since $F_\nabla$ is right $\alg(\real_\theta^4)$-linear, it may be regarded as an element
$$
F_\nabla\ \in\ \Hom_{\alg(\real_\theta^4)} 
\big( \bun\,,\, \bun \otimes_{\alg(\real_\theta^4)}\Omega^2_{\real_\theta^4}\big) \ .
$$
If $\nabla$ and $\nabla'$ are gauge equivalent via the gauge transformation $g\in\Aut_{\alg(\real_\theta^4)}(\bun)$, then the corresponding curvatures are easily found to be related as $F_\nabla=\tilde g\circ F_{\nabla'}\circ g^{-1}$.

To define the instanton connection, let $\iota:\bun\to
V_0\otimes\alg(\real_\theta^4)$ denote the natural inclusion, and
$\dd:\alg(\real_\theta^4)\to \Omega^1_{\real_\theta^4}$ the
differential introduced above. We then take $\nabla$ to be the
Grassmann connection associated to the projection $P$, which is
defined by the composition
$$
\nabla\,:\, \bun~ \xrightarrow{\ \iota\ }~ V_0\otimes\alg(\real_\theta^4)~ \xrightarrow{\Id_{V_0}\otimes\dd}~ V_0\otimes\Omega^1_{\real_\theta^4}~ \xrightarrow{P\otimes\Id}~ \bun\otimes_{\alg(\real_\theta^4)}\Omega^1_{\real_\theta^4} \ ,
$$
or $\nabla=P\circ\dd\circ\iota$. It is easily seen to be compatible
with the $\dag$-involution, since
$$
\nabla\big(\sigma^\dag\,\big)=P\,
\dd\big(\sigma^\dag\,\big)=P\,(\dd\sigma)^\dag= (P\, \dd\sigma)^\dag=
\nabla(\sigma)^\dag
$$
for $\sigma\in\bun$. Its curvature is given by
$$
F_\nabla=\nabla^2 = P~(\dd P)^2 \ .
$$
\begin{lemma}
The gauge equivalence class of the instanton connection $\nabla=P\circ\dd\circ\iota$ depends only on the class of the noncommutative ADHM data $(B,I,J)$ in the moduli variety $\Mcal_\theta^{\rm tw}(W,V)$.
\label{instconngauge}\end{lemma}
\Proof{
Let $g_0:\bun\to\bun'$ be the $\alg(\real_\theta^4)$-module isomorphism induced by $\tilde g\oplus\Id_W\in\U(V_0)$ in (\ref{monadisog}). Let $\iota':\bun'\to V_0\otimes\alg(\real_\theta^4)$ be the inclusion, and let $P':V_0\otimes\alg(\real_\theta^4)\to\bun'$ be the projection defined as in (\ref{instbunproj}). Then $\iota'=g_0\circ\iota\circ g_0^{-1}$ and $P'=g_0\circ P\circ g_0^{-1}$. Thus
\begin{eqnarray*}
\nabla' &=& P'\circ\dd\circ\iota' \\[4pt]
&=& \big(g_0\circ P\circ g_0^{-1}\big)\circ\dd\big(g_0\circ\iota\circ g_0^{-1}\big) \\[4pt]
&=& \big(g_0\circ P\circ g_0^{-1}\big)\circ\big((\dd g_0)\circ\iota\circ g_0^{-1} + g_0\circ\dd(\iota\circ g_0^{-1})\big) \\[4pt]
&=&\tilde g_0\circ\big(P\circ\dd\circ\iota\big)\circ g_0^{-1} \= \tilde g_0\circ\nabla\circ g_0^{-1} \ ,
\end{eqnarray*}
where we have used the Leibniz rule together with the fact that $g_0$ acts as the identity on $\alg(\real_\theta^4)$ so that $\dd g_0=0=\dd g_0^{-1}$.
}

Similarly, the coequivariant differential calculus
$\big(\Omega^\bullet_{\widetilde\real_\theta^4}\,,\,\dd \big)$ over the affine variety
$\widetilde\real_\theta^4$ is generated by $\zeta_i,\bar\zeta_i$ and
$\dd\zeta_i,\dd\bar\zeta_i$ with $i=1,2$. The relations between
$\zeta_i,\bar\zeta_i$ are given in Proposition~\ref{tildeR4prop} with
the torus coaction in \eqref{DeltaLR4tilde}. By coequivariance, one has
\begin{eqnarray*}
\Delta_L(\dd a)&=& (\Id\otimes \dd)\Delta_L(a) \ , \\[4pt]
a\blacktriangleright_\theta \Delta_L(\omega) \ =\ \Delta_L(a\blacktriangleright_\theta\omega) \ &,& \ \Delta_L(\omega)\blacktriangleleft_\theta a \ = \ \Delta_L(\omega\blacktriangleleft_\theta a)
\end{eqnarray*}
for $a\in\alg(\widetilde\real_\theta^4)$ and $\omega\in\Omega^1_{\widetilde\real_\theta^4}$, where here $\alg(\widetilde\real_\theta^4)$ acts on $\alg(\widetilde\real_\theta^4)\otimes\Omega^1_{\widetilde\real_\theta^4}$ in the tensor product representation. Thus the
differentials $\dd\zeta_i,\dd\bar\zeta_i$ for $i=1,2$ behave algebraically in exactly the same way as
$\zeta_i,\bar\zeta_i$, and so we have
the relations
\begin{eqnarray}
\zeta_i~\dd\zeta_j\= q_{ij}^{-2}~\dd\zeta_j~\zeta_i \quad &,& \quad
\bar\zeta_i~\dd\bar\zeta_j\=
q_{i+1\,j+1}^{-2}~\dd\bar\zeta_j~\bar\zeta_i \ , \nonumber \\[4pt]
\zeta_i~\dd\bar\zeta_j\= q_{i\,j+1}^{-2}~\dd\bar\zeta_j~\zeta_i \quad &,& \quad
\bar\zeta_i~\dd\zeta_j\= q_{i+1\,j}^{-2}~\dd\zeta_j~\bar\zeta_i \ , \nonumber \\[4pt]
\dd\zeta_i\wedge \dd\zeta_j\= -q_{ij}^{-2}~\dd\zeta_j\wedge\dd\zeta_i \quad &,& \quad
\dd\bar\zeta_i\wedge \dd\bar\zeta_j\= -q_{i+1\,j+1}^{-2}~\dd\bar\zeta_j\wedge\dd\bar\zeta_i \ , \nonumber \\[4pt] 
\dd\zeta_i\wedge \dd\bar\zeta_j &=& -q_{i\,j+1}^{-2}~\dd\bar\zeta_j\wedge\dd\zeta_i
\label{dzetacommrels}\end{eqnarray}
for $i,j=1,2$. 
The instanton connection
$\widetilde\nabla=\widetilde P\circ\dd\circ\tilde\iota:\tilde\bun\to
\Omega^1_{\widetilde\real_\theta^4}\otimes_{\alg(\widetilde\real_\theta^4)}
\tilde\bun$ on the projective left
$\alg(\widetilde\real_\theta^4)$-module $\tilde\bun$ is similarly defined as in the previous case, 
with curvature $F_{\widetilde\nabla}=\widetilde P~(\dd\widetilde P)^2$ as an
element of $\Hom_{\alg(\widetilde\real_\theta^4)}\big(\tilde\bun\,,\, \Omega^2_{\widetilde\real_\theta^4} \otimes_{\alg(\widetilde\real_\theta^4)}
\tilde\bun\big)$.

We need to check consistency between the connections
$\nabla_{\tilde\rho}$ and ${}_\rho\widetilde\nabla$ on the adjoinment
bundles $\bun[\tilde\rho\, ]$ and $[\rho]\tilde\bun$, respectively, using
the module isomorphism $G_\bun:\bun[\tilde\rho\, ]\to [\rho]\tilde\bun$
induced by the commutative diagram (\ref{commdiaginst}). For this,
consider the diagram
\beq
&&
\xymatrix{
\bun[\tilde\rho\, ] ~ \ar[r]^{\!\!\!\!\!\!\!\!\!\!\!\!\!\!\!\!\!\!\! \iota} \ar[d]_{G_\bun} & ~
V_0\otimes \alg\big(\real_\theta^4\big)[\tilde\rho\,] ~ \ar[r]^{\ \ \ \Id_{V_0}\otimes\dd}
\ar[d]_{\Id_{V_0}\otimes G}  & ~ \ 
V_0\otimes\Omega^1_{\real_\theta^4}[\tilde\rho\, ] ~
\ar[r]^{\!\!\!\!\!\!\! P}
\ar[d]^{\Id_{V_0}\otimes G^{(1)}} & ~ \bun\otimes_{\alg(\real_\theta^4)}
\Omega^1_{\real_\theta^4}[\tilde\rho\, ] \ar[d]^{G_\bun^{(1)}} \\
[\rho]\tilde\bun ~ \ar[r]_{\!\!\!\!\!\!\!\!\!\!\!\!\!\!\!\!\!\!\! \tilde\iota} & ~ V_0\otimes
[\rho]\alg\big(\widetilde\real_\theta^4\big) ~ \ar[r]_{\ \ \ \Id_{V_0}\otimes
  \dd} & \ ~V_0\otimes[\rho] \Omega^1_{\widetilde\real_\theta^4} ~
\ar[r]_{\!\!\!\!\!\!\! \tilde P} & ~ [\rho] \Omega_{\widetilde\real_\theta^4}^1\otimes_{\alg(\widetilde\real_\theta^4)}
\tilde\bun
} \ .
\label{connectiondiagram}\eeq
The commutativity of the first square follows from commutativity of
the diagram in (\ref{commdiaginst}). 
Next, we can lift $G$ to the bimodule isomorphism
$G^{(1)}:\Omega^1_{\real_\theta^4}[\tilde\rho\, ]\to
[\rho]\Omega_{\widetilde\real_\theta^4}^1$ through the intertwining
relation defined by commutativity of the second square.
Finally, with this definition we can extend $G^{(1)}$ to the bundle
isomorphism 
$$
G_\bun^{(1)}\, :\, \bun\otimes_{\alg(\real_\theta^4)}
\Omega^1_{\real_\theta^4}[\tilde\rho\, ] ~\longrightarrow~
[\rho]\Omega_{\widetilde\real_\theta^4}^1\otimes_{\alg(\widetilde\real_\theta^4)}
 \tilde\bun
$$
by demanding that the third square be commutative; it is also
compatible with the $\dag$-involutions. Then the instanton connections are related by the gauge transformation
\beq
{}_\rho\widetilde\nabla=G_\bun^{(1)}\circ\nabla_{\tilde\rho}\circ
G_\bun^{-1} \ ,
\label{nablarhotransf}\eeq
as desired. This defines the instanton connection on the noncommutative sphere~$S_\theta^4$.

\subsection{Anti-selfduality equations}

We will now demonstrate that the Grassmann connections constructed in
\S\ref{sec:instconn} define noncommutative instantons in analogy with
the classical case, i.e. that they satisfy some form of anti-selfduality
equations. For this, we need to construct suitable versions of the
classical Hodge duality operator acting on two-forms. The euclidean
metric on $\real^4$ induces a Hodge duality operator which on
two-forms is the linear map
$*:\Omega^2_{\real^4}\to\Omega^2_{\real^4}$ defined by
\begin{eqnarray}
*(\dd\xi_1\wedge\dd\bar\xi_2) = -\dd\xi_1\wedge\dd\bar\xi_2 \quad &,&
\quad *(\dd\xi_1\wedge\dd\xi_2) = \dd\xi_1\wedge\dd\xi_2 \ ,
\nonumber \\[4pt] 
*(\dd\xi_1\wedge\dd\bar\xi_1) = -\dd\xi_2\wedge\dd\bar\xi_2 \quad &,&
\quad *(\dd\xi_2\wedge\dd\bar\xi_2) = -\dd\xi_1\wedge\dd\bar\xi_1 \ ,
\nonumber \\[4pt]
*(\dd\bar\xi_1\wedge\dd\bar\xi_2) = \dd\bar\xi_1\wedge\dd\bar\xi_2
\quad &,& \quad *(\dd\xi_2\wedge\dd\bar\xi_1) = -\dd\xi_2\wedge\dd\bar\xi_1 \ .
\label{Hodgeclass}\end{eqnarray}

In contrast to the case of isospectral
deformations~\cite{cl,LPRvS,BL,BL1}, the coaction (\ref{DeltaLR4}) of
the torus algebra $\hil=\alg((\complex^\times)^2)$ on $\alg(\real^4)$
is \emph{not} isometric. However, it does coact by conformal
transformations on $\alg(\real^4)$, and hence preserves the Hodge
operator. This is easily checked using (\ref{Hodgeclass}) and
(\ref{DeltaLR4}) which shows that the operator
$*:\Omega^2_{\real^4}\to\Omega^2_{\real^4}$ is coequivariant,
$$
\Delta_L(*\omega)=(\Id\otimes *)\Delta_L(\omega) \qquad \mbox{for} \quad
\omega\in\Omega^2_{\real^4} \ ,
$$
and hence defines a morphism of the category ${}^\hil\Module$. Since
the vector space $\Omega_{\real_\theta^4}^2$ coincides with its
classical counterpart $\Omega^2_{\real^4}$, and since the quantization
functor $\mathscr{F}_\theta:
{}^{\hil}\Module\to {}^{\hil_\theta}\Module$ acts as the identity
  on objects and morphisms of ${}^{\hil}\Module$, we can define a
  Hodge duality operator $*_\theta:\Omega_{\real_\theta^4}^2\to
  \Omega_{\real_\theta^4}^2$ by the same formula (\ref{Hodgeclass}),
  which by construction is a morphism in the category
  ${}^{\hil_\theta}\Module$. In particular, it satisfies
  $*_\theta^2=\Id$ and the $\hil_\theta$-coequivariant decomposition of the right
  $\alg(\real_\theta^4)$-module
$$
\Omega_{\real_\theta^4}^2 =\Omega_{\real_\theta^4}^{2,+} \oplus
\Omega_{\real_\theta^4}^{2,-}
$$
into submodules, corresponding to eigenvalues $\pm\,1$ of $*_\theta$, is
identical as a vector space to that of the classical
case. Hence the eigenmodules of selfdual and anti-selfdual
two-forms are given respectively by
\begin{eqnarray}
\Omega_{\real_\theta^4}^{2,+}&=&
\alg\big(\real_\theta^4\big)\big\langle
\dd\xi_1\wedge\dd\xi_2\,,\, \dd\bar\xi_1\wedge\dd\bar\xi_2\,,\,
\dd\xi_1\wedge\dd\bar\xi_1- \dd\xi_2\wedge\dd\bar\xi_2 \big\rangle \ ,
\nonumber \\[4pt]
\Omega_{\real_\theta^4}^{2,-} &=&
\alg\big(\real_\theta^4\big)\big\langle \dd\xi_1\wedge\dd\bar\xi_2\,,\,
\dd\xi_2\wedge\dd\bar\xi_1\,,\, \dd\xi_1\wedge\dd\bar\xi_1 +
\dd\xi_2\wedge\dd\bar\xi_2 \big\rangle \ .
\label{OmegapmR4}\end{eqnarray}

In a completely analogous way, by using the $\hil$-coaction
(\ref{DeltaLR4tilde}) one constructs a morphism $\tilde *_\theta:\Omega_{\widetilde\real_\theta^4}^2\to
  \Omega_{\widetilde\real_\theta^4}^2$ in the category
  ${}^{\hil_\theta}\Module$ with the same formula (\ref{Hodgeclass})
  but with affine coordinates $\xi_i$ substituted with $\zeta_i$, and
  with overall changes of sign reflecting the change of
  ``orientation''. There is an analogous $\hil_\theta$-coequivariant decomposition $$
\Omega_{\widetilde\real_\theta^4}^2 =\Omega_{\widetilde\real_\theta^4}^{2,+} \oplus
\Omega_{\widetilde\real_\theta^4}^{2,-}
$$
of two-forms into selfdual and anti-selfdual
two-forms with
\begin{eqnarray}
\Omega_{\widetilde\real_\theta^4}^{2,+}&=&
\big\langle \dd\zeta_1\wedge\dd\bar\zeta_2\,,\,
\dd\zeta_2\wedge\dd\bar\zeta_1\,,\, \dd\zeta_1\wedge\dd\bar\zeta_1 +
\dd\zeta_2\wedge\dd\bar\zeta_2 \big\rangle\alg\big(\widetilde\real_\theta^4\big) \ ,
\nonumber \\[4pt]
\Omega_{\widetilde\real_\theta^4}^{2,-} &=&
\big\langle
\dd\zeta_1\wedge\dd\zeta_2\,,\, \dd\bar\zeta_1\wedge\dd\bar\zeta_2\,,\,
\dd\zeta_1\wedge\dd\bar\zeta_1- \dd\zeta_2\wedge\dd\bar\zeta_2 \big\rangle\alg\big(\widetilde\real_\theta^4\big) \ .
\label{OmegapmR4tilde}\end{eqnarray}
The consistency condition for the corresponding morphisms
on the adjoinment bimodules
$*_\theta\,^{\tilde\rho}:\Omega_{\real_\theta^4}^2[\tilde\rho\, ]\to
  \Omega_{\real_\theta^4}^2[\tilde\rho\, ]$ and ${}^\rho\,\tilde *_\theta:[\rho]\Omega_{\widetilde\real_\theta^4}^2\to
  [\rho]\Omega_{\widetilde\real_\theta^4}^2$ are given by lifting the
  isomorphism $G^{(1)}$ of (\ref{connectiondiagram}) to the bimodule isomorphism
$G^{(2)}:\Omega^2_{\real_\theta^4}[\tilde\rho\, ]\to
[\rho]\Omega_{\widetilde\real_\theta^4}^2$ through the intertwining
relation defined by commutativity of the diagram
$$
\xymatrix{
\Omega^1_{\real_\theta^4}[\tilde\rho\, ] ~
\ar[r]^{\!\!\!\!\!\!\! \dd}
\ar[d]_{\tilde\Jscr_2\circ G^{(1)}} & ~ 
\Omega^2_{\real_\theta^4}[\tilde\rho\, ] \ar[d]^{G^{(2)}} \\
[\rho] \Omega^1_{\widetilde\real_\theta^4} ~
\ar[r]_{\!\!\!\!\!\!\! \dd} & ~ [\rho] 
\Omega_{\widetilde\real_\theta^4}^2
} \ ,
$$
where $\tilde\Jscr_2:\Omega_{\widetilde\real_\theta^4}^\bullet \to
\Omega_{\widetilde\real_\theta^4}^\bullet$ is induced by graded extension of the map
$(\zeta_1,\zeta_2)\mapsto (\zeta_1,q^2\,\bar\zeta_2)$.
One then has
\beq
{}^\rho\,\tilde *_\theta = G^{(2)}\circ *_\theta\,^{\tilde\rho}\circ
G^{(2)}\,^{-1} \ ,
\label{Hodgerhotransf}\eeq
which defines the $\hil_\theta$-coequivariant Hodge operator
on the noncommutative sphere~$S_\theta^4$.

\begin{proposition}
The curvatures $F_\nabla$ and $F_{\widetilde\nabla}$ of the instanton
connections are anti-selfdual, i.e. as two-forms they obey the anti-selfduality equations
$$
*_\theta F_\nabla=-F_\nabla \ , \qquad \tilde*_\theta
F_{\widetilde\nabla}= -F_{\widetilde\nabla} \ .
$$
\label{selfdualprop}\end{proposition}
\Proof{
Using $\Dcal(\sigma)=0$ for $\sigma\in\bun$, so that
$\Dcal(\dd\sigma)=-(\dd\Dcal)(\sigma)$ by the Leibniz rule, and $P({\rm
  im}(\Dcal^\dag\,))=0$, the action of the curvature $F_\nabla=P\,(\dd P)^2\in \Hom_{\alg(\real_\theta^4)} 
\big( \bun\,,\, \bun
\otimes_{\alg(\real_\theta^4)}\Omega^2_{\real_\theta^4} \big)$
is given by
$$
F_\nabla(\sigma)=P\,\big(\dd(\Id_{V_0}-\Dcal^\dag\circ\triangle^{-1}
\circ\Dcal)\wedge\dd \sigma\big)=
P\,\big(\dd\Dcal^\dag\circ\triangle^{-1} \wedge\dd\Dcal(\sigma)\big) \
.
$$
Using the commutation relations of Proposition~\ref{R4prop} and (\ref{CP1Koszulrels}), the real
structures (\ref{starxi}) and (\ref{Jscrcheckw}), and the ADHM
equations (\ref{NCADHMeqs}) and (\ref{realADHMeq}), one finds that the
Laplace operator (\ref{Laplaceop}) assumes the block diagonal form
$$
\triangle=\begin{pmatrix} \delta & 0 \\ 0 & \delta \end{pmatrix} \ ,
$$
where the isomorphism
$\delta\in\End_{\alg(\real_\theta^4)}\big(V\otimes
\alg(\real_\theta^4) \big)$
is given by
\begin{eqnarray*}
\delta= B_1\, B_1^\dag+q^{-2}\, B_2\,
B_2^\dag\,+I\, I^\dag+q^{-1}\,
\rho -q^{-1}\, B_1\, \bar\xi_1-B_1^\dag\,\xi_1+ q^{-1}\, B_2\, \bar\xi_2-B_2^\dag\, \xi_2 \ .
\end{eqnarray*}
It follows that $F_\nabla$ is proportional to $\dd\Dcal^\dag\wedge\dd \Dcal$ as a two-form. The exterior derivative of the Dirac operator (\ref{Diracopexpl}) and its adjoint are easily computed to be
$$
\dd\Dcal = \begin{pmatrix} -q^{-1}\, \dd\xi_1\otimes\check w_1-q\, \dd\xi_2\otimes\check w_2 & 0 \\[6pt] -q^{-1}\, \dd\bar\xi_2\otimes\check w_1-q\, \dd\bar\xi_1\otimes\check w_2 & 0 \end{pmatrix} \ , \qquad \dd\Dcal^\dag = \begin{pmatrix} -q^{-2}\, \dd\bar\xi_1\otimes\check w_1 & \dd\xi_2\otimes\check w_1 \\[6pt] \dd\bar\xi_2\otimes\check w_2 & -q^2\, \dd\xi_1\otimes\check w_2 \\[6pt] 0 & 0 \end{pmatrix} \ .
$$
Using the commutation relations (\ref{R4skewcomm}) we then find
$$
\dd\Dcal^\dag\wedge \dd\Dcal = \begin{pmatrix} \ -q^{-1}\,\big(\dd\xi_1\wedge\dd\bar\xi_1+ \dd\xi_2\wedge\dd\bar\xi_2\big) & -\big(q+q^{-1}\big)\, \dd\xi_2\wedge\dd\bar \xi_1 & 0 \ \\[6pt] \big(q+q^{-1}\big)\, \dd\xi_1\wedge\dd\bar \xi_2 & q^3\,\big(\dd\xi_1\wedge\dd\bar \xi_1+\dd\xi_2\wedge\dd\bar\xi_2\big) & 0 \ \\[6pt] 0 & 0 & 0 \ \end{pmatrix} \ .
$$
Comparing with (\ref{OmegapmR4}), we see that each entry of $F_\nabla$ belongs to the submodule $\Omega_{\real_\theta^4}^{2,-}$.

The anti-selfduality equation for the curvature
$F_{\widetilde\nabla}=\tilde P~(\dd\tilde P)^2\in \Hom_{\alg(\widetilde\real_\theta^4)}\big(\tilde\bun\,,\,\Omega^2_{\widetilde\real_\theta^4} \otimes_{\alg(\widetilde\real_\theta^4)}
\tilde\bun\big)$ follows analogously. This time
we use Proposition~\ref{tildeR4prop} and (\ref{starzeta}) to write the
Laplace operator (\ref{Laplaceoptilde}) in the block diagonal form
$$
\tilde \triangle=\begin{pmatrix} \, \tilde\delta & 0 \\ 0 & \tilde\delta \, \end{pmatrix} \ ,
$$
where the isomorphism
$\tilde\delta\in\End_{\alg(\widetilde\real_\theta^4)}\big(V\otimes
\alg(\widetilde\real_\theta^4) \big)$
is given by
\begin{eqnarray*}
\tilde\delta= q^{-1}\big(B_1\, B_1^\dag+q^{-2}\, B_2\,
B_2^\dag\,+I\, I^\dag\, \big) \,
\tilde\rho -B_1\, \bar\zeta_1-q^{-1}\, B_1^\dag\,\zeta_1+ q^{-2}\,
B_2\, \bar\zeta_2-q\, B_2^\dag\, \zeta_2 +1 \ .
\end{eqnarray*}
The exterior derivative of the Dirac operator (\ref{Diracoptilde}) and its adjoint have the form
\begin{eqnarray*}
\dd\tilde\Dcal &=& \begin{pmatrix} \begin{matrix} {
      -q\, \big(B_1\, \dd\bar\zeta_1+B_2^\dag\,\dd\zeta_2\big)\otimes
\check w_1} \\ {-\,q\, \big(B_2\,\dd\bar\zeta_1-B_1^\dag\, \dd
\zeta_2\big)\otimes\check w_2} \end{matrix} & \ \
-q\,\big(I\,\dd\bar\zeta_1 +J^\dag\, \dd\zeta_2\big) \\~\\ \begin{matrix} {
      q^{-1}\, \big(B_1\, \dd\bar\zeta_2+B_2^\dag\,\dd\zeta_1\big)\otimes
\check w_1} \\ { +\,q^{-1}\, \big(B_2\,\dd\bar\zeta_2-B_1^\dag\, \dd
\zeta_1\big)\otimes\check w_2} \end{matrix} & \ \
q^{-1}\,\big( I\,\dd\bar\zeta_2+J^\dag\, \dd\zeta_1\big)
\end{pmatrix} \ , \\[4pt]
\dd\tilde\Dcal^\dag&=&\begin{pmatrix}
-\big(B_1^\dag\, \dd\zeta_1-q^{-2}\, B_2\, \dd\bar\zeta_2\big)\otimes\check w_1 & -\big(q^2\,B_1^\dag\, \dd\zeta_2+B_2\,\dd\bar\zeta_1\big)\otimes\check w_1 \\[6pt] -\big(B_2^\dag\,\dd\zeta_1+q^{-2}\, B_1\,\dd\bar\zeta_2\big)\otimes\check w_2 & -\big(q^2\, B_2^\dag\, \dd\zeta_2-B_1\,\dd\bar\zeta_1\big)\otimes\check w_2 \\[6pt] -I^\dag\, \dd\zeta_1+q^{-2}\, J\,\dd\bar\zeta_2 & -q^2\, I^\dag\, \dd\zeta_2+J\, \dd\bar\zeta_1
\end{pmatrix} \ ,
\end{eqnarray*}
and using (\ref{dzetacommrels}) we find
\begin{eqnarray*}
\dd\tilde\Dcal^\dag\wedge\dd\tilde\Dcal &=& q\ \Fcal^{1,1}\ \big(\dd\zeta_1\wedge\dd\bar\zeta_1-\dd\zeta_2\wedge\dd\bar\zeta_2\big) \\ && +\, \big(q+q^3\big)\ \Fcal^{2,0}\ \dd\zeta_1\wedge\dd\zeta_2+\big(q +q^{-1}\big)\ \Fcal^{0,2}\ \dd\bar\zeta_1\wedge\dd\bar\zeta_2
\end{eqnarray*}
where
\begin{eqnarray*}
\Fcal^{1,1} &=& \begin{pmatrix}
\, B_1^\dag\,B_1+B_2\, B_2^\dag & B_1^\dag\, B_2-B_2\, B_1^\dag & B_1^\dag\, I+ B_2\, J^\dag\, \\[6pt] B_2^\dag\,B_1-B_1\, B_2^\dag & B_1\, B_1^\dag+B_2^\dag\, B_2 & B_2^\dag\, I-B_1\, J^\dag \\[6pt] I\, B_1-J\, B_2^\dag & I^\dag\, B_2+J\, B_1^\dag & I^\dag\, I-J\, J^\dag
\end{pmatrix} \ , \\[4pt]
\Fcal^{2,0} &=& \begin{pmatrix}
\, B_1^\dag\,B_2^\dag & -\big(B_1^\dag\,\big)^2 & B_1^\dag\, J^\dag \, \\[6pt] \big(B_2^\dag\,\big)^2 & -B_2^\dag\, B_1^\dag & B_2^\dag\, J^\dag \\[6pt] I^\dag\, B_2^\dag & -I^\dag\, B_1^\dag & I^\dag\, J^\dag
\end{pmatrix} \ , \\[4pt]
\Fcal^{0,2} &=& \begin{pmatrix}
\, -B_2\, B_1 & -B_2^2 & -B_2\, I \, \\[6pt] B_1^2 & B_1\, B_2 & B_1\, I \\[6pt] J\, B_1 & J\, B_2 & J\, I
\end{pmatrix} \ .
\end{eqnarray*}
Comparing with (\ref{OmegapmR4tilde}), we see that $F_{\widetilde\nabla}$ belongs to $\Omega_{\widetilde\real_\theta^4}^{2,-}$ as a two-form.
}

It is easy to see that the anti-selfduality equations of Proposition~\ref{selfdualprop} are consistent with the overlap relations (\ref{nablarhotransf}) and (\ref{Hodgerhotransf}). This defines the instanton equations on the noncommutative sphere $S_\theta^4$.

\newsection{Construction of instanton moduli spaces\label{Instmodsp}}

\subsection{Instanton moduli functors\label{Instmodulifunctor}}

In this section we will exploit the fact that our noncommutative variety
$\CP_\theta^2$ is a
member of one of the general classes of noncommutative projective
planes considered in~\cite{NS,DNVdB}; hence we can straightforwardly utilize their
moduli space constructions, and in the following we will freely borrow from
their results. In the notation of~\cite{NS},
our $\CP_\theta^2$ is of type ${\bf S}_1$, associated to a curve
$E\subset\CP^2$ which is isomorphic to a triangle (union of three
lines $\CP^1$) such that each component is stabilized by an
automorphism $\sigma$. Note that this is \emph{very} different from
the projective planes considered in~\cite{KKO}, which are each of type
${\bf S}_1'$ with $E$ isomorphic to the union of a line and a conic (and in
particular are not given by toric noncommutative deformations of
$\CP^2$). Recall from \S\ref{subsec:Toricquant} that the noncommutative toric variety $\CP_\theta^2$ occurs
in the (universal) flat family $\alg=\alg(\CP_\theta^2)$ parametrized
by the commutative unital algebra $\alg(\complex^\times)=\complex(q)$ over
$\complex$ dual to the smooth irreducible curve
$\bigwedge^2T\cong\complex^\times$. This family
includes the commutative polynomial algebra
$\alg(\CP^2):=\complex[w_1,w_2,w_3]$ (for $q=1$ or $\theta=0$). The
moduli spaces constructed by Nevins and Stafford in~\cite{NS} all behave well in this
family, and are $\complex$-schemes in the usual sense; they are constructed as
geometric invariant theory quotients of subvarieties of products of grassmannians.

In~\cite{NS} it is shown that there exists a projective coarse moduli
space $\Mcal_{\CP_\theta^2}(r,c_1,\chi)$ for semistable torsion free
modules in $\coh(\CP_\theta^2)$ of rank $r\geq1$, first Chern class $c_1$,
and Euler characteristic $\chi$. This moduli space behaves well in
families. In particular, there exists a quasi-projective $T$-scheme
$\Mcal_T(r,c_1,\chi)\to\bigwedge^2 T$ which is smooth over
$\bigwedge^2 T$ and whose fibre over
$q=\exp(\ii\theta)\in\bigwedge^2T$ is precisely the moduli space
$\Mcal_{\CP_\theta^2}(r,c_1,\chi)$; it follows that
$\Mcal_{\CP_\theta^2}(r,c_1,\chi)$ is smooth. We will show that the
variety $\widehat\Mcal_\theta^{\rm ADHM}(r,k)$ of noncommutative (complex) ADHM data
is a fine moduli space for framed sheaves of rank $r$, first Chern
class $c_1=0$, and Euler characteristic $\chi=r-k$ on
$\Open(\CP_\theta^2)$. This isomorphism induces the bijections
of~\S\ref{Instcounting}. Recall that by a ``fine'' moduli space
$\Mcal$ here we mean that there exists a \emph{universal}
framed sheaf $\hat\bun$, i.e. a family of framed sheaves parametrized
by $\Mcal$ such that for any other family of framed sheaves
$\bun$ parametrized by a $\complex$-scheme $S$, there exists a unique
morphism $\alpha:S\to\Mcal$ and an isomorphism
$\bun\cong(\Id\times\alpha^*)(\hat\bun)$ preserving the framing
isomorphisms. In this case, $\Mcal_{\CP_\theta^2}(r,0,r-k)$ is a smooth
quasi-projective $\complex$-scheme for all $r\geq1$ and for all $k\in\nat_0$; when
non-empty it has dimension
$$
\dim_\complex\big(\Mcal_{\CP_\theta^2}(r,0,r-k)\big)=2\, r\,k-r^2+1
$$
as in the classical case, and the tangent space at a point $[E]$ is
the vector space $$T_{[E]}\Mcal_{\CP_\theta^2}(r,0,r-k) = \Ext^1(E,E)
\ . $$

To construct our instanton moduli spaces, we look not for a set of
objects as before but rather for a space which parametrizes those objects. For
this, we consider functors
from the category $\Alg$ of commutative, unital noetherian
$\complex$-algebras to the category $\Set$ of sets. Let $A$ be a unital
noetherian $\complex$-algebra. We write $\CP_{\theta,A}^n$ for the
noncommutative variety dual to the algebra $\alg(\CP_{\theta,A}^n):=
A\otimes\alg(\CP_\theta^n)$. We consider such \emph{families} of
algebras $\alg(\CP_{\theta,A}^n)$ in order to study the \emph{global}
structure of our moduli spaces, and endow them with geometric
structure.
\begin{definition}
\begin{itemize}
\item[(1)]
A \emph{family of
  $(W,V)$-framed sheaves parametrized by the algebra $A$} is an $A$-flat torsion
free module $\bun$ of rank $r$ on $A\otimes\alg$ such that
$H^1\big(\CP_{\theta,A}^2 \,,\, \bun(-1)\big)\cong A\otimes V$ and
$i^\bullet(\bun) \cong W\otimes A\otimes \sheaf_{\CP_{\theta}^1}$. \\
\item[(2)] Two families of $(W,V)$-framed sheaves $\bun$ and $\bun'$ are
  \emph{isomorphic} if they are isomorphic as $(W,V)$-framed modules
  on $A\otimes\alg$.
\end{itemize}
\label{sheaffamilydef}\end{definition}

For $A=\complex$, i.e. for a family parametrized by a one-point space,
this definition reduces to Definition~\ref{modspdef}.
\begin{definition}
\begin{itemize}
\item[(1)]
The \emph{instanton moduli functor} is the covariant functor
$$
\Module_{\CP_\theta^2}^{\rm inst}(W,V) \,:\, \Alg \ \longrightarrow \ 
\Set
$$
that associates to every algebra $A$ the set
$\Module_{\CP_\theta^2}^{\rm inst}(W,V)(A)$ of isomorphism classes of
families $\bun$ of $(W,V)$-framed sheaves parametrized by $A$, and to every algebra morphism
$f:A\to B$ associates the pushforward $\Module_{\CP_\theta^2}^{\rm inst}(W,V)(f)$ sending families
parametrized by $A$ to families parametrized by $B$. When bases have been 
fixed for the vector spaces $V\cong\complex^k$ and $W\cong\complex^r$,
we will denote this moduli functor by~$\Module_{\CP_\theta^2}^{\rm inst}(r,k)$. \\
\item[(2)]
A \emph{fine instanton moduli space} is a variety dual to a universal object
representing the instanton moduli functor, i.e. a pair $(\hat A,\hat\bun)$,
where $\hat A$ is an object of the category $\Alg$ and $\hat\bun\in
\Module_{\CP_\theta^2}^{\rm inst}(W,V)(\hat A)$, such that for any pair
$(A,\bun)$ with $A$ an object of $\Alg$ and
$\bun\in\Module_{\CP_\theta^2}^{\rm inst}(W,V)(A)$ there exists a
unique morphism $\alpha\in \Hom_{\Alg}(\hat A,A)$ such that
$\Module_{\CP_\theta^2}^{\rm inst}(W,V)(\alpha)(\hat\bun)\cong
\bun$.
\end{itemize}
\label{instmodulidef}\end{definition}

By Yoneda's lemma, a fine instanton moduli space given by $(\hat A,\hat\bun)$ corresponds
bijectively to a representation of
the instanton moduli functor via an
isomorphism of functors
$$
\Homfun_{\hat A} := \Hom_\Alg(\hat A,-) \ \xrightarrow{ \ \approx \ } \
\Module_{\CP_\theta^2}^{\rm inst}(W,V) \ ,
$$
with universal sheaf represented by $\hat\bun=\Homfun_{\hat
  A}(\Id_{\hat A})$.
For example, if $c_1=0$ and $(r,k)$ are coprime integers,
then $\Mcal_{\CP_\theta^2}(r,0,r-k)$ is a fine moduli space (when non-empty)
by~\cite[Prop.~7.15]{NS}.

\begin{proposition}
For any $r\geq1$ and $k\geq0$, the quotient $\widehat\Mcal_\theta^{\rm
  ADHM}(r,k)$ is a fine instanton moduli space.
\label{functorrepprop}\end{proposition}
\Proof{
We show that the coordinate algebra $\alg\big(\widehat\Mcal_\theta^{\rm
  ADHM}(r,k)\big)$ is a universal object in $\Alg$ which represents
the instanton moduli functor $\Module_{\CP_\theta^2}^{\rm inst}(r,k)$.
As detailed computations
have been carried out in an analogous context in~\cite{NS}
(see~\cite[Prop.~8.13]{NS}), we only outline the main steps
and skip many details. Given $(B,I,J)\in \Mcal_\theta^{\rm
  ADHM}(r,k)$ and a commutative $\complex$-algebra $A$, we use
Theorem~\ref{BIJmonadthm} to define a monad complex
$\underline{\calg}\,_\bullet(B,I,J)_A$ in $\coh(\CP_{\theta,A}^2)$ as
in (\ref{sigmaBIJ})--(\ref{BIJchain}). By Theorem~\ref{BIJmonadthm}
and~\cite[Thm.~5.8]{NS}, the cohomology sheaf
$H^0\big(\,\underline{\calg}\,_\bullet(B,I,J)_A\big)$ is an $A$-flat
family of torsion free sheaves in $\coh(\CP_{\theta,A}^2)$ with
$i^\bullet\big(H^0(\,\underline{\calg}\,_\bullet(B,I,J)_A)\big) =
H^0\big(i^\bullet(\, \underline{\calg}\,_\bullet(B,I,J)_A)\big)$, and
so $H^0\big(\,\underline{\calg}\,_\bullet(B,I,J)_A\big)$ is
$(W,V)$-framed. This defines a natural transformation of functors
$\Homfun_{\alg(\, \widehat\Mcal_\theta^{\rm
  ADHM}(r,k))} \to \Module_{\CP_\theta^2}^{\rm inst}(r,k)$, which is injective
by (\ref{monadisog}) and~\cite[Lem.~5.11]{NS}. Applying
Theorem~\ref{sheafmonadthm} and Theorem~\ref{BIJmonadthm} to $A$-flat
families of $(W,V)$-framed torsion free objects of
$\coh(\CP_{\theta,A}^2)$ shows that this transformation is surjective. Finally,
to demonstrate universality we use the proof of
Lemma~\ref{stablem} to observe that the quotient map $\Mcal_\theta^{\rm
  ADHM}(r,k) \to \widehat\Mcal_\theta^{\rm
  ADHM}(r,k)$ is a principal $\GL(k)$-bundle in the \'etale topology. The group functor
$\GL(k)$ acts on $\Homfun_{\alg(\Mcal_\theta^{\rm
  ADHM}(r,k))}$, hence using
Lemma~\ref{framedstablelemma} and~\cite[Thm.~4.3]{NS} one argues as
in the proof of~\cite[Prop.~8.13]{NS} that the map of functors $\Homfun_{\alg(\Mcal_\theta^{\rm
  ADHM}(r,k))} \to \Module_{\CP_\theta^2}^{\rm inst}(r,k)$ is also a principal
$\GL(k)$-bundle, and hence that $\alg\big(\widehat\Mcal_\theta^{\rm
  ADHM}(r,k)\big)$ represents $\Module_{\CP_\theta^2}^{\rm inst}(r,k)$.
}

We may summarize the main consequences of the results described thus far in
this section as follows.
\begin{theorem}
Let $(r\geq1,k\geq0)$ be coprime integers. Then the instanton moduli
space $\Mcal_\theta(r,k)$, when non-empty, is a smooth
quasi-projective variety of dimension
$$
\dim_\complex\big(\Mcal_\theta(r,k)\big)=2\, r\,k \ .
$$
The tangent space to $\Mcal_\theta(r,k)$ at a point $[E]$ is canonically
the vector space
$$
T_{[E]}\Mcal_\theta(r,k)=\Ext^1\big(E\,,\, E(-1) \big) \ .
$$
As an element of the Grothendieck group of the abelian category
$\coh\big(\CP_{\theta, \alg(\Mcal_\theta(r,k))}^2\big)$, the universal module $\hat\bun$ on
$\alg\big(\Mcal_\theta(r,k)\big)\otimes \alg(\CP_\theta^2)$ is
isomorphic to the virtual vector bundle whose fibre over $[E]$ is the
virtual bundle in the K-theory of locally free coherent sheaves on
$\Open(\CP_\theta^2)$ given by
$$
\hat\bun_{[E]}=W\otimes\sheaf_{\CP_\theta^2} \ \oplus \ V\otimes \big(
(\alg_\ell^!)_1\otimes\sheaf_{\CP_\theta^2}\ \ominus\
\sheaf_{\CP_\theta^2}(-1)\ \ominus \ \sheaf_{\CP_\theta^2}(1)\big)
$$
with $V=H^1\big(\CP_\theta^2\,,\,E(-1)\big)$ and
$W=H^0\big(\CP_\theta^1\,,\,i^\bullet(E)\big)$.
\label{modulispthm}\end{theorem}
\Proof{
By Lemma~\ref{framedstablelemma} the moduli space $\Mcal_\theta(r,k)$
is an open subscheme of the moduli space
$\Mcal_{\CP_\theta^2}(r,0,r-k)$ which figures in~\cite[Cors.~8.3--8.4
and Lem.~8.5]{NS}, and it represents the corresponding moduli
functor. Via Proposition~\ref{CP1equivprop}, its deformation theory
may thus be embedded into the more general
theory of framed modules developed by Huybrechts and
Lehn~\cite{HLframed}. The Zariski tangent space to $\Mcal_\theta(r,k)$
at a point corresponding to a framed
sheaf $E$ on $\Open(\CP_\theta^2)$ is isomorphic to the
cohomology group $\Ext^1(E,E(-1))$, and there is an appropriate
obstruction theory with values in
$\Ext^2(E,E(-1))$. We will show that
\beq
\Hom\big(E\,,\,E(-1)\big) \ = \ 0 \ = \ \Ext^2\big(E\,,\, E(-1)\big) \
.
\label{0obstruction}\eeq

For this, we use the short exact sequence (\ref{EWcanexseq}) in the
category $\coh(\CP_\theta^2)$ to induce long exact cohomology sequences for $k\in\zed$ which
start at
$$
0\ \longrightarrow\ \Hom\big(E\,,\,E(k-1)\big)\ \longrightarrow \ 
\Hom\big(E\,,\, E(k)\big)\ \longrightarrow\ \Hom\big(E\,,\,
i_\bullet\, i^\bullet(E(k))\big) \ .
$$
Since $i^\bullet(E)=W\otimes\sheaf_{\CP_\theta^1}$, the
$\mu$-semistability of $E$ implies $\Hom\big(E\,,\,
i_\bullet\, i^\bullet(E(k))\big)=0$ and hence
$$
\Hom\big(E\,,\, E(k-1)\big) \ \cong\ \Hom\big(E\,,\, E(k)\big)
$$
for all $k\in\zed$. In particular, $\Hom(E,E(-1))\cong\Hom(E,E(-3))$
and the latter vector space is trivial by~\cite[Lem.~7.14]{NS}. Hence
$\Hom(E,E(-1))=0$. By~\cite[Prop.~2.4~(2)]{NS}, $\Ext^2(E,E(-1))$ is
dual to the vector space $\Hom(E,E_\alpha(-2))$ for some automorphism
$\alpha\in{\rm Aut}(\alg)$, where if $E=\pi(M)$ for a graded
$\alg$-module $M\in\gr(\alg)$,
then $E_\alpha=\pi(M_\alpha)$ with $M_\alpha$
the right $\alg$-module $M$ twisted by $\alpha$, i.e. the same
underlying vector space $M_\alpha=M$ with right $\alg$-module
structure $v\triangleleft_\alpha f:= v\triangleleft \alpha(f)$ for
$f\in\alg$ and $v\in M$. Since $\alpha(f)=0$ if and only if $f=0$,
it is easy to see that the restriction functor $i^\bullet$ commutes
with $\alpha$-twisting,
i.e. $i^\bullet(E_\alpha)=i^\bullet(E)_\alpha$, and hence so does
framing. Thus one can apply the short exact sequence
(\ref{EWcanexseq}) to the twisted sheaf $E_\alpha$, and then repeat
the previous argument verbatim to deduce
$\Hom(E,E_\alpha(-2))\cong\Hom(E,E_\alpha(-3))$, where the latter
vector space is again trivial by~\cite[Lem.~7.14]{NS}. Hence $\Ext^2(E,E(-1))=0$.

Using (\ref{0obstruction}) and~\cite[Cor.~6.2]{NS}, the dimension of
the tangent space is given by
\begin{eqnarray*}
\dim_\complex\Ext^1\big(E\,,\, E(-1)\big) &=& {\rm rank}(E)\,\Big({\rm
  rank}\big(E(-1)\big) -\chi\big(E(-1)\big)\Big)\\ && +\, c_1(E)\, \Big(3\,{\rm
  rank}\big(E(-1)\big)+ c_1\big(E(-1) \big)\Big) \\ && -\,
\chi(E)\, \rank\big(E(-1)\big) \ .
\end{eqnarray*}
Using Corollary~\ref{framedtopinvscor}, together with $c_1(E(-1))=-r$
by (\ref{c1Ek}) and $\chi(E(-1))=h_E(-1)=-k$ by (\ref{hEsCP2}), this
number is equal to $2\, r\, k$. Finally, the expression for the virtual
class of the universal sheaf $\hat\bun$ follows from the proof of
Proposition~\ref{functorrepprop} and~\cite[Prop.~6.8~(1)]{CLSI}.
}

To incorporate the construction of instanton gauge bundles and
connections on the noncommutative sphere $S_\theta^4$ from
\S\ref{Instconn}, we may also wish to extend the definition of the
instanton moduli functor according to the twistor construction of
\S\ref{scmonad}. Rather than going through all these details, which
are not needed in this paper, we simply
axiomatize the extension of the instanton moduli space that one
finds. This extension appropriately reduces the dimension of the
moduli space of Theorem~\ref{modulispthm}.

\begin{definition}
A \emph{family of instantons over $S_\theta^4$ parametrized by a unital
  $*$-algebra $A$} is a quintuple
$(\bun,\tilde\bun,\nabla,\widetilde\nabla, G_\bun)$ consisting of:
\begin{itemize}
\item[(1)] Finitely generated
  projective right and left modules $\bun$ and $\tilde\bun$ over the
  respective algebras
  $A\otimes\alg(\real_\theta^4)$ and
  $A\otimes\alg(\widetilde\real_\theta^4)$ with dualizing anti-involutions
  which are compatible with the $*$-algebra structures; \\
\item[(2)] Connections
\begin{eqnarray*}
\nabla\,:\, \bun & \longrightarrow &
\bun\otimes_{A\otimes\alg(\real_\theta^4)}
\big(A\otimes\Omega^1_{\real_\theta^4}\big) \cong \bun\otimes_{\alg(\real_\theta^4)}
\Omega^1_{\real_\theta^4} \ , \\[4pt]
\widetilde\nabla\,:\, \tilde\bun & \longrightarrow &
\big(A\otimes\Omega^1_{\widetilde\real_\theta^4}\big)\otimes_{A\otimes\alg(\widetilde\real_\theta^4)}
\tilde\bun \cong \Omega^1_{\widetilde\real_\theta^4}\otimes_{\alg(\widetilde\real_\theta^4)}
\tilde\bun
\end{eqnarray*}
which are compatible with the dualizing anti-involutions and whose
curvatures $F_\nabla=\nabla^2$ and $F_{\widetilde\nabla}=\widetilde\nabla{}^2$
obey the anti-selfduality equations
$$
(\Id\otimes *_\theta) F_\nabla=-F_\nabla \ , \qquad (\Id\otimes \tilde*_\theta)
F_{\widetilde\nabla}= -F_{\widetilde\nabla}
$$
in $$\Hom_{A\otimes\alg(\real_\theta^4)}\big(\bun\,,\,\bun\otimes_{A\otimes\alg(\real_\theta^4)}
(A\otimes\Omega^2_{\real_\theta^4})\big)\cong
\Hom_{A\otimes\alg(\real_\theta^4)}\big(\bun\,,\,\bun \otimes_{\alg(\real_\theta^4)}
\Omega^2_{\real_\theta^4}\big)$$ and in
$$\Hom_{A\otimes\alg(\widetilde\real_\theta^4)}\big(\tilde\bun\,,\,
(A\otimes\Omega^2_{\widetilde\real_\theta^4}) \otimes_{A\otimes\alg(\widetilde\real_\theta^4)} 
\tilde\bun\big)\cong
\Hom_{A\otimes\alg(\widetilde\real_\theta^4)}\big(\tilde\bun\,,\, \Omega^2_{\widetilde\real_\theta^4}
\otimes_{\alg(\widetilde\real_\theta^4)}
\tilde\bun\big) \ , $$ respectively; and \\
\item[(3)] An isomorphism
$$
G_\bun\,:\, \begin{matrix} \bun\otimes_{A\otimes\alg(\real_\theta^4)}
\big(A\otimes\alg(\real_\theta^4)[\tilde\rho\,]\big) \\ \parallel \\
\bun[\tilde\rho\,] \end{matrix} \ \longrightarrow \ \begin{matrix}
\big(A\otimes[\rho]\alg(\widetilde\real_\theta^4)\big) \otimes_{A\otimes\alg(\widetilde\real_\theta^4)}
\tilde\bun \\ \parallel \\ [\rho]\tilde\bun \end{matrix}
$$
which is compatible with the dualizing anti-involutions, and obeys
$G_\bun(\sigma\triangleleft a) = (\Id\otimes G(a))\triangleright
G_\bun(\sigma)$ for $\sigma\in\bun[\tilde\rho\,]$,
$a\in\alg(\real_\theta^4)[\tilde\rho\,]$ and ${}_\rho\widetilde\nabla\circ
G_\bun=G_\bun\circ\nabla_{\tilde\rho}$.
\end{itemize}
\label{NCinstfamilydef}\end{definition}

\begin{definition}
Two families of instantons $(\bun,\tilde\bun,\nabla,\widetilde\nabla,
G_\bun)$ and $(\bun',\tilde\bun',\nabla',\widetilde\nabla', G_\bun')$ are
\emph{equivalent} if there exist isomorphisms of projective modules
$\varrho:\bun \xrightarrow{\ \approx \ } \bun'$ and $\tilde\varrho:\tilde\bun
  \xrightarrow{\ \approx \ } \tilde\bun'$ together with commutative
    diagrams
$$
\xymatrix{
\bun ~
\ar[r]^{\!\!\!\!\!\!\!\!\!\!\!\!\!\!\! \nabla}
\ar[d]_{\varrho} & ~ 
\bun\otimes_{\alg(\real_\theta^4)}\Omega^1_{\real_\theta^4}
\ar[d]^{\varrho\otimes \Id} \\
\bun' ~
\ar[r]_{\!\!\!\!\!\!\!\!\!\!\!\!\!\!\! \nabla'} & ~ \bun'\otimes_{\alg(\real_\theta^4)}
\Omega^1_{\real_\theta^4} 
} \ , \qquad
\xymatrix{
\tilde\bun ~
\ar[r]^{\!\!\!\!\!\!\!\!\!\!\!\!\!\!\! \widetilde\nabla}
\ar[d]_{\tilde\varrho} & ~ 
\Omega^1_{\widetilde\real_\theta^4}\otimes_{\alg(\widetilde\real_\theta^4)} \tilde\bun
\ar[d]^{\Id\otimes \tilde\varrho} \\
\tilde\bun' ~
\ar[r]_{\!\!\!\!\!\!\!\!\!\!\!\!\!\!\! \widetilde\nabla'} & ~
\Omega^1_{\widetilde\real_\theta^4} \otimes_{\alg(\widetilde\real_\theta^4)}
\tilde\bun'
} \ ,
$$
and
$$
\xymatrix{
\bun[\tilde\rho\,] ~
\ar[r]^{\!\!\!\!\! G_\bun}
\ar[d]_{\varrho\otimes \Id} & ~ 
[\rho]\tilde\bun\ar[d]^{\Id \otimes \tilde\varrho} \\
\bun'[\tilde\rho\,] ~
\ar[r]_{\!\!\!\!\! G_\bun'} & ~ [\rho]\tilde\bun'
} \ .
$$
\label{equivfamdef}\end{definition}

The corresponding moduli functor $\Alg\to\Set$ assigns to each unital
$*$-algebra $A$ the set of equivalence classes of families of
instantons parametrized by $A$. We may also restrict the target of this functor to
families of instantons with gauge bundles of rank $r\geq1$ and Euler
characteristic $\chi= r-k$. The construction of \S\ref{Instconn} yields a
universal object representing this functor. In \S\ref{sec:gaugetheory}
we shall also
restrict the sources of these functors to the subcategory
${}^{\hil_\theta}\Alg$ consisting of left $\hil_\theta$-comodule
algebras. Below we consider in detail some explicit instances of these
moduli space constructions.

\subsection{Rank~$0$ instantons\label{Rank0}}

The case $r=0$ ($W=0$) is somewhat degenerate; it is not covered by
the general analysis of
\S\ref{Instmodulifunctor} and must be dealt with separately. Sheaves $E$
of rank~$0$ are given by $E=\pi(M)$ for some \emph{torsion} module
$M\in\gr(\alg)$, i.e. every element of $M$ is annihilated by a
non-zero element of the algebra $\alg=\alg(\CP_\theta^2)$. The
deformation theory of~\cite[\S7--\S8]{NS} does not apply to the moduli
space of such sheaves, which is only
set-theoretic, i.e. it does not corepresent the instanton moduli
functor of \S\ref{Instmodulifunctor}. Nevertheless, we will now demonstrate
that the moduli space of instantons of rank~$0$ is still a coarse
moduli space for some functor, which identifies it as the moduli space
of finite-dimensional representations of an algebra dual
to an affine noncommutative toric variety.

In terms of noncommutative ADHM data, in this case one has $I=J=0$ and
the braided ADHM equation (\ref{NCADHMeqs}) reduces to $B\wedge_\theta B=0$,
or equivalently
\beq
B_1\, B_2=q^{-2}\ B_2\, B_1 \ .
\label{B1B2qcomm}\eeq
Thus the datum $B\in\Mcal_\theta^{\rm ADHM}(0,k)$ defines a $k$-dimensional representation of the
affine coordinate algebra $\alg(\complex_\theta^2)$ of the complex
algebraic Moyal plane $\complex_\theta^2$~\cite[\S3.2]{CLSI}, i.e. the
polynomial algebra $\complex[z_1,z_2]$ in two generators modulo the
relation $z_1\,z_2=q^2\, z_2\, z_1$. By the stability condition of Definition~\ref{NCADHMvar}, this
representation is irreducible. Thus by~\cite[\S6.2]{HL}, $\Mcal_\theta^{\rm ADHM}(0,k)$ is
the affine algebraic $\complex$-scheme representing the moduli functor
$\Alg\to\Set$ which sends a $\complex$-algebra $A$ to the set of
simple $A\otimes \alg(\complex_\theta^2)$-module structures on
$A^{\oplus k}$; for $A=\complex$ this set consists of irreducible
representations of $\alg(\complex_\theta^2)$ on
$\complex^k$. The natural $\GL(k)$-action corresponds to changes of
basis. By~\cite[Prop.~6.3]{HL}, the quotient
$\widehat\Mcal_\theta^{\rm ADHM}(0,k)$ corepresents the moduli functor
which sends a $\complex$-scheme $S$ to the set of isomorphism classes
of $S$-families of simple $k$-dimensional
$\alg(\complex_\theta^2)$-modules, i.e. locally free coherent
$\sheaf_S$-modules $\Fcal_S$ of rank $k$, together with $\complex$-algebra
homomorphisms $\varrho_S:\alg(\complex_\theta^2)\to
\End_{\coh(S)}(\Fcal_S)$, such that $\Fcal_S$ contains no proper subsheaves
invariant under $\varrho_S(z_i)$ for $i=1,2$. 

The classical case
$\theta=0$ is covered by~\cite[Prop.~6.4]{HL}. In this instance the scheme $\widehat\Mcal_{\theta=0}^{\rm
  ADHM}(0,k)$ is isomorphic to the affine quotient $\big(\widehat\Mcal_{\theta=0}^{\rm
  ADHM}(0,1)\big)^k\,\big/\, S_k$ since all simple modules over a commutative
algebra are one-dimensional, and $\widehat\Mcal_{\theta=0}^{\rm
  ADHM}(0,1)=\complex^2$. It follows that
the moduli space $\widehat\Mcal_{\theta=0}^{\rm
  ADHM}(0,k)$ is canonically isomorphic to the $k$-th symmetric
product $\Sym^k(\CP^2):=(\complex^2)^k/S_k$ of the commutative affine
plane; it parametrizes coherent sheaves on $\Open(\CP^2)$ which
have zero-dimensional
support of length $k$ contained in $\complex^2\cong \CP^2\setminus \CP^1$. Hence we write $\Sym_\theta^k(\CP_\theta^2):= \widehat\Mcal_\theta^{\rm
  ADHM}(0,k)$ for all $\theta\in\complex$. This construction
generalizes~\cite[Prop.~2.10]{Nakajima}. For generic deformation parameters
$\theta\in\complex$ this moduli space is
relatively small; when $q\in\complex$ is not a root of unity of order
$2k$, the space $\Sym_\theta^k(\CP_\theta^2)$ only parametrizes
representations of $\alg(\complex_\theta^2)$ wherein one of the
matrices $B_1$ or $B_2$ is singular.

\subsection{Rank~$1$ instantons\label{Rank1}}

A torsion free sheaf $E\in\coh(\CP_\theta^2)$ has rank~$1$ if and only
if $M=\Gamma(E)\in\gr(\alg)$ is isomorphic to a shift $\Ical(m)$ of a
right ideal $\Ical\subset \alg$~\cite[\S4.3]{CLSI}. By~\cite[Cor.~6.6~(1)]{NS}, there exists a smooth, projective fine moduli space
$\Mcal_{\CP_\theta^2}(1,0,k)$ of dimension $2k$ for torsion free $\alg$-modules in
$\coh(\CP_\theta^2)$ with rank $r=1$, first Chern class $c_1=0$, and
Euler characteristic $\chi=1-k$. This moduli space
also behaves well in the family $\alg=\alg(\CP^2_\theta)$, in the
sense described in \S\ref{Instmodulifunctor}. In particular,
$\Mcal_{\CP_\theta^2}(1,0,k)$ is irreducible, hence connected, and is
a \emph{commutative} deformation of the Hilbert scheme of points
$\Hilb^k(\CP^2)$ parametrizing zero-dimensional subschemes of degree
$k$ in $\CP^2$. We thus write
$\Hilb_\theta^k(\CP_\theta^2):=\Mcal_{\CP_\theta^2}(1,0,k)$ for all $\theta\in\complex$; it is non-empty
for all $k\geq0$. As such sheaves are
automatically (semi)stable, we may identify
$\Hilb_\theta^k(\CP_\theta^2)$ with the instanton moduli space
$\Mcal_\theta(1,k)$ in this case.

De~Naeghel and Van~den~Bergh~\cite{DNVdB} describe the deformation
$\Hilb_\theta^k(\CP_\theta^2)$ as the scheme parametrizing torsion free
graded $\alg$-modules $\Ical=\bigoplus_{n\geq0}\, \Ical_n$ of projective
dimension one such that
\beq
\dim_\complex(\alg_n)-\dim_\complex(\Ical_n)=k \qquad \mbox{for} \quad
n\gg0 \ .
\label{DNVdBHilb}\eeq
In particular, it follows from~\cite[Lem.~3.3.1]{DNVdB} that $\Ical$ has
rank $r=1$ as an $\alg$-module and invariants $c_1(\Ical)=0$,
$\chi(\Ical)=1-k$. Thus $\Ical$ corresponds to an \emph{ideal sheaf} in the
sense of~\cite[\S4.3]{CLSI}, and hence corresponds to a closed
subscheme of $\CP_\theta^2$ by~\cite[Thm.~4.10]{CLSI}. A
stratification of $\Hilb_\theta^k(\CP_\theta^2)$ by Hilbert series is described
in~\cite[Thm.~6.1]{DNVdB}. There is a bijective correspondence between the set $\Dcal_k$ of integer partitions of $k$ with distinct parts and Hilbert series of objects in $\Hilb_\theta^k(\CP_\theta^2)$. Let $\Dcal=\bigcup_{k\geq 0}\, \Dcal_k$ be the set of all integer partitions $\lambda=(\lambda_i)_{i\geq 1}$ with distinct parts. It is a classical result~\cite{andrews} that the generating function for elements of $\Dcal$ is given by
\beq
Z_\Dcal(Q)=\sum_{\lambda\in\Dcal}\, Q^{|\lambda|}=
\prod_{n=1}^\infty\, \big(1+ Q^n\big) \ ,
\label{Dcalgenfn}\eeq
where $Q$ is a formal variable and $|\lambda|=\sum_i\, \lambda_i$ is the weight of the partition
$\lambda$. This formula will be used in conjunction with instanton
partition functions in \S\ref{sec:gaugetheory}.

In this case there is a bijection between
the set of ideals of codimension $k$ in the coordinate algebra $\alg(\complex_\theta^2)$
of the complex algebraic Moyal plane and the set of triples
$(B_1,B_2,I)\in \End_\complex(V)^{\oplus
  2}\oplus\Hom_\complex(\complex,V)$ satisfying (\ref{B1B2qcomm}) such
that no proper $B_i$-invariant subspaces of $V$ contain the image of
$I$ for $i=1,2$; the proof is essentially a step by step repetition of that in the
classical case $\theta=0$~\cite[Thm.~1.9]{Nakajima}. In the classical
situation one shows that $J=0$ in the rank~$1$
case~\cite[Prop.~2.9]{Nakajima} and thus directly establishes an
isomorphism between the Hilbert scheme of $k$ points in $\complex^2$ and the ADHM moduli space
$\widehat\Mcal_{\theta=0}^{\rm ADHM}(1,k)$. However, we will see
directly below that the linear map $J\neq0$ in general when
$\theta\neq0$, in agreement with what we found in \S\ref{Rank0}. This
reflects the fact that the noncommutative algebra $\alg$
contains very few ideals, or equivalently that $\CP_\theta^2$ for
generic $\theta$ has very few zero-dimensional noncommutative
subschemes~\cite[\S4.3]{CLSI}. We can regard the scheme $\Sym_\theta^k(\CP_\theta^2)$
of \S\ref{Rank0} and $\Hilb_\theta^k(\CP_\theta^2)$ 
simultaneously as a commutative deformation of the resolution of the
singularity $\Sym^k(\CP^2)$ provided by the Hilbert--Chow morphism
$\Hilb^k(\CP^2)\to \Sym^k(\CP^2)$, which sends an ideal to its support. Noncommutative deformations of
this kind are constructed in~\cite{GS,KR} using the covering by
cotangent bundles $T^*U$ provided by the symplectic resolution.

\subsection{Charge~$1$ instantons}

Consider now the case $k=\dim_\complex(V)=1$. The projective plane is
rigid against commutative deformations. 
Hence there are no commutative deformations of the Hilbert scheme
$\Hilb^1(\CP^2)=\CP^2$, and so
$$
\Hilb_\theta^1\big(\CP_\theta^2\big)\ \cong \ \CP^2
$$
for all $\theta\in\complex$. To see this directly, we note that in
this case the morphisms
of the ADHM data $(B_1,B_2,I,J)$ act via multiplication by scalars
$( b_1, b_2,i,j)\in\complex^4$. The stability condition of
Definition~\ref{NCADHMvar} implies
that $i\neq0$, while invariance under the action (\ref{GLVaction}) of
$\GL(1)=\complex^\times$ means that we can rescale so that $i=1$. The
braided ADHM equation (\ref{NCADHMeqs}) can then be used to solve for
$j\in\complex$ as
$$
j=\big(q^{-2}-1\big)\  b_1\,  b_2 \ ,
$$
which for $\theta\neq0$ is non-zero in general. Thus the moduli space
$\widehat\Mcal_{\theta}^{\rm ADHM}(1,1)$ is coordinatized by the quadruples
$( b_1, b_2,1,(q^{-2}-1)\,  b_1\,  b_2)\in\complex^4$,
representing an affine patch $( b_1, b_2)\in\complex^2$ of the
projective plane $\CP^2$. Similarly, one has (set-theoretically)
$$
\Sym_\theta^1\big(\CP_\theta^2\big)\ \cong \ \complex^2
$$
for all $\theta\in\complex$ as the one-dimensional representations
over $\complex$ of the
algebra relation (\ref{B1B2qcomm}) necessarily have either $B_1=0$ or
$B_2=0$ when $\theta\neq0$.

In the higher rank cases $r\geq2$, we can regard the morphisms $I$ and
$J$ as vectors 
$$ I=(i_1,\dots,i_r)\ \in \ W^* \ , \qquad J=\begin{pmatrix} \, j_1 \,
  \\ \vdots \\
  \, j_r \, \end{pmatrix}\ \in \ W
$$ in a chosen orthonormal basis of
$W\cong\complex^r$. The braided ADHM equation (\ref{NCADHMeqs}) now
defines a quadric in $\complex^{2r+2}$ given by
\beq
\big(1-q^{-2}\big)\  b_1\,  b_2+\sum_{l=1}^r\, i_l\, j_l =0 \ .
\label{ADHMquadric}\eeq
Stability is now equivalent to $i_l\neq0$ for all $l=1,\dots,r$, showing that the moduli space $\widehat\Mcal_{\theta}^{\rm ADHM}(r,1)$
is quasi-projective. An
element $t\in\GL(1)=\complex^\times$ acts trivially on
$ b_1, b_2$, and as multiplication by $t$ on $i_l$ and by
$t^{-1}$ on $j_l$ for each $l=1,\dots,r$. We can use this scaling
symmetry to set $i_r=1$, and then use (\ref{ADHMquadric}) to eliminate
$j_r\in\complex$. This coordinatizes 
$\widehat\Mcal_{\theta}^{\rm ADHM}(r,1)$ as a patch
$\complex^2\times \complex^{r-1}\times(\complex^\times)^{r-1}$. Again this construction
is identical to that of the classical case $\theta=0$, giving the
charge~$1$ instanton moduli space
$$
\Mcal_{\theta}(r,1) \ \cong\ \CP^2\times T^*\CP^{r-1}
$$
for all $\theta\in\complex$ and $r\geq1$. For $k\geq2$, the moduli spaces of noncommutative instantons are
generically different from their classical counterparts at $\theta=0$.

\subsection{Instanton deformation complex}

We will now give an alternative proof of Theorem~\ref{modulispthm}
which provides a much more powerful description of the geometry of the
instanton moduli spaces. For this, we will view the noncommutative ADHM equation (\ref{NCADHMeqs}) on the
affine space of triples (\ref{BIJsubvar}) as the
zero locus of the map
$\mu_c:\Xcal(W,V)\to\gl(V)^*\otimes(\alg_\ell^!)_2$ defined by
\beq
\mu_c(B,I,J):= B\wedge_\theta B+I\circ J \ .
\label{momentmap}\eeq
The restriction of this map to stable elements of $\Xcal(W,V)$ (in the
sense of Definition~\ref{NCADHMvar}) is denoted $\tilde\mu_c$. Then the moduli space (\ref{MhatADHM}) can be represented as
the reduction $$
\widehat\Mcal_\theta^{\rm ADHM}(W,V)=\tilde\mu_c^{-1}(0)\,
\big/\,\GL(V) \ . $$ Fixing bases of the complex vector spaces $V$ and
$W$ naturally induces a basis for each fibre of the tangent bundle $T\Xcal(W,V)$,
with dual basis denoted $(\dd B,\dd I,\dd J)$ at a point
$x=(B,I,J)\in\Xcal(W,V)$. The differential
$$
\dd\mu_c=\dd B\wedge_\theta B+B\wedge_\theta\dd B+\dd I\circ J+I\circ\dd
J
$$
is the linearization of the braided ADHM quiver relations
(\ref{ADHMquiverrels}) with
\beq
\dd\mu_c(b,i,j) =b\wedge_\theta B+B\wedge_\theta b+ i\circ J+I\circ j
\ .
\label{dmuc}\eeq

Let $\varphi:\GL(V)\to \Xcal(W,V)$ be the orbit map $g\mapsto
g\triangleright(B,I,J)$ defined by (\ref{GLVaction}). Its restriction
to stable elements of $\Xcal(W,V)$ is denoted $\tilde\varphi$. The differential
$$
\dd\varphi(\xi)=\big([B,\xi]\,,\, (\xi\otimes\Id_{\alg_\ell^!})\,I\,,\, -J\,\xi\big)
$$
is the linearization of the action of the gauge group $\GL(V)$ on
$\Xcal(W,V)$. It is easy to compute that
$\dd\mu_c\circ \dd\varphi=0$ for all $\theta\in\complex$.

\begin{theorem}
The tangent space $T_{[E]}\Mcal_\theta(r,k)$ to the instanton moduli
space at a closed point $[E]= [(B,I,J)]$ is isomorphic to the
cohomology group $H^1(\,\underline{\Iscr}\,_\bullet(E))$ of the complex
$$
\underline{\Iscr}\,_\bullet(E) \,:\, 0\ \longrightarrow\ 
\End_\complex(V)~
\xrightarrow{\dd\tilde\varphi}~\begin{matrix}
  \big(\alg_\ell^!\big)_1\otimes\End_\complex(V)
  \\[4pt] \oplus \\[4pt] \big(\alg_\ell^!\big)_2\otimes\Hom_\complex(W,V)\\[4pt] \oplus \\[4pt] \Hom_\complex(V,W) \end{matrix}
~ \xrightarrow{\dd\tilde\mu_c}~\big(\alg_\ell^!\big)_2\otimes
\End_\complex(V) \ \longrightarrow\ 0 \, .
$$
\label{defcomplexthm}\end{theorem}
\Proof{
We know that $[E]\in \Mcal_\theta(r,k)$ is quasi-isomorphic in the
category $\coh(\CP_\theta^2)$ to the monad complex
$\underline{\calg}\,_\bullet(B,I,J)$ defined in (\ref{BIJchain}). The tangent-obstruction complex of the
moduli space $\Mcal_\theta(r,k)$ can then be determined by a standard
calculation in deformation theory, using the cohomology calculation of~\cite[Prop.~6.8]{CLSI}. In particular, the $p$-th
hypercohomology group $\mathbb{H}^p$ of the complex $\underline{\Dcal}\,_\bullet(E):= \hil
om(\,\underline{\calg}\,_\bullet(E)\,,\,
\underline{\calg}\,_\bullet(E(-1)))^\bullet$ computes
$\Ext^p(E,E(-1))$, and the complex $\underline{\Iscr}\,_\bullet(E)$ is
the standard hypercohomology spectral sequence for $\underline{\Dcal}\,_\bullet(E)$. The infinitesimal deformation space is
$H^1(\,\underline{\Iscr}\,_\bullet(E))= \hyper^1(\underline{\Dcal}\,_\bullet(E))$ and the obstruction space is
$H^2(\,\underline{\Iscr}\,_\bullet(E))=
\hyper^2(\underline{\Dcal}\,_\bullet(E))$. We will show that
$$
H^0\big(\,\underline{\Iscr}\,_\bullet(E)\big) \ = \ 0 \ = \
H^2\big(\,\underline{\Iscr}\,_\bullet(E)\big) \ .
$$

For this, we consider the dual of the differential
$\dd\tilde\mu_c:\underline{\Iscr}\,_1(E) \to \underline{\Iscr}\,_2(E)$
given by
$$
\dd\tilde\mu_c^* \,:\, 
\End_\complex(V) \ \longrightarrow \ \begin{matrix}
  \End_\complex(V)\otimes \big(\alg_\ell^!\big)_1
  \\[4pt] \oplus \\[4pt]
  \Hom_\complex(W,V)\\[4pt] \oplus
  \\[4pt] \Hom_\complex(V,W) \end{matrix}
$$
with 
$$
\dd\tilde\mu_c^*(\psi)=\big([\psi,B_1]_\theta\otimes\check
w_1+[B_2,\psi]_\theta\otimes\check w_2 \,,\, \psi\, I\,,\, J\,
\psi\big) \ ,
$$
where we use the identification
$(\alg_\ell^!)_2\cong\complex$. Suppose that
$\dd\tilde\mu_c^*(\psi)=0$, and consider the subspace
$V'=\ker(\psi)\subset V$. Then it is easy to see that
$V'$ is $B$-invariant and contains the image of $I$, whence by the
stability condition either $V'=0$ or $V'=V$. But if $\psi\in\End_\complex(V)$ is
injective then $I=0$, contradicting stability, and so $V'=V$. It
follows that the differential $\dd\tilde\mu_c$ is surjective, and thus
$H^2(\,\underline{\Iscr}\,_\bullet(E)) =0$. The proof that the
differential $\dd\tilde\varphi:\underline{\Iscr}\,_0(E) \to
\underline{\Iscr}\,_1 (E)$ is injective, and hence that
$H^0(\,\underline{\Iscr}\,_\bullet(E)) =0$, is carried out in exactly
the same way.
}

\subsection{Braided symplectic reduction}

By analogy with the classical case, it is natural to hope that the noncommutative deformation $\CP^2\to \CP_\theta^2$ induces a (commutative) deformation $\Hilb^k(\CP^2)\to
\Hilb_\theta^k(\CP_\theta^2)$ (as constructed in~\S\ref{Rank1}) that also
carries a Poisson structure and a (holomorphic) symplectic
structure. In particular, the Hilbert scheme $\Hilb^k(\complex^2)$ has
an algebraic symplectic structure induced by the hyper-K\"ahler
metric. On the other hand, in~\cite[\S9]{KKO} it is pointed out that the braided
ADHM equations (\ref{NCADHMeqs}) and (\ref{realADHMeq}) are not
hyper-K\"ahler moment map equations, whence one cannot construct
hyper-K\"ahler or even symplectic structures on the quotient space
using standard hyper-K\"ahler or symplectic quotient techniques. We
will now briefly describe the sense in which our instanton moduli spaces may
be regarded as symplectic quotients.

For this, define a braided symplectic form on
$\Xcal(W,V)$ by
$$
\omega\big((B,I,J)\,,\, (B',I',J'\,)\big):=\Tr(B\wedge_\theta
B'+I\circ J'- I'\circ J) \ ,
$$
where we use the identification $(\alg_\ell^!)_2\cong\complex$, and
$\Tr$ is the usual trace on $\End_\complex(V)$ which agrees with the quantum
trace $\Tr_q$ of~\cite[Prop.~9.3.5]{Majid1} for our class of
deformations. We use the same notation for the form induced on the
tangent bundle $T\Xcal(W,V)$. We will study conditions under which the
map $\mu_c$ defined by (\ref{momentmap}) is a braided (complex) moment
map; this will restrict the allowed hamiltonian vector fields.

Firstly, it is easy to check that $\mu_c$ is $\GL(V)$-equivariant,
i.e. $\mu_c(g\triangleright x)={\rm Ad}^*_{g^{-1}}\mu_c(x)$. Secondly, we
require compatibility with the braided symplectic form $\omega$, i.e. for
any $v\in T\Xcal(W,V)$ and $\xi\in\gl(V)$, we want
\beq
\big\langle \dd\mu_c(v)\,,\,\xi\big\rangle = \omega(\hat\xi,v)
\label{muomegacomp}\eeq
with $\hat\xi$ the vector field in $T\Xcal(W,V)$ associated to the linearized action
of $\xi\in\gl(V)$ on $\Xcal(W,V)$. We embed the two-torus
$T=(\complex^\times)^2$ in $\GL(V)$; then the compatibility condition
(\ref{muomegacomp}) can be formulated as a compatibility requirement
between the ``geometric'' $T$-action and the ``internal'' action of
the gauge group
$\GL(V)$ on $\Hom_\complex(V,V\otimes(\alg_\ell^!)_1)$.
\begin{proposition}
The map $\mu_c$ is compatible with the braided symplectic form $\omega$
for coequivariant morphisms $\xi\in\End_\complex(V)$ with left
$\hil_\theta$-coaction of the form
$$
\Delta_L(\xi)=\xi^{(-1)}\otimes \xi \ ,
$$
where $\xi^{(-1)}\in\hil_\theta$ obeys
$$
F_\theta\big(t_1\,,\,\xi^{(-1)}\big)= q^{-1} =
F_\theta\big(\xi^{(-1)}\,,\, t_2\big) \ .
$$
\label{momentmapprop}\end{proposition}
\Proof{
We compute both sides of (\ref{muomegacomp}) for $v=(v_1\otimes\check
w_1+v_2\otimes\check w_2,v_I,v_J)$ and $\hat\xi=(\hat\xi_1\otimes\check
w_1+\hat\xi_2\otimes\check w_2,\hat\xi_I,\hat\xi_J)$. Using (\ref{dmuc})
we find that the left-hand side of (\ref{muomegacomp}) is given by
$$
\big\langle \dd\mu_c(v)\,,\,\xi\big\rangle =
\Tr\big([v_1,B_2]_\theta\,+[B_1,v_2]_\theta+ v_I\, J+I\,
v_J\big) \, \xi \ .
$$
This is equal to the right-hand side of (\ref{muomegacomp}) provided
that the image of $\xi\in\gl(V)$ under the tangent of the orbit map of
$x=(B,I,J)$ is given by
\beq
&& \hat\xi_1=[\xi,B_1]_\theta \ , \qquad \hat\xi_2=[\xi,B_1]_{-\theta} \
, \qquad \hat\xi_I=(\xi\otimes\Id_{\alg_\ell^!})\, I \ , \qquad
\hat\xi_J =-J\, \xi \ .
\label{hatxiq}\eeq
The linearization of the $\GL(V)$-action (\ref{GLVaction}), with
$g=\Id_V+\xi$, on $B$ is given by
$$
\big((\Id_V+\xi)\otimes\Id_{\alg_\ell^!}\big) \, \big(B_1\otimes\check
w_1+ B_2\otimes \check w_2\big)\,
\big((\Id_V-\xi)\otimes\Id_{\alg_\ell^!}\big) \ ,
$$
which we compose using the braided tensor product
$$
(B_i\otimes\check w_i)\,(\xi\otimes\Id_{\alg_\ell^!}) = B_i\,
F_\theta^{-2}\big(\check w_i^{(-1)}\,,\, \xi^{(-1)}\big)\,
\xi^{(0)}\otimes \check w_i^{(0)}
$$
with the usual Sweedler notation
$\Delta_L(\xi)=\xi^{(-1)}\otimes\xi^{(0)}$, and so on. Using the
coaction (\ref{DeltaLdual}) the conjugate action on $B$ becomes
$$
\big(\xi\, B_1-B_1\, F_\theta^2(t_1,\xi^{(-1)}) \,\xi^{(0)}\big)
\otimes\check w_1 + \big(\xi\, B_2-B_2\, F_\theta^{-2}(\xi^{(-1)},t_2)\,\xi^{(0)}\big)
\otimes\check w_2 \ .
$$
The requisite $\hil_\theta$-coequivariance conditions now follow by
comparing with (\ref{hatxiq}).
}

One can take, for example, $\xi^{(-1)}=t_1^{-1}+t_2^{-1}$ in
Proposition~\ref{momentmapprop}; this condition is of course an
identity for $\theta=0$, in which case there is no restriction on the
allowed hamiltonian vector fields. In a similar fashion, one can treat a
real moment map $\mu_r:\Xcal(W,V)\to\gl(V)^*\otimes(\alg_\ell^!)_2$
whose zero locus is the noncommutative real ADHM
equation~(\ref{realADHMeq}).

\newsection{Gauge theory partition functions\label{sec:gaugetheory}}

\subsection{Torus actions}

In instanton computations one is
interested in equivariant characteristic classes with respect to the
natural action of a torus $\widetilde T$ on the instanton moduli space~\cite{nekrasov}. A combinatorial
formula for the instanton counting functions can be computed
explicitly by classifying the $\widetilde T$-fixed points in the instanton
moduli space $\Mcal_\theta(r,k)$, and computing the Euler
characteristic classes of the equivariant normal bundles to the
fixed loci {of the torus action}. We first describe this torus action explicitly.

For $r\geq1$, let $\widetilde T=T \times(\complex^\times)^{r-1}$ with
$T= (\complex^\times)^2$ the ``geometrical'' torus used for the
deformation of $\CP^2$ and the instanton moduli below. The canonical generators of the coordinate algebra of $\widetilde T$ are denoted $$z= (t_1,t_2,\rho_1,\dots,\rho_{r}) \ , $$
where we identify $(\complex^\times)^{r-1}$ with the maximal torus of ${\rm SL}(r)$ given as the hypersurface
$$
\prod_{l=1}^r\, \rho_l=1
$$
in $(\complex^\times)^r$. Using this presentation of the torus $(\complex^\times)^{r-1}$, we denote its characters by $m=\sum_l\, m_l\, g^*_l= (m_1,\dots,m_r)\in\zed^r$. For any $[E]\in\Mcal_\theta(r,k)$, there is a natural coaction of the Hopf algebra $\hil^{(r)}:=\complex(\rho_1,\dots,\rho_r)$ on the framing module $i^\bullet(E)\cong W\otimes\sheaf_{\CP_\theta^1}$ obtained by fixing a basis $w_1,\dots,w_r$ for the complex vector space $W$ and defining
$$
\Delta_L^{(r)}\, :\, i^\bullet(E)\ \longrightarrow \ \hil^{(r)}\otimes i^\bullet(E)
$$
on $f=\sum_l\, w_l\otimes f_l\in i^\bullet(E)$ by
$$
\Delta_L^{(r)}(f)= \sum_{l=1}^r\, \rho_l\otimes w_l\otimes f_l \ .
$$

The coaction of the Hopf algebra $\hil_\theta$ on the moduli space $\widehat\Mcal_\theta^{\rm ADHM}(W,V)$ is given by (\ref{DeltaLBIJ}). To describe the coaction of $\hil^{(r)}$, consider the corresponding dual basis $w_1^*,\dots,w_r^*$ for $W^*$. Similarly, introduce a basis $v_1,\dots,v_k$ for the complex vector space $V$ with corresponding dual basis $v_1^*,\dots,v_k^*$ for $V^*$. With respect to these bases, we decompose the linear maps $I$ and $J$ as $I=\sum_{a,l}\,v_a \otimes I_{a,l}\otimes w_l^*$ with $I_{a,l}\in\big(\alg_\ell^!\big)_2$ and $J=\sum_{l,a}\, J_{l,a}\, w_l\otimes v_a^*$ with $J_{l,a}\in\complex$. Then the left $\hil^{(r)}$-coaction on the noncommutative ADHM data $(B,I,J)$ is given by
$$
\Delta_L^{(r)}(B,I,J) = \Big(1\otimes B\,,\, \mbox{$ \sum\limits_{a=1}^k\ \sum\limits_{l=1}^r\, \rho_l^{-1}\otimes v_a\otimes I_{a,l} \otimes w_l^*\,,\, \sum\limits_{l=1}^r\ \sum\limits_{a=1}^k\, J_{l,a}\, \rho_l\otimes w_l\otimes v_a^*$}\Big) \ .
$$
By construction, the isomorphism $\Mcal_\theta(W,V)\xrightarrow{\
  \approx\ } \widehat\Mcal_\theta^{\rm ADHM}(W,V)$ is $\widetilde T$-coequivariant.

\subsection{Torus fixed points\label{subsec:Tfixedpoints}}

A fixed point $[E]\in\Mcal_\theta(r,k)^{\widetilde T}$ is an isomorphism
class of a coequivariant sheaf $E$ (see \S\ref{subsec:coeq-sheaves})
which is equipped with a natural coaction of the Hopf algebra $\widetilde \hil_\theta:=\hil_\theta\otimes\hil^{(r)}$; hence $E$ decomposes into a finite direct sum of torsion free $\alg$-modules graded by the character lattice of the torus $\widetilde T$ as
\beq
E=\bigoplus_{p\in L^*}\ \bigoplus_{m\in\zed^r}\, E(p,m) \ .
\label{Edecomp}\eeq
The left $\widetilde\hil_\theta$-coactions $\widetilde\Delta_L:E(p,m)\rightarrow\widetilde\hil_\theta\otimes
E(p,m)$
are given for $f\in E(p,m)$ by
$\widetilde\Delta_L(f)=t^p\otimes\rho^m\otimes f$.

The coaction of $\widetilde\hil_\theta$ on the moduli space naturally
makes the vector spaces $V$ and $W$ into objects of the monoidal category
${}^{\widetilde\hil_\theta}\Module$. By definition, as $\widetilde\hil_\theta$-comodules there are
decompositions $W=\bigoplus_l\, W_l$ with $W_l=\complex w_l$ and
\beq
V= \bigoplus_{p\in L^*}\ \bigoplus_{m\in\zed^r}\, V(p,m) \ .
\label{Vfixedptdecomp}\eeq
The maps (\ref{BIJsubvar}) are morphisms in the category
${}^{\widetilde\hil_\theta}\Module$, i.e. there are commutative
diagrams
$$
\xymatrix{
V ~
\ar[r]^{\!\!\!\!\!\!\!\!\!\!\!\!\!\!\! B}
\ar[d]_{\widetilde\Delta_L} & ~ 
V\otimes \big(\alg_\ell^!\big)_1
\ar[d]^{\widetilde\Delta_L} \\
\widetilde\hil_\theta\otimes V
\ar[r]_{\!\!\!\!\!\!\!\!\!\!\!\!\!\!\! \Id\otimes B} & ~
\widetilde\hil_\theta\otimes V\otimes\big(\alg_\ell^!\big)_1
} \ , \qquad
\xymatrix{
W ~
\ar[r]^{\!\!\!\!\!\!\!\!\!\!\!\!\!\!\! I}
\ar[d]_{\widetilde\Delta_L} & ~ 
V\otimes \big(\alg_\ell^!\big)_2
\ar[d]^{\widetilde\Delta_L} \\
\widetilde\hil_\theta\otimes W ~
\ar[r]_{\!\!\!\!\!\!\!\!\!\!\!\!\!\!\! \Id\otimes I} & ~
\widetilde\hil_\theta\otimes V\otimes \big(\alg_\ell^!\big)_2
} \ ,
$$
and
$$
\xymatrix{
V ~
\ar[r]^{\!\!\!\!\! J}
\ar[d]_{\widetilde\Delta_L} & ~ 
W \ar[d]^{\widetilde\Delta_L} \\
\widetilde \hil_\theta\otimes V ~
\ar[r]_{\!\!\!\!\! \Id\otimes J} & ~ \widetilde \hil_\theta\otimes W
} \ .
$$
From these commutative diagrams it follows that the only non-trivial
components of the linear maps $B_1,B_2,I,J$ with respect to the
character decomposition (\ref{Vfixedptdecomp}) are given by
\bea
B_i(p,m)& : & V(p,m) \ \longrightarrow \ V(p-e_i^*,m) \ , \nonumber
\\[4pt]
I_l & : & W_l\ \longrightarrow \
V(-e_1^*-e_2^*,g_l^*)\otimes\big(\alg_\ell^!\big)_2 \ , \nonumber
\\[4pt]
J_l & : & V(0,g_l^*)\ \longrightarrow \ W_l
\label{BIJchardecomp}\eea
for $i=1,2$, $l=1,\dots,r$, $p=p_1\, e_1^*+ p_2\, e_2^*\in L^*$, and $m=\sum_l\, m_l\,
g_l^*\in\zed^r$. In particular, we have
\beq
I(W)\ \subseteq \ \bigoplus_{l=1}^r\, V(-e_1^*-e_2^*,g_l^*)
\otimes\big(\alg_\ell^!\big)_2 \ .
\label{IWsubset}\eeq
Moreover, from the braided ADHM equation (\ref{NCADHMeqs}) we have the
relations
\beq
B_2(p-e_1^*,g_l^*)\circ B_1(p,g_l^*)= q^2\ B_1(p-e_2^*,g_l^*)\circ
B_2(p,g_l^*)
\label{B1B2pqcomm}\eeq
in $\Hom_\complex\big(V(p,g_l^*)\,,\, V(p-e_1^*-e_2^*,g_l^*)\big)$ for
any $p\neq0$ and $l=1,\dots,r$.

Using the isomorphism $\big(\alg_\ell^!\big)_2\cong\complex$, we
define the subspace $V'\subseteq V$ by
\beq
V'=\bigoplus_{(i_1,\dots,i_n)}\, B_{i_1\dots i_n}\big({\rm im}(I)\big) \ ,
\label{VprimeBI}\eeq
where the sum runs through all finite ordered collections $(i_1,\dots,i_n)$
of indices $i_a=1,2$ of arbitrary length, and $B_{i_1\dots i_n}:=B_{i_n}\,
B_{i_{n-1}}\cdots B_{i_1}$; to the empty collection we
assign $B_\emptyset:=\Id_V$. Then $B_i(V'\, )\subseteq V'${, $i=1,2$,} and ${\rm
  im}(I)\subseteq V'$. From (\ref{BIJchardecomp}) and (\ref{IWsubset}) it
follows that
$$
V'\ \subseteq\ \bigoplus_{p_1,p_2\leq -1}\ \bigoplus_{l=1}^r\,
V(p,g_l^*) \ .
$$
Hence by the stability condition, $V(p,m)=0$ in (\ref{Vfixedptdecomp})
if $p_1\geq0$ or $p_2\geq0$ or $m\neq g_l^*$, for otherwise $V'$ would
be a proper destabilizing subspace of $V$. It follows from
(\ref{BIJchardecomp}) that $J=0$ for any fixed
point $[(B,I,J)] \in\widehat\Mcal_\theta^{\rm ADHM}(W,V)^{\widetilde T}$,
and the character decomposition (\ref{Vfixedptdecomp}) truncates to
\beq
V=\bigoplus_{p_1,p_2\leq-1}\ \bigoplus_{l=1}^r\, V_l(p)
\label{Vtruncdecomp}\eeq
with $V_l(p):=V(p,g_l^*)$ and $V=V'$. By \S\ref{Rank1}, points of
$\widehat\Mcal_\theta^{\rm ADHM}(W,V)^{\widetilde T}$ correspond
bijectively to ideals of codimension $k=\dim_\complex(V)$ in the
affine coordinate algebra $\alg(\complex_\theta^2)$. In the following
we denote the finite set of lattice points
$\lambda^l:=\big\{(p_1,p_2)\in\nat^2\ \big|\ V_l(-p)\neq0\big\}$ for
each $l=1,\dots,r$.

Since $\dim_\complex(W_l)=1$ for $l=1,\dots,r$, using
(\ref{B1B2pqcomm}) and $\complex$-linearity the above argument also
implies
$$
V_l(0)=\complex \, I_l(1) \ , \qquad V_l(-p)=\complex\, B_{1,l}^{p_1}\, B_{2,l}^{p_2}\,
I_l(1)
$$
for any $p=(p_1,p_2)\in\lambda^l$ and $l=1,\dots,r$, where we have
equated (\ref{VprimeBI}) with (\ref{Vtruncdecomp}) and set
\begin{eqnarray*}
B_{1,l}^{p_1}&:=& B_1\big(-p_1\, e_1^*-p_2\, e_2^*,g_l^*\big)\circ B_1
\big(-(p_1-1)\, e_1^*-p_2\, e_2^*,g_l^*\big) \\ && \qquad \qquad
\qquad \circ \cdots \circ
B_1\big(-e_1^*-p_2\, e_2^*,g_l^*\big) \ , \\[4pt]
B_{2,l}^{p_2}&:=& B_2\big(-e_1^*-p_2\, e_2^*,g_l^*\big)\circ B_2
\big(-e_1^*-(p_2-1)\, e_2^*,g_l^*\big) \circ \cdots \circ
B_2\big(-e_1^*-e_2^*,g_l^*\big) \ .
\end{eqnarray*}
In particular, $\dim_\complex\big(V_l(-p)\big)=1$ for all
$p\in\lambda^l$ and $l=1,\dots,r$. Combined with the expression for
the universal sheaf $\hat\bun$ given by Theorem~\ref{modulispthm}, it
follows that the
character decomposition (\ref{Edecomp}) truncates to
$$
E=\bigoplus_{l=1}^r\, \Ical_l \ ,
$$
where $\Ical_l=\bigoplus_{p\in L^*}\, E(p,g_l^*)$ for each
$l=1,\dots,r$ is an $\hil_\theta$-coequivariant torsion free sheaf of
rank one on $\Open(\CP_\theta^2)$. It thus suffices to focus on the
case $r=1$ and look for a combinatorial characterization of the finite
lattice $\lambda\subset\nat^2$; in this instance $\widetilde T=T=
(\complex^\times)^2$.

We first note that by applying $B_i$ to vectors of the form
$B_i^{p_i}\, I(1)$, we may conclude that
$$
\dim_\complex\big(V(-p_1,0)\big) \geq
\dim_\complex\big(V(-p_1-s_1,0)\big) \ , \qquad
\dim_\complex\big(V(0,-p_2)\big) \geq
\dim_\complex\big(V(-p_2-s_2)\big)
$$
for all $s_1,s_2\geq1$, where these dimensions are all either zero or
one. Given $p\in\lambda$, so that $V(-p)\cong\complex$, consider now
the (\emph{not} commutative) diagram
$$
\xymatrix{
V(-p) ~
\ar[r]^{\!\!\!\!\!\!\!\!\!\!\!\!\!\!\! B_1}
\ar[d]_{B_2} & ~ 
V(-p-e_1^*)
\ar[d]^{B_2} \\
V(-p-e_2^*)
\ar[r]_{\!\!\!\!\!\!\!\!\!\!\!\!\!\!\! B_1} & ~
V(-p-e_1^*-e_2^*)
}
$$
and use the braided commutation relation $B_2\,B_1(1)=q^2\ B_1\,
B_2(1)$ to deduce the allowed configurations of non-trivial vector
spaces around each such square. For example, the configurations with
$V(-p-e_1^*)=0$, $V(-p-e_2^*)\cong\complex \cong V(-p-e_1^*-e_2^*)$
and $V(-p-e_2^*)=0$, $V(-p-e_1^*)\cong\complex \cong
V(-p-e_1^*-e_2^*)$ are forbidden by the braided commutation
relation. On the other hand, configurations with
$V(-p-e_i^*)\cong\complex$ for $i=1,2$ consistently allow for
$V(-p-e_1^*-e_2^*)$ to have dimension zero or one. It follows from
these conditions that the finite lattice $\lambda\subset\nat^2$
defines a Young diagram oriented as in~\cite{Nakajima},
i.e. if $p=(p_1,p_2)\in\lambda$, then $p'\in\lambda$ for all integral
points $p'=(p_1',p_2')$ with $1\leq p_1'\leq p_1$ and $1\leq p_2'\leq
p_2$. The total number of points in $\lambda$ is denoted $|\lambda|=
\sum_i\, \lambda_i$, where $\lambda_i$ is the number of points in the
$i$-th column of $\lambda$. In the general case $r\geq1$, we have thus
proven the following result.
\begin{proposition}
The $\widetilde T$-fixed locus $\Mcal_\theta(r,k)^{\widetilde T}$ is a finite
set of points in bijective correspondence with length $r$ sequences
$\vec\lambda=(\lambda^1,\dots,\lambda^r)$ of Young diagrams
$\lambda^l$ of size $|\,\vec\lambda\, |=k$, where
$$
|\,\vec\lambda\, | := \sum_{l=1}^r\, |\lambda^l| \ .
$$
\label{tildeTfixed}\end{proposition}

This result coincides with that of the classical case
$\theta=0$~\cite[Prop.~2.9]{NakYosh}. It can be understood in terms of
the noncommutative toric geometry as
follows. By~\cite[Prop.~4.15]{CLSI}, $\hil_\theta$-coequivariant ideal
sheaves on $\Open(\CP_\theta^2)$ are in bijective correspondence with
$L^*$-graded subschemes of $\CP_\theta^2$, where $L^*$ is the
character lattice of $T=(\complex^\times)^2$; in particular,
irreducible subschemes correspond to prime ideals in the spectrum of
the homogeneous coordinate algebra $\alg$. Moreover, the
$\hil_\theta$-coinvariant ideals $\Ical \subset\alg$ are \emph{monomial} ideals. If $\Ical$ obeys the condition (\ref{DNVdBHilb}), then it determines a finite partition $\lambda_{\Ical}(k)$ of $k$ by considering lattice points corresponding to monomials \emph{not} contained in $\Ical$.

\subsection{Instanton partition functions}

In the classical case, instanton partition functions of topologically twisted supersymmetric
gauge theories are given by integrating suitable
characteristic classes over the instanton moduli spaces. The
equivariant partition function is then the generating function for the
integral
$$
Z_{\rm inst}(r;Q, z)=\sum_{k=0}^\infty\, Q^k\
\int_{\Mcal_\theta(r,k)^{\widetilde T}}\, \omega(z) \ ,
$$
where $Q$ is a formal variable and $\omega(z)$ is an equivariant cohomology class depending
on the canonical generators $z$ of the coordinate algebra of $\widetilde T$. The integral is evaluated formally by applying the localization
theorem in equivariant cohomology, hence
$\int_{\Mcal_\theta(r,k)^{\widetilde T}}\, \omega(z)$ is a rational
function in the coordinate algebra $\alg(\widetilde T)$.

From \S\ref{subsec:Tfixedpoints} it follows that the equivariant
characters of the $\widetilde\hil_\theta$-comodules $V$ and $W$ are given
by
$$
\ch_{\widetilde T}(V)=\sum_{l=1}^r\
\sum_{p\in\lambda^l}\,\rho_l\ t_1^{1-p_1}\, t_2^{1-p_2} \ , \qquad
\ch_{\widetilde T}(W)=\sum_{l=1}^r\, \rho_l \ ,
$$
as in the classical case~\cite{NakYosh}. By (\ref{DeltaLdual}), the restriction
$\underline{\Iscr}\,_\bullet(E)\big|_{\vec\lambda}$ of the complex of
Theorem~\ref{defcomplexthm} to a $\widetilde T$-fixed point $\vec\lambda$
is a complex in the category ${}^{\widetilde\hil_\theta}\Module$. By
Proposition~\ref{tildeTfixed}, the computation of the $\widetilde T$-equivariant character of the tangent bundle over the instanton
moduli space thus proceeds exactly as in the classical case
(see~\cite{FlumePog} and~\cite[Thm.~2.11]{NakYosh}). At a fixed point
parametrized by a length $r$ sequence
$\vec\lambda=(\lambda^1,\dots,\lambda^r)$ of Young diagrams with
$|\,\vec\lambda\,|=k$, one has
\bea
\ch_{\widetilde T}\big(T_{\vec\lambda}\Mcal_\theta(r,k) \big) =
\sum_{l,l'=1}^r\, \rho_l^{-1}\, \rho_{l'}\, \Big(\,
\sum_{p\in\lambda^l}\, t_1^{p_1-(\lambda^{l'})^t_{p_2}}\,
t_2^{\lambda_{p_1}^l-p_2+1} +
\sum_{p'\in\lambda^{l'}}\, t_1^{(\lambda^l)_{p_2'}^t-p_1'+1}\,
t_2^{p_2'-\lambda_{p_1'}^{l'}}\, \Big) \nonumber
\eea
where $\lambda_j^t$ denotes the number of points in the $j$-th row of
the Young diagram $\lambda$. In particular, the corresponding
equivariant Euler class of the normal bundle to the fixed point is
given by
\bea
\mbox{$\bigwedge_{-1}$}T_{\vec\lambda}^*\Mcal_\theta(r,k) &=&
\prod_{l,l'=1}^r\ \prod_{p\in\lambda^l}\, \Big( 1-\rho_l\, \rho_{l'}^{-1}\, t_1^{(\lambda^{l'})^t_{p_2}-p_1}\,
t_2^{p_2-\lambda_{p_1}^l-1} \Big) \nonumber\\ && \qquad \qquad \times\
\prod_{p'\in\lambda^{l'}}\, \Big( 1-\rho_l\, \rho_{l'}^{-1}\, t_1^{p_1'-(\lambda^l)_{p_2'}^t-1}\,
t_2^{\lambda_{p_1'}^{l'}-p_2'} \Big) \ . \nonumber
\eea

With these ingredients one can now write down the standard equivariant
instanton partition functions for supersymmetric gauge theories
(without matter fields). The equivariant cohomology class $\omega(z)$ is, by
the localization theorem, given by the pullback of a class
$\tilde\omega$ on $\Mcal_\theta(r,k)$ which descends from an
equivariant class, evaluated at the fixed point
$\vec\lambda$, and divided by the Euler character
$\bigwedge_{-1}T_{\vec\lambda}^*\Mcal_\theta(r,k)$. For example, for
$\tilde\omega=1$ we reproduce Nekrasov's partition
function~\cite{nekrasov} for pure $\Ncal=2$ gauge theory
$$
Z_{\rm inst}^{\Ncal =2}(r;Q,z)= \sum_{k=0}^\infty\
\sum_{\vec\lambda\, :\, |\,\vec\lambda\,|=k}\
\frac{Q^{|\,\vec\lambda\,|}}{\mbox{$\bigwedge_{-1}$}
    T_{\vec\lambda}^*\Mcal_\theta(r,k)} \ .
$$
Another standard example is obtained by taking $\tilde\omega$ to be
the Euler class of the tangent bundle over $\Mcal_\theta(r,k)$; in
this case $\omega(z)=1$ (independently of the equivariant parameters~$z$) and the localization
integral simply counts the fixed points of the $\widetilde T$-action on
the instanton moduli spaces. This results in the Vafa--Witten partition function~\cite{VafaWitten}
for $\Ncal=4$ gauge theory
\beq
Z_{\rm inst}^{\Ncal=4}(r;Q)=\sum_{\vec\lambda}\, Q^{|\,\vec\lambda\,|}
= \prod_{n=1}^\infty\, \frac1{\big(1-Q^n\big)^r} \ .
\label{VWpartfn}\eeq
For $r=1$, we can compare the ``bosonic'' partition function
(\ref{VWpartfn}) enumerating torus fixed points with the ``fermionic'' partition function
(\ref{Dcalgenfn}) which counts Hilbert series stratifications of the
instanton moduli spaces. It would be interesting to combine these two
partition functions in the non-equivariant case in order to truly
capture the generic differences between the moduli spaces
$\Mcal_\theta(r,k)$ for $\theta=0$ and $\theta\neq0$ away from the
torus fixed points.

\vfill\eject

\end{document}